\RequirePackage{fix-cm}

\documentclass[smallextended]{svjour3}       

\smartqed  

\usepackage{graphicx}
\usepackage{amsmath}
\usepackage{listings}
\usepackage{mdframed}
\usepackage{multirow}
\usepackage{tikz}
\usetikzlibrary{arrows.meta, calc, shapes}
\usepackage{makecell}
\usepackage{subcaption}
\usepackage{rotating}
\usepackage{tcolorbox}
\usepackage{enumitem}
\usepackage{adjustbox}
\usepackage{xspace}
\usepackage{xcolor}
\usepackage{pifont}
\usepackage{booktabs}
\usepackage{url}
\usepackage[colorlinks=true,linkcolor=blue,citecolor=blue,urlcolor=blue]{hyperref}
\usepackage{natbib}

\definecolor{codegreen}{rgb}{0,0.6,0}
\definecolor{codegray}{rgb}{0.5,0.5,0.5}
\definecolor{codepurple}{rgb}{0.58,0,0.82}
\definecolor{codeblue}{rgb}{0.0, 0.0, 0.55}

\newcommand{\header}[1]{\smallskip\noindent\textbf{#1}}
\newcommand{\rqi}{The Exploratory Analysis: How does LLM-generated code differ from human-written code across models and granularities?\xspace}
\newcommand{\rqii}{The Detection: How does detectability vary across LLMs and code granularities?\xspace}
\newcommand{\rqiii}{The Explanatory Analysis: Which features distinguish LLM-generated from human-written code, and how do detection patterns generalize?\xspace}
\newcommand{\rqiv}{Adversarial Robustness of Structural Classifiers: To what extent can the structural detectors be evaded by adversarial transformation of their dominant features?\xspace}
\newcommand{\etal}{\textit{et al.}}

\journalname{Empirical Software Engineering}

\begin{document}

\title{Automatic Detection of LLM-Generated Code: A Comparative Case Study of Contemporary Models Across Function and Class Granularities}

\author{Musfiqur Rahman \and 
        SayedHassan Khatoonabadi \and 
        Ahmad Abdellatif \and 
        Emad Shihab}

\institute{Musfiqur Rahman \at
           Concordia University, Montr\'{e}al, QC, Canada \\
           \email{musfiqur.rahman@mail.concordia.ca}
           \and
           SayedHassan Khatoonabadi \at
           Concordia University, Montr\'{e}al, QC, Canada \\
           \email{sayedhassan.khatoonabadi@concordia.ca}
           \and
           Ahmad Abdellatif \at
           University of Calgary, Calgary, AB, Canada \\
           \email{ahmad.abdellatif@ucalgary.ca}
           \and
           Emad Shihab \at
           Concordia University, Montr\'{e}al, QC, Canada \\
           \email{emad.shihab@concordia.ca}}

\date{Received: date / Accepted: date}

\maketitle

\begin{abstract}
The widespread adoption of Large Language Models (LLMs) for code generation introduces risks of incorporating suboptimal or vulnerable code into software systems. While detection mechanisms exist, they face two critical limitations: existing detectors are typically optimized for specific LLMs without systematic cross-model validation, and they operate as black boxes that identify machine-generated code without explaining the underlying structural reasons. To address these challenges, we present a comparative study examining the detectability of Python code generated by four distinct LLMs: GPT-3.5, Claude 3 Haiku, Claude Haiku 4.5, and GPT-OSS.

We analyzed 14,485 Python functions and 11,913 classes from the CodeSearchNet dataset, generating corresponding code with all four LLMs at both function and class granularities. Using 18 function-level and 39 class-level interpretable software metrics, we trained CatBoost classifiers for each model-granularity configuration. Our analysis reveals that granularity effects dominate model differences by a factor of 8.6, with negligible feature overlap between levels. This indicates that function-level and class-level detection rely on fundamentally disjoint structural signatures.

We discover critical granularity-dependent inversions: while modern models (Claude models, GPT-OSS) are more detectable at the class level, GPT-3.5 is an anomaly that uniquely excels at the function level. Through SHapley Additive exPlanations (SHAP), we identify the \emph{Comment to Code Ratio} as the sole universal discriminator. However, its predictive magnitude varies drastically across models, explaining why detectors trained to detect code generated by a specific LLM fail to generalize. Targeted adversarial perturbation of the dominant comment features causes one of the eight classifiers (GPT-OSS at the class level) to collapse to near-random performance while the other seven remain unaffected or improved, indicating that vulnerability to feature-targeting attacks is also configuration-specific.

Our findings reassess current practices by demonstrating that GPT-3.5's exceptional detectability (AUC-ROC 0.96) is unrepresentative of other contemporary models (AUC-ROC $\approx$ \textcolor{blue}{0.68--0.83}). We conclude that robust detection requires moving beyond single-model studies to account for the substantial diversity in structural fingerprints across architectures and granularities, and for the configuration-specific adversarial vulnerabilities that arise from feature-importance concentration.

\keywords{Large Language Models \and Comparative Analysis \and Code Detection \and Multi-Model Study \and Code Granularity \and Adversarial Robustness \and Machine Learning \and Empirical Software Engineering}
\end{abstract}

\section{Introduction}

Large Language Models (LLMs) have become integral to modern software engineering, demonstrating capabilities across diverse tasks including code generation~\citep{chen2021evaluating,austin2021program}, code completion~\citep{fried2022incoder,brown2024identifying}, program repair~\citep{fan2023automated,xia2023keep}, test generation~\citep{schafer2023empirical,fan2023large}, code review~\citep{vijayvergiya2024ai}, code documentation~\citep{dvivedi2024comparative}, and code summarization~\citep{ahmed2022few}. Their widespread adoption in development workflows has fundamentally transformed how software is created~\citep{karpathy2017software,dilhara2021understanding,carbin2019overparameterization}. However, this growing reliance on LLM-generated code introduces critical risks: existing literature reports that LLMs can provide vulnerable code~\citep{pearce2022asleep,toth2024llms,wang2023enhancing,khoury2023secure,siddiq2023generate,wang2024your,tihanyi2025secure}, and the ownership of LLM-generated code remains legally ambiguous~\citep{kahveci2023attribution,srivastava2025legal,amon2024uncertain,lucchi2024chatgpt,manolakev2017works}. With many open source software (OSS) packages such as \textit{npm} and \textit{PyPI} being incorporated into commercial applications, the risk of unintentional inclusion of LLM-generated code becomes a pressing concern. These issues necessitate robust mechanisms for accurate detection of LLM-generated code to facilitate code review and maintain software quality.

Recent work has begun addressing LLM-generated code detection, with approaches such as CodeGPTSensor~\citep{xu2025distinguishing}, GPTSniffer~\citep{nguyen2024gptsniffer}, and DetectCodeGPT~\citep{shi2025between} targeting ChatGPT-generated code, while other studies have examined detection in competitive programming contexts~\citep{idialu2024whodunit}. However, these existing studies share a common limitation: they focus on individual LLMs without systematically comparing detectability across models, which is a critical gap because (1) different LLMs may exhibit fundamentally different generation patterns that affect detection strategies, and (2) detectors optimized for one model may not generalize to others~\citep{suh2024empirical}, limiting practical deployment. Furthermore, while LLMs are used to generate both individual functions~\citep{chen2021evaluating,austin2021program} and entire classes~\citep{du2023classeval,rahman2025beyond}, existing detection studies have primarily focused on function-level code rather than examining how code granularity affects detection patterns in real-world software projects.

The practical significance of this granularity gap follows from the fact that different software-development workflows encounter LLM-generated code at different granularities, each of which demands granularity-appropriate detection. Pull-request and commit-level code review, the most common context for LLM-detection in industrial practice, typically operates at the function or method level: individual changes are naturally scoped to the function level, and reviewer attention is allocated function-by-function. Architectural code reviews and codebase audits, in contrast, operate at the class or file level, where the unit of analysis is the design of an abstraction rather than the implementation of a specific behaviour. Educational and academic-integrity workflows span both granularities, since student submissions may consist of standalone functions or full class implementations. A detection system that performs well at one granularity but poorly at the other is therefore not a partial solution but a workflow-specific one; practitioners need empirical evidence about how detection performance and the relevant structural signals vary with granularity in order to deploy classifiers that match their review workflow. This consideration motivates our explicit investigation of function-level versus class-level detection.

To bridge these gaps, our objective in this work is to conduct a systematic comparative study of LLM-generated Python code detection using interpretable structural software metrics. Specifically, we aim to understand: (1) whether detection patterns generalize across different LLMs or require model-specific approaches, (2) how code granularity (functions vs. classes) affects detectability, (3) which structural features enable interpretable detection that explains \emph{why} code appears LLM-generated rather than merely identifying it, and (4) to what extent the structural detectors can be evaded by adversarial transformation of their dominant detection features. To address these questions, we conduct our analysis across four established LLMs: GPT-3.5~\citep{openaiOpenAIPlatform}, GPT-OSS~\citep{openaiIntroducingGptoss}, Claude 3 Haiku~\citep{anthropicClaudeHaiku}, and Claude Haiku 4.5~\citep{anthropicIntroducingClaude}. These specific LLMs are selected based on their popularity~\citep{nguyen2023snippet,nguyen2024gptsniffer,idialu2024whodunit,xu2025distinguishing} and demonstrated code generation capabilities~\citep{swebenchSWEbenchLeaderboards}. By including multiple models and examining both function-level and class-level Python code from real-world open-source projects (sourced from CodeSearchNet~\citep{husain2019codesearchnet}), we perform a multi-dimensional analysis to determine how detection patterns vary across these factors.

Our restriction to Python is a deliberate methodological choice, elaborated in Section~\ref{sec:dataset}. Our granularity-comparison design requires a language in which functions and classes can be analyzed as syntactically independent code units, which rules out both class-mandatory languages (where every function is a method of an enclosing class, e.g., Java and C\#) and class-less languages (where no class construct exists, e.g., C). Python satisfies this requirement and is the dominant language in the LLM code-generation research community, keeping our results directly comparable with prior detection studies that uniformly target Python.

We extract structural software metrics from human-written and LLM-generated code at both granularities. We train binary classifiers to distinguish human-written from LLM-generated code for each model-granularity combination and evaluate their detection performance. We then analyze which features most strongly discriminate between human and LLM-generated code and examine whether these discriminative patterns generalize across models and granularities through feature importance and overlap analysis. Finally, we test the adversarial robustness of the trained classifiers by removing comment features---the dominant discriminator identified through feature importance analysis---from LLM-generated samples and measuring detection performance on the modified samples. In summary, we aim to answer the following research questions (RQs):

\begin{itemize}

    \item[\textbf{RQ1:}] \textbf{How does LLM-generated code differ from human-written code across models and granularities?}
    We perform statistical analysis comparing distributions of structural features between human-written and LLM-generated code across all model-granularity configurations. We find that LLM-generated code exhibits statistically significant differences ($p < 0.001$) in 72-89\% of features depending on configuration, with particularly strong divergence with medium to strong effect sizes in stylometric features such as the ratio of comment lines to code lines and blank line usage. However, the magnitude and direction of these differences vary substantially across both models and granularities, indicating that each configuration produces distinct structural signatures.
    
    \item[\textbf{RQ2:}] \textbf{How does detectability vary across LLMs and code granularities?} We train classifiers for each of the eight model-granularity configurations and evaluate their detection performance. Detection accuracy ranges from \textcolor{blue}{0.68} to 0.96 AUC-ROC, with GPT-3.5 exhibiting exceptional detectability (0.96 for functions, 0.89 for classes) compared to other models (\textcolor{blue}{0.68-0.83}). We discover granularity-dependent ranking inversions: three models (Claude 3 Haiku, Claude Haiku 4.5, GPT-OSS) show improved detectability at class-level, while GPT-3.5 uniquely achieves superior performance at function-level. This \textcolor{blue}{28.0} percentage-point performance gap between GPT-3.5 and other models challenges assumptions about detector generalization.
    
    \item[\textbf{RQ3:}] \textbf{Which features distinguish LLM-generated from human-written code, and how do detection patterns generalize?} We analyze feature importance through SHAP values and examine feature overlap across configurations. \emph{Comment to Code Ratio} emerges as the sole universal discriminator, appearing in the top-3 features for 7 of 8 configurations, though its SHAP importance varies dramatically between 0.178 and 3.795 across configurations. Feature overlap analysis reveals that granularity dominates model effects: overlap is 8.6 times higher across models at the same granularity than within models across different granularities. Cross-granularity Jaccard similarity is only 0.099, indicating that function-level and class-level detection rely on almost entirely different structural signatures.

    \item[\textbf{RQ4:}] \textbf{To what extent can the structural detectors be evaded by adversarial transformation of their dominant features?} We evaluate the trained classifiers under a post-hoc adversarial transformation that removes all comment features---the dominant discriminator identified in RQ3---from LLM-generated samples. We find that adversarial robustness is highly asymmetric across the eight model-granularity configurations: seven of the eight classifiers are either unaffected or made more accurate by comment stripping, while one (GPT-OSS at the class level) collapses from AUC 0.808 to 0.490, statistically indistinguishable from random. The asymmetry indicates that adversarial vulnerability is a configuration-specific property that does not align with baseline detection performance.
\end{itemize}

\textcolor{blue}{These four research questions form a single connected investigation rather than four independent analyses. RQ1 asks whether LLM-generated code differs structurally from human-written code at all, and if so, in what direction, for each model. RQ2 builds on RQ1 by asking whether any such differences are large enough to support automated detection. RQ3 goes a level deeper: rather than only measuring how well detection works, it asks which specific structural features are responsible, and whether those features generalize across models and across granularities. RQ4 asks the natural adversarial follow-up to RQ3: if detection depends on identifiable features, can a developer who knows which features matter defeat the detector by manipulating them? Together, the four questions trace a single arc: whether a difference exists, whether it is usable, why it is usable, and whether that usability can be defeated.}

Our findings demonstrate that established LLMs generate code with unique structural characteristics that enable detection, but detectability, discriminative patterns, and adversarial robustness vary significantly across models and granularities. We organize our analysis around three interconnected dimensions that we revisit throughout each research question: (1) how detection signatures differ across models, (2) how granularity affects these signatures, and (3) whether universal detection across models and granularities is achievable. This recurring analytical framework allows us to build a comprehensive understanding of detection challenges by examining each research question through these three fundamental lenses.

\header{Our Contributions.} We make the following contributions:

\begin{itemize}[leftmargin=*]
    \item To the best of our knowledge, this is the first comparative study examining LLM-generated code detection across multiple established models and code granularities (function and class) using interpretable structural software metrics.
    
    \item We provide empirical evidence that GPT-3.5's exceptional detectability is not representative of other established models, with performance gaps up to \textcolor{blue}{28.0} percentage points.
    
    \item We discover that code granularity affects detection signatures more strongly than model architecture differences.
    
    \item We demonstrate that testing detectors on unseen LLMs without retraining measures robustness to distribution shift rather than intrinsic detector quality.

    \item We empirically characterize the adversarial robustness of structural LLM-code detectors under feature-targeting attacks, identifying an asymmetric vulnerability pattern governed by feature-dominance and the direction of LLM-human structural differences.
    
    \item To promote reproducibility and facilitate future research, we publicly share our dataset, analysis scripts, and comprehensive results (see the~\nameref{sec:replication-package} section).
\end{itemize}

\header{Paper Organization.}
The rest of this paper is organized as follows. Section~\ref{sec:dataset} describes our data curation, code generation, and feature extraction. Sections~\ref{sec:rqi} and~\ref{sec:rqii} present our analysis of code differences and detection performance. Section~\ref{sec:rqiii} examines feature importance and generalization patterns. Section~\ref{sec:rqiv} evaluates the adversarial robustness of the trained classifiers under a feature-targeting attack. Section~\ref{sec:discussion} discusses implications, while Sections~\ref{sec:limitations},~\ref{sec:threats}, and~\ref{sec:related_work} address limitations, validity threats, and related work. Section~\ref{sec:conclusion} concludes the paper.

\section{Dataset and Methodology}
\label{sec:dataset}

To compare human-written code with code generated by LLMs, we need a data source containing code already authored by human programmers so that we can generate corresponding code using different LLMs for the same tasks. In this section, we explain the scope of this paper, how we choose our data source, generate corresponding code from four contemporary LLMs, and ensure fair cross-model comparison through careful data curation.

\subsection{Scope of the Study}
Our restriction to Python is motivated by three considerations: the structural requirement of a clean granularity comparison, the technical constraints of our metric extraction tool, and the practical requirement of comparability with the existing LLM-detection literature. We detail each below.

\textbf{Structural requirement.} Our study compares detection at the function level and at the class level, treating the two granularities as comparable but distinct units of analysis. This requires both constructs to exist as syntactically independent code units in the language under study, and not as one nested inside the other. Java, C\#, and similar object-oriented languages have classes but lack standalone functions: every function is a method of some class, so function-level samples cannot be isolated from class context without violating the language's syntactic constraints, whereas C and similar procedural languages lack a class construct entirely. 

\textbf{Technical constraint of metric extraction.} The structural requirement above is sharpened by how structural-metric tools handle method-versus-function ambiguity. \textit{Understand} by \textit{SciTools}~\citep{scitoolsUnderstandSoftware}, the metric extractor we use, computes function-level metrics cleanly when the unit is a standalone function. When the unit is a method of a class, several of the metrics extracted at the method level depend on the enclosing class context, for example, coupling-related metrics that count interactions with other methods or instance variables of the enclosing class. A Java or C\# method analyzed through \textit{Understand}, therefore, yields a metric set whose computation is inherently entangled with its class, even when we intend to characterize the method as an isolated unit. This makes a clean function-versus-class granularity comparison difficult to construct in object-oriented languages: the function-level samples would not be truly isolated, and the comparison with class-level samples would be confounded by overlapping context-dependent features.

\textbf{Comparability with existing detection literature.} Several modern languages do satisfy the structural requirement above: JavaScript and TypeScript support both module-level functions and ES6 classes~\citep{ecmainternationalECMAScript2015}; Go and Rust support functions and methods on structs. These languages are technically viable for our study. The decisive factor is comparability with the existing LLM-detection literature. Python is the dominant language in the LLM code-generation research community~\citep{shi2025between,idialu2024whodunit,nguyen2024gptsniffer} and beyond~\citep{gonzalez2020state,rahman2025impact}: it is the most-targeted language by contemporary code-generating LLMs (the GPT, Claude, Llama, Qwen, and DeepSeek model families all benchmark Python performance first), the most heavily represented language in standard code-generation benchmarks (HumanEval~\citep{liu2023your}, MBPP~\citep{austin2021program}, ClassEval~\citep{du2023classeval}), and the language used in essentially all prior LLM-detection studies that we build upon. Restricting our scope to Python aligns the protocol of our study with the established research paradigm while satisfying the structural and technical constraints described above.

\subsection{Data Source}
As mentioned before, we break our analyses into two levels: function-level and class-level. In the following, we explain how we prepare the dataset for these two levels of code.

\subsubsection{Function-level:}
In this work, we choose CodeSearchNet \citep{husain2019codesearchnet} as our data source for function-level code. Widely adopted by prior research \citep{wang2023codet5+,guo2022unixcoder,zhang2023unifying,neelakantan2022text,saberi2023utilization,ahmed2022multilingual,gong2022multicoder,pavsek2022mqdd}, this corpus enables us to study real-life development through over two million functions mined from open-source GitHub repositories across six programming languages. We specifically utilize the dataset's \emph{(comment, code)} pairs, where \emph{code} denotes the human-written function body and \emph{comment} denotes the corresponding top-level docstring, as illustrated in Listing~\ref{lst:example_comment_code_function}. 

\begin{lstlisting}[
    language=Python,
    basicstyle=\ttfamily\small,
    breaklines=true,
    caption={Example of a standalone function from {\tt pysubs2} \citep{githubGitHubTkarabelapysubs2} with its docstring.},
    label=lst:example_comment_code_function,
    captionpos=t,
    commentstyle=\color{codegray},
    keywordstyle=\color{codeblue}\bfseries,
    stringstyle=\color{codegreen},
    showstringspaces=false,
    showspaces=false,
    numbers=none
]
def timestamp_to_ms(groups):
    ``""
    Convert groups from :data:`pysubs2.time.TIMESTAMP` match to milliseconds.
    Example:
    >>> timestamp_to_ms(TIMESTAMP.match("0:00:00.42").groups())
    420
    ``""
    h, m, s, frac = map(int, groups)
    ms = frac * 10**(3 - len(groups[-1]))
    ms += s * 1000
    ms += m * 60000
    ms += h * 3600000
    return ms
\end{lstlisting}

CodeSearchNet has over $500,000$ Python functions with docstrings. For our study, we focus exclusively on standalone functions. These functions have no dependencies on other functions or classes within the same module. We adopt this constraint because, through preliminary experimentation, we observed that when functions have contextual dependencies (e.g., calling other module-level functions or referencing class attributes), older LLMs (such as GPT-3.5 and Claude 3 Haiku) often fail to generate complete implementations and instead return placeholder code or incomplete snippets. This issue is not unique to our study; Xu \etal~\citep{xu2025distinguishing} reported similar challenges when generating code with contextual dependencies. From the available standalone functions, we randomly sample $20,000$ functions to limit processing time and cost associated with code generation across multiple LLMs.

\subsubsection{Class-level:}

Functions and classes are not only fundamentally different software artifacts, but they are also the natural units of analysis in different software-development workflows, such as function-level analysis for pull-request and commit-level review, class-level analysis for architectural reviews and codebase audits, and both for educational and academic-integrity contexts. Detection systems deployed in these workflows must discriminate LLM-generated code at the corresponding granularity, and a detector that works at one granularity does not transfer to the other without empirical validation. To enable a direct comparison across granularities, we curate distinct function-level and class-level datasets that mirror each other's construction protocol as closely as possible while preserving the structural distinctions between the two artifacts. The empirical question of whether detection strategies must be tailored to a specific granularity, or can be generalized across them, is then directly answerable from these two datasets, with practical implications for how detectors should be trained and deployed in different review workflows.

To perform a comparative analysis between function-level and class-level LLM-generated code, we curate our class-level dataset by extracting standalone classes from the same OSS projects that were used in curating the CodeSearchNet dataset. A class is considered standalone when it does not inherit from other classes, and no other classes inherit from it. We apply the same standalone constraint for two main reasons. First, when there is a hierarchical relationship between classes due to inheritance, an LLM needs to be prompted with not only the class definition but also extensive context about the class hierarchy. This makes input prompts arbitrarily long, resulting in higher costs and processing time for code generation. Second, by choosing only standalone classes, we maintain consistency with our function-level dataset, ensuring that both datasets have comparable characteristics and dependency-related complexity.

Following is an example of a standalone class with its docstring from our dataset:

\begin{lstlisting}[
    language=Python,
    basicstyle=\ttfamily\small,
    breaklines=true,
    caption={Example of a standalone class from {\tt pysubs2} \citep{githubGitHubTkarabelapysubs2} with its docstring.},
    label=lst:example_comment_code_class,
    captionpos=t,
    commentstyle=\color{codegray},
    keywordstyle=\color{codeblue}\bfseries,
    stringstyle=\color{codegreen},
    showstringspaces=false,
    showspaces=false,
    numbers=none
]
class Notification:
    ``""Wrapper for notifications.

    In order to listen for notifications, call `activate(callback)`
    with a coroutine to be called when a notification is received.
    ``""

    def __init__(self, endpoint, switch_method, payload):
        ``""Notification constructor.

        :param endpoint: Endpoint.
        :param switch_method: `Method` for switching this notification.
        :param payload: JSON data containing name and available versions.
        ``""
        self.endpoint = endpoint
        self.switch_method = switch_method
        self.versions = payload["versions"]
        self.name = payload["name"]
        self.version = max(x["version"] for x in self.versions if ``version" in x)

        _LOGGER.debug("notification payload: %s", pf(payload))

    def asdict(self):
        ``""Return a dict containing the notification information."""
        return {"name": self.name, ``version": self.version}

    async def activate(self, callback):
        ``""Start listening for this notification.

        Emits received notifications by calling the passed `callback`.
        ``""
        await self.switch_method({"enabled": [self.asdict()]}, _consumer=callback)

    def __repr__(self):
        return ``<Notification {}, versions={}, endpoint={}>".format(
            self.name,
            self.versions,
            self.endpoint,
        )
\end{lstlisting}

Consistent with our function-level approach, we randomly sample 20,000 standalone classes from the same CodeSearchNet projects to perform LLM-based code generation. Note that the selected functions are not part of the classes chosen because any function (technically, a method) part of these classes does not satisfy the condition of being standalone. Hence, our function-level and class-level datasets are independent, allowing us to examine granularity effects without confounding from structural relationships between samples.

\subsection{Code Generation with Multiple LLMs}

We selected four established LLMs for our comparative study: GPT-3.5~\citep{openaiOpenAIPlatform} and GPT-OSS~\citep{openaiIntroducingGptoss} from OpenAI, and Claude 3 Haiku~\citep{anthropicClaudeHaiku} and Claude Haiku 4.5~\citep{anthropicIntroducingClaude} from Anthropic. These models are extensively used to perform various software engineering tasks such as code completion~\citep{li2024evaluating}, program repair~\citep{yu2025patchagent,fruntke2025automatically}, and test generation~\citep{schafer2023empirical}, making them representative of real-world LLM deployment. This selection enables us to analyze detection patterns across two major LLM providers and examine how model evolution (for example, comparing Claude 3 Haiku and Claude Haiku 4.5) affects detectability. 

We use function and class docstrings as part of the prompt sent to each model, and the response received from each model is the corresponding LLM-generated code. We format our prompt as follows:

\noindent{%
    \fbox{\parbox {\linewidth}{%
        \small You're an expert Python programmer. Please generate a Python [FUNCTION or CLASS] from the given docstring. Do not explain the code. The docstring is as follows:
        \newline
        \newline
        \{the [FUNCTION or CLASS] docstring\}
    }%
    }%
}

To reduce the cost of generating code, we added the ``Do not explain the code'' instruction as part of the prompt so that the generated response does not get unnecessarily long. For each of the $20,000$ functions and $20,000$ classes, we generate code using all four LLMs with identical prompts, resulting in four versions of generated code, ideally for each human-written snippet.

\subsection{Global Intersection Methodology}

A fundamental challenge in comparative LLM studies is ensuring that differences in detection performance reflect actual model characteristics rather than artifacts of unequal dataset composition—that is, if Model A successfully generates code for easy tasks while Model B handles harder tasks, lower detectability for Model B might simply reflect task difficulty rather than generation quality. To address this, we implement a strict intersection-based sampling strategy. We explain the strategy below.

Not all code snippets are generated with valid, complete implementations by all four LLMs. Some models may return placeholder code (e.g., \texttt{pass} statements with comments like ``implementation goes here") or incomplete snippets. We define successful generation as cases where the LLM API returns a complete code implementation with valid Python syntax. We verify completeness by: (1) parsing the generated code with Python's \texttt{ast}~\citep{pythonAbstractSyntax} module to confirm syntactic validity and (2) checking that the implementation contains executable statements beyond placeholders using regular expressions.

Through our generation process, we observe varying success rates across models. For example, suppose a particular function is successfully generated by Claude 3 Haiku, GPT-3.5, and GPT-OSS, but Claude Haiku 4.5 fails to generate it (returning placeholder code). In that case, we exclude this function from our final dataset. This ensures that we only retain code snippets that were successfully generated by all four LLMs. This intersection-based approach is essential for fair comparison: if we used different sets of functions or classes for different LLMs, the inherent characteristics of the snippets themselves would introduce an additional layer of variability, making it impossible to attribute detection performance differences to model-specific patterns.

Table~\ref{fig:data_collection_flowchart} illustrates our data collection and filtering process. After applying the intersection filter, our final dataset consists of $14,485$ functions and $11,913$ classes, where each snippet has five versions: one human-written version and four LLM-generated versions (one from each model). This results in a total of $72,425$ function-level code snippets and $59,565$ class-level code snippets across all configurations.

\begin{table}[t]
    \centering
    \caption{Data collection and filtering process.}
    \label{fig:data_collection_flowchart}
    \begin{tabular}{|c||c|c|}
        \hline
        \textbf{Stage} & \textbf{Function-Level} & \textbf{Class-Level} \\
        \hline
        \textbf{1. Data Source} & CodeSearchNet Dataset & Classes from same OSS Projects \\
        \hline
        \textbf{2. Initial Sample} & 20,000 standalone functions & 20,000 standalone classes \\
        \hline
        \textbf{3. Code Generation} & \multicolumn{2}{c|}{Generate with 4 LLMs: GPT-3.5, GPT-OSS,} \\
         & \multicolumn{2}{c|}{Claude 3 Haiku, Claude Haiku 4.5} \\
        \hline
        \textbf{4. Validation} & \multicolumn{2}{c|}{Verify valid Python syntax using \texttt{ast} parsing} \\
        \hline
        \textbf{5. Intersection Filter} & \multicolumn{2}{c|}{Keep only snippets successfully generated by ALL 4 LLMs} \\
        \hline
        \textbf{6. Final Dataset} & \textbf{14,485} functions & \textbf{11,913} classes \\
        \hline
    \end{tabular}
\end{table}

This data curation process ensures that any differences we observe in detection performance or feature importance across models can be attributed to genuine model-specific patterns rather than differences in the difficulty or characteristics of the code snippets themselves.

\subsection{Metrics Extraction}

Existing works on program comprehension reveal that software metrics can be a valuable source of information for understanding the properties of a piece of software \citep{curtis1979measuring,zuse1993criteria,sneed1995understanding}. Building on top of this existing finding, we leverage software metrics from the point of view of distinguishing between human-written and LLM-generated code, because the LLMs might exhibit different code quality characteristics in their generated code~\citep{idialu2024whodunit}. We used \textit{Understand} by \textit{SciTools} \citep{scitoolsUnderstandSoftware} to extract software metrics from the functions and classes in our dataset. \textit{Understand} is an industry-standard tool for software analytics with support for all popular programming languages.

We extracted metrics from two main categories: code stylometry metrics (measuring coding style and structure) and code complexity metrics (measuring algorithmic complexity). The specific metrics differ by granularity to capture characteristics relevant at each level. Note that all generated code consists of syntactically valid Python confirmed through \texttt{ast} parsing, requiring no further processing before metric extraction.

\subsubsection{Theoretical Motivation for Structural Metric Selection}
\label{sec:metric-motivation}

\textcolor{blue}{In this subsubsection, we make explicit the theoretical basis for our metric selection. The last three paragraphs below correspond to the two broad categories used to organize the metric lists later in this section, under \textbf{Function-Level Metrics} and \textbf{Class-Level Metrics}: \emph{Comment-related stylometry metrics} and \emph{Structural-count stylometry metrics} together make up the \textbf{Code Stylometry} category enumerated there, while \emph{Complexity metrics} corresponds directly to the \textbf{Code Complexity} category.}

\paragraph{Why LLM-generated code is expected to differ structurally from human-written code.} Contemporary code-generating LLMs are next-token predictors trained on heterogeneous corpora that mix natural-language text with code from many sources: well-documented tutorial repositories, lightly commented industrial code, formal documentation, instructional Q\&A sites, and academic snippets. The training procedure averages over this diversity. As a result, generated code tends to reflect a \emph{modal} structural profile of the training distribution: comment density that splits the difference between tutorial-style and production-style code, complexity profiles that cluster around canonical idiomatic patterns, and verbosity that follows the central tendency of training examples. Human-written code, particularly from production projects, departs from this modal profile in directions specific to the project's domain and the author's experience. The metrics we extract are intended to capture the structural dimensions along which the modal LLM profile and the diverse human profile may diverge. We articulate the expected mechanism for each metric category below.

\paragraph{Comment-related stylometry metrics (Comment Lines, Average Comment Lines, Comment to Code Ratio).} These metrics measure documentation density. Comments are the most direct channel through which training-data heterogeneity is inherited by generated code: LLMs encounter both heavily-documented tutorial code and sparsely-commented production code during training, and produce code with a commenting density that averages across these styles. This average is unlikely to match the practice of any specific human author, particularly authors of the under-documented industrial code that dominates real production. We therefore expect comment-related metrics to be among the most discriminative features (as reported in previous studies~\citep{idialu2024whodunit,shi2025between}\textcolor{blue}{)}, with the \emph{direction} of the difference (over- or under-commenting relative to humans) varying by LLM as a function of which subsets of the training corpus dominate each model's training mix.

\paragraph{Complexity metrics (Cyclomatic Complexity, Modified Cyclomatic Complexity, Strict Cyclomatic Complexity, Strict Modified Cyclomatic Complexity, Essential Complexity, Maximum Nesting Depth, Paths, Logarithmic Paths, and their class-level aggregates).} These metrics measure the algorithmic complexity of control flow. Our theoretical expectation is rooted in the next-token-prediction objective: LLMs assign higher probability to common patterns in their training distribution, and common code patterns have characteristic complexity profiles. Human authors, particularly experienced ones, may diverge from these idioms when domain knowledge or context requires non-standard control flow. \textcolor{blue}{Cyclomatic Complexity, the canonical metric in this category, was originally proposed as a basis for bounding test-path coverage and modularization decisions rather than as a validated defect predictor~\citep{mccabe1976complexity}. Subsequent large-scale empirical work has found that complexity metrics do correlate with post-release defects in commercial systems, though the specific correlating subset is project-dependent rather than universal across systems~\citep{nagappan2006mining}. This pattern echoes our own RQ3 finding that no single structural feature generalizes uniformly across LLMs. This gives complexity metrics additional practical relevance: to the extent that they distinguish LLM-generated from human-written code, that distinction plausibly tracks real, if context-dependent, differences in maintainability and defect-proneness.}

\paragraph{Structural-count stylometry metrics (Lines, Code Lines, Blank Lines, Declarative Code Lines, Executable Code Lines, Statements, Declarative Statements, Executable Statements, and their class-level averages).} These metrics measure the verbosity profile of a code unit. Different LLMs exhibit systematic differences in verbosity: some are tuned to produce elaborate docstrings, explicit type annotations, and intermediate variables, while others favour concise expressions. These differences arise from a combination of training-data composition and post-training fine-tuning procedures (instruction tuning, reinforcement learning from human feedback) that emphasize different stylistic targets. \textcolor{blue}{This mechanism has direct empirical support outside the code domain: \citet{singhal2023long} show that Reinforcement Learning from Human Feedback (RLHF) systematically increases output length across multiple task settings, and that improvements in reward during RLHF are driven in large part by shifting the distribution over output length rather than by other features. Insofar as code-generating LLMs are post-trained with similar RLHF pipelines, differences in this training stage across models provide a plausible mechanism for the verbosity differences we measure, even though Singhal et al.'s own study does not examine code specifically.} \textcolor{blue}{Within the code domain, \citet{idialu2024whodunit} likewise report stylometric features as the primary discriminators between human-written and GPT-4-generated functions, which supports the premise that the structural and stylistic properties of code carry detectable signal about its origin. Their study is limited to competitive programming problems and a single LLM, gaps that our multi-model, two-granularity design addresses.} These structural-count metrics may potentially be able to discriminate not only LLM-generated code from human-written code, but also LLMs from each other, which is a property that is relevant to our RQ3 analysis of cross-model feature overlap.

\begin{table}[t]
\centering
\caption{\textcolor{blue}{Metric categories and the supporting literature for each.}}
\label{tab:metric_rationale}
\textcolor{blue}{
\resizebox{\textwidth}{!}{
\begin{tabular}{p{3cm}p{6cm}p{7cm}}
\toprule
\textbf{Category} & \textbf{Metrics} & \textbf{Supporting literature} \\
\midrule
Comment-related stylometry & Comment Lines, Average Comment Lines, Comment to Code Ratio & Reported as discriminative in prior work~\citep{idialu2024whodunit,shi2025between}. \\
\midrule
Complexity & Cyclomatic, Modified, Strict, and Strict Modified Cyclomatic Complexity, Essential Complexity, Maximum Nesting Depth, Paths, Logarithmic Paths & Reported as a measure of control-flow complexity~\citep{mccabe1976complexity} and as a project-dependent predictor of post-release defects~\citep{nagappan2006mining}. \\
\midrule
Structural-count stylometry & Lines, Code Lines, Blank Lines, Declarative and Executable Code Lines, Statements, Declarative and Executable Statements & Reported as discriminative in prior work~\citep{idialu2024whodunit}; RLHF reported to increase output length outside code~\citep{singhal2023long}. \\
\bottomrule
\end{tabular}}}
\end{table}

In summary, the above taxonomy organizes our metric selection along three theoretical axes (comment density, control-flow complexity, and verbosity). \textcolor{blue}{Table~\ref{tab:metric_rationale} summarizes, for each category, the metrics it contains and the supporting literature. The comment-density and verbosity axes both fall under Code Stylometry below, and the control-flow-complexity axis corresponds to Code Complexity below.} In the following subsubsections, we enumerate the specific metrics extracted at each granularity, organized by the broad stylometry/complexity split introduced above. \textcolor{blue}{We return to this taxonomy empirically in Section~\ref{sec:rqiii}, where we test whether the three categories' relative predictive contributions differ systematically across models and granularities.}

\subsubsection{Function-Level Metrics}
For function-level analysis, we used 18 metrics covering:

\textbf{Code Stylometry (10 metrics):} Lines, Blank Lines, Code Lines, Declarative Code Lines, Executable Code Lines, Comment Lines, Statements, Declarative Statements, Executable Statements, Comment to Code Ratio

\textbf{Code Complexity (8 metrics):} Cyclomatic Complexity, Modified Cyclomatic Complexity, Strict Cyclomatic Complexity, Strict Modified Cyclomatic Complexity, Essential Complexity, Maximum Nesting Depth, Paths, Logarithmic Paths

\subsubsection{Class-Level Metrics}
For class-level analysis, we used 39 features covering:

\textbf{Code Stylometry (18 metrics):} Lines, Blank Lines, Code Lines, Declarative Code Lines, Executable Code Lines, Comment Lines, Statements, Declarative Statements, Executable Statements, Comment to Code Ratio, Average Lines, Average Blank Lines, Average Code Lines, Average Comment Lines, Instance Methods, Instance Variables, Methods, All Methods

\textbf{Code Complexity (21 metrics):} Average Cyclomatic Complexity, Average Modified Cyclomatic Complexity, Average Strict Cyclomatic Complexity, Average Strict Modified Cyclomatic Complexity, Average Essential Complexity, Maximum Cyclomatic Complexity, Maximum Modified Cyclomatic Complexity, Maximum Strict Cyclomatic Complexity, Maximum Strict Modified Cyclomatic Complexity, Maximum Essential Complexity, Sum Cyclomatic Complexity, Sum Modified Cyclomatic Complexity, Sum Strict Cyclomatic Complexity, Sum Strict Modified Cyclomatic Complexity, Sum Essential Complexity, Maximum Nesting Depth, Maximum Inheritance Tree, Base Classes, Derived Classes, Coupled Classes, Coupled Classes Modified

\textcolor{blue}{We use all class-level metrics reported by \textit{Understand}, with the values it produces and without any further filtering.} In the rest of the paper, we use the terms `feature' and `metric' interchangeably where necessary to follow machine learning nomenclature \citep{langley1994selection}.

\section{RQ1: \rqi}
\label{sec:rqi}

\subsection{Objective}

As a first step toward classifying code as human-written or LLM-generated, we must understand the characteristics of LLM-generated code. To put these characteristics into perspective, we compare them against human-written code to identify systematic differences. This exploratory analysis serves as the foundation for our subsequent research questions: if we can identify measurable differences in code characteristics, these differences can potentially be leveraged as signals for automated detection (RQ2). Furthermore, by examining how these differences vary across different LLMs and code granularities, we can understand whether detection patterns are model-specific or universal.

We investigate this question across eight configurations: four contemporary LLMs (GPT-3.5, GPT-OSS, Claude 3 Haiku, and Claude Haiku 4.5) at two granularities (function-level and class-level). We use a \emph{global intersection approach} as described in Section~\ref{sec:dataset} to eliminate task difficulty confounds and enable a fair comparison across models.

\subsection{Approach}

\subsubsection{Statistical Analysis}
We first extracted all software metrics from both human-written and LLM-generated code for each configuration. After obtaining these metrics, we investigated differences between human and LLM-authored code by performing pairwise comparisons for every metric, repeating the process for each LLM-granularity configuration. Our analysis involved three components:

\paragraph{1. Mann-Whitney U Test}
We employed the Mann-Whitney U test~\citep{mann1947test}, a non-parametric statistical test that assesses whether two independent samples come from the same distribution. We chose this test over parametric alternatives (e.g., t-test) because:
\begin{itemize}[leftmargin=*]
    \item Software metrics often exhibit non-normal distributions (e.g., skewed distributions for complexity metrics)
    \item The test is robust to outliers, which are common in code metrics
    \item It makes no assumptions about the underlying distributions
\end{itemize}

The null hypothesis for each test is that human and LLM-generated code have the same distribution for a given feature. We set our significance threshold at $\alpha = 0.01$ to reduce Type I errors given the large number of tests performed.

\paragraph{2. Cliff's Delta Effect Size}
Statistical significance alone does not indicate practical importance. For example, a large sample size can yield significant p-values for trivial differences. To measure the \emph{magnitude} of differences, we calculated Cliff's Delta \citep{cliff1993dominance}, a non-parametric effect size measure ranging from -1 to +1. Cliff's Delta quantifies the degree of overlap between two distributions:

\begin{equation}
\delta = \frac{\#(X > Y) - \#(X < Y)}{n_X \times n_Y}
\end{equation}

where, in our context, $X$ represents LLM-generated values, $Y$ represents human-written values, and $\#(X > Y)$ counts pairs where an LLM value exceeds a human value.

We interpret effect sizes using established thresholds \citep{hess2004robust,romano2006exploring}:
\begin{itemize}
    \item $|\delta| < 0.147$: Negligible
    \item $0.147 \leq |\delta| < 0.33$: Small
    \item $0.33 \leq |\delta| < 0.474$: Medium
    \item $|\delta| \geq 0.474$: Large
\end{itemize}

A positive $\delta$ indicates LLM-generated code has higher values than human-written code for that metric, while a negative $\delta$ indicates lower values.

\paragraph{3. Holm-Bonferroni Correction}
Testing multiple features creates a multiple comparisons problem: with many tests, some will appear significant by chance alone. To control the family-wise error rate, we applied the Holm-Bonferroni correction \citep{holm1979simple}.

We applied the correction \emph{per configuration} rather than globally across all tests. This decision reflects that:
\begin{itemize}[leftmargin=*]
    \item Each configuration represents an independent comparison between human and LLM-generated code
    \item Function and class levels use different feature sets
    \item Per-configuration correction maintains adequate statistical power while controlling false discoveries within each analysis
\end{itemize}

For function-level configurations, we corrected across 18 tests; for class-level configurations, across 39 tests.

\subsubsection{Significance Criteria}
We consider a feature to exhibit a \emph{meaningful difference} only when it satisfies \emph{both} criteria:
\begin{enumerate}
    \item \textbf{Statistical significance:} $p < 0.01$ (after Holm-Bonferroni correction)
    \item \textbf{Non-negligible effect size:} $|\delta| \geq 0.147$ (Small, Medium, or Large)
\end{enumerate}

This dual criterion ensures we report only differences that are both statistically reliable and practically meaningful.

\subsection{Findings}

\subsubsection{Pervasive Statistical Differences with Variable Effect Sizes}

Table~\ref{tab:rq1_summary} presents the count of features showing significant differences for each LLM-granularity configuration. After applying Holm-Bonferroni correction, the proportion of features with statistically significant differences is uniformly high at the function level (94\% to 100\% across the four LLMs) but more variable at the class level (54\% to 92\%).

\begin{table}[h]
\centering
\caption{Count of features with significant differences between LLM-generated and human-written code after Holm-Bonferroni correction. ``Non-Negligible" indicates features that are both statistically significant and have effect sizes $\geq$ Small.}
\label{tab:rq1_summary}
\resizebox{\textwidth}{!}{
\begin{tabular}{|c|c|l||cc|}
\hline
\textbf{Granularity} & \textbf{Total Features} & \textbf{Model} & \textbf{p $<$ 0.01} & \textbf{\& Non-Negligible} \\
\hline
\multirow{4}{*}{Function} & \multirow{4}{*}{18} & Claude 3 Haiku & 18 (100\%) & 3 (17\%) \\
 & & Claude Haiku 4.5 & 17 (94\%) & 13 (72\%) \\
 & & GPT-3.5 & 18 (100\%) & 17 (94\%) \\
 & & GPT-OSS & 18 (100\%) & 18 (100\%) \\
\hline
\multirow{4}{*}{Class} & \multirow{4}{*}{39} & Claude 3 Haiku & 36 (92\%) & 23 (59\%) \\
 & & Claude Haiku 4.5 & 21 (54\%) & 2 (5\%) \\
 & & GPT-3.5 & 36 (92\%) & 29 (74\%) \\
 & & GPT-OSS & 36 (92\%) & 9 (23\%) \\
\hline
\end{tabular}}
\end{table}

As stated before, statistical significance does not always translate to practical importance. When we apply the stricter criterion of \emph{non-negligible effect sizes}, the picture varies substantially across configurations. At the function level, the proportion of features with non-negligible effect sizes ranges from 17\% (Claude 3 Haiku) to 100\% (GPT-OSS), a six-fold range across LLMs at a single granularity. At the class level, the variance is even more pronounced: from 5\% (Claude Haiku 4.5) to 74\% (GPT-3.5), a fifteen-fold range. This heterogeneity is itself a key finding: the structural fingerprint of LLM-generated code is not uniform across LLMs but reflects model-specific generation tendencies. Configurations with low effect-size coverage (such as Claude 3 Haiku at the function level and Claude Haiku 4.5 at the class level, the latter with only two features showing non-negligible differences out of 39) are those for which detection is intrinsically more difficult; configurations with high coverage (such as GPT-OSS at the function level and GPT-3.5 at the class level) are correspondingly easier to detect.

\subsubsection{Model-Specific Patterns: Directional Divergence}

An interesting finding emerges regarding the direction of differences, splitting the models into two distinct groups. \textbf{Claude 3 Haiku} and \textbf{GPT-3.5} consistently generate shorter and simpler code across both granularities, with GPT-3.5 exhibiting the strongest downward trend (e.g., function-level \emph{Lines}: $\delta=-0.68$, class-level \emph{Lines}: $\delta=-0.58$). In contrast, \textbf{Claude Haiku 4.5} and \textbf{GPT-OSS} produce significantly longer and more complex outputs at both granularities, characterized by higher blank line usage and increased structural verbosity. At the function level, both models show medium positive effects for Blank Lines (Claude Haiku 4.5: $\delta=+0.37$, GPT-OSS: $\delta=+0.36$). At the class level, this verbosity manifests differently: Claude Haiku 4.5 emphasizes spacing (\emph{Average Blank Lines}: $\delta=+0.32$), while GPT-OSS prioritizes documentation (\emph{Average Comment Lines}: $\delta=+0.42$). This directional split implies that newer models may prioritize code completeness and readability through explicit structure over the conciseness favoured by earlier iterations.

\subsubsection{GPT-3.5: The Most Distinctive Pattern}

GPT-3.5 exhibits the most profound systematic deviation from human coding patterns, distinguishing code snippets generated by it as the most easily detectable. This is evident in the magnitude of differences: 94\% of significant function-level features and 81\% of class-level features show non-negligible effect sizes.

This divergence is driven by drastic reductions in code volume and documentation. At the function level, \emph{Comment Lines} show a large negative effect ($\delta = -0.81$), with the LLM averaging just 1.22 comment lines per function compared to the human average of 9.57 (an 87\% reduction). Similar large reductions are observed in \emph{Total Lines} ($\delta = -0.68$) and \emph{Executable Code} ($\delta = -0.51$). At the class level, this pattern persists with large negative effect of \emph{Executable Statements} ($\delta = -0.51$), \emph{Executable Code Lines} ($\delta = -0.50$), and complexity-related metrics such as \emph{Maximum Nesting} ($\delta = -0.53$) indicating that GPT-3.5 generates classes with fewer nesting and less code lines than human implementations. These findings suggest that GPT-3.5 prioritizes extreme brevity across both granularities, creating a unique structural signature that is significantly simpler than human-written code.

\subsubsection{Claude Haiku 4.5 Class-Level Anomaly}

An unexpected anomaly appears in Claude Haiku 4.5's class-level results: while 21 (54\%) features show statistical significance, only 2 (5\%) have non-negligible effect sizes. This indicates that while the model differs from human code in many dimensions, the magnitude of these deviations is subtle. This pattern contrasts sharply with its function-level performance (72\% non-negligible), suggesting that Claude Haiku 4.5 mimics human coding architecture significantly better at the class level than at the function level.

\subsubsection{The Rarity of Truly Universal Differences}

A critical insight from our analysis is the distinction between \emph{statistical significance} and \emph{practical importance}. While many features are statistically significant across all models, the number of features that are truly universal in terms of practical significance is incredibly small.

\begin{itemize}
    \item \textbf{Function-Level:} Only 2 of 18 features (11\%) are universal: \emph{Total Lines} and \emph{Executable Code Lines}.
    \item \textbf{Class-Level:} \textbf{Zero} features meet the criteria across all four models.
\end{itemize}

This scarcity reveals that \emph{universal statistical significance does not imply universal practical significance}. Even features that appear significant everywhere often vary wildly in effect size. For instance, while \emph{Comment to Code Ratio} is statistically significant for all models, its magnitude ranges from Negligible for Claude 3 Haiku ($\delta = -0.138$) to Large for GPT-3.5 ($\delta = -0.557$) on the function-level. A similar pattern can be observed on the class-level for this metric, with magnitude ranging from Negligible for Claude Haiku 4.5 ($\delta=+0.001$) to Medium for GPT-OSS ($\delta=-0.358$).

It is worth noting that even the two ``universal" function-level features cannot support a simple, generalized detector because of \textbf{opposing directions}.
\begin{itemize}
    \item \textbf{Simpler Models (Claude 3 Haiku, GPT-3.5):} Generate significantly \emph{shorter} code than humans (↓).
    \item \textbf{Complex Models (Claude Haiku 4.5, GPT-OSS):} Generate significantly \emph{longer} code than humans (↑).
\end{itemize}

This directional split is fatal for threshold-based detection. A rule designed to ``flag short functions" would correctly identify GPT-3.5 but would fail for GPT-OSS. Consequently, detection strategies must be model-specific; a detector trained on the brevity of GPT-3.5 or Claude 3 Haiku will likely fail to recognize the verbosity of GPT-OSS or Claude Haiku 4.5. This limitation motivates our subsequent investigation into cross-model generalization (RQ3).

\subsubsection{Granularity Effects: The Inversion of Structural Divergence}

A comparison of function-level and class-level results reveals a critical ``granularity effect" where the uniqueness of LLM-generated code diverges along two opposing trajectories.

\paragraph{Trajectory 1: Amplified Divergence (Claude 3 Haiku).}
For this model, structural deviation from human code \textbf{increases} significantly at the class level. The proportion of non-negligible features rises from 17\% at the function level to 59\% at the class level. This suggests that while Claude 3 Haiku generates functions that statistically resemble human code, its high-level architectural choices, such as how methods and state are organized, deviate sharply from human developers' programming norms.

\paragraph{Trajectory 2: Diminished Divergence (Claude Haiku 4.5, GPT-OSS, and GPT-3.5)}
Conversely, these models show a dramatic \textbf{convergence} toward human patterns when moving to the class level.
\begin{itemize}
    \item \textbf{GPT-OSS:} Drops from 100\% non-negligible distinct features at the function level to just 23\% at the class level.
    \item \textbf{Claude Haiku 4.5:} Drops from 72\% at the function level to just 5\% at the class level.
    \item \textbf{GPT-3.5:} Although the diminishing divergence effect is not as sharp as the previous two models, this model's non-negligible feature count drops from 94\% at the function level to 74\% at the class level.
\end{itemize}
This collapse in distinctiveness implies that while their individual functions are structurally unique (differing significantly from human metrics), their aggregate class structures closely mimic human architectural patterns. These opposing shifts demonstrate that the ``uniqueness" of LLM-generated code is not a fixed property but is dependent on granularity. A model may appear distinct at the micro-level (functions) yet statistically indistinguishable at the macro-level (classes). Consequently, characterizing the structural signature of these models requires analyzing them at the specific granularity of interest rather than generalizing across structural units.

\subsection{Interpretation}

Our findings establish four definitive conclusions regarding the structural uniqueness of LLM-generated code:

\paragraph{1. LLM-Generated Code Is Statistically Distinct from Human Code.}
The prevailing assumption that LLMs generate ``human-like" code is statistically unfounded with respect to structural metrics. Across all eight configurations, LLM-generated code deviates significantly from human baselines in the majority of features. These are not trivial variances; the prevalence of Medium-to-Large effect sizes confirms that current models fail to replicate the nuanced structural distributions of human-authored software.

\paragraph{2. GPT-3.5 Exhibits the Most Extreme Structural Deviation.}
GPT-3.5 is the structural outlier, exhibiting the largest statistical distance from human code across almost all dimensions. Its signature is defined by extreme brevity and a near-total absence of documentation. This systematic simplification makes GPT-3.5 the most structurally distinct model in our study, suggesting it operates under a uniquely aggressive optimization for token minimization that newer models tend not to follow.

\paragraph{3. Divergence is Model-Specific and Granularity-Dependent.}
Our analysis effectively dismantles the feasibility of a ``universal" LLM detector. We identified two contradictory phenomena that prevent generalization. First, a \textbf{Directional Split} divides models into opposing archetypes: \textbf{Reductive Models} (Claude 3 Haiku, GPT-3.5) deviate from their corresponding human-written code snippets through excessive simplicity, while \textbf{Expansive Models} (Claude Haiku 4.5, GPT-OSS) deviate from theirs through excessive complexity. A feature flagging one model would inadvertently validate the other. Second, a granularity-driven \textbf{Inversion Effect} flips distinctiveness based on the observation window. Distinctiveness is \textbf{Amplified} at the class level for some models (Claude 3 Haiku) yet \textbf{Diluted} for others (for example, GPT-OSS). Consequently, any effective detection strategy must be strictly conditioned on both the specific model family and the structural granularity, as no single ``LLM signature" exists across these dimensions.

\begin{tcolorbox}[colback=gray!10, colframe=black, boxrule=0.5pt, arc=2mm]
\textbf{Answer to RQ1:} LLM-generated code differs systematically from human-authored code (54--100\% significant features), yet the nature of these differences is strictly model-specific. Deviations range from extreme (GPT-3.5, 94\% non-negligible) to subtle (Claude Haiku 4.5, 5\%), and ``universal" features are virtually non-existent (2 at function level and 0 at class level). Furthermore, the few shared metrics exhibit opposing directions (reductive vs. expansive). This contradictory structural landscape demonstrates that no single ``LLM signature" exists, necessitating \emph{model and granularity-specific} detection strategies.
\end{tcolorbox}

\section{RQ2: \rqii}
\label{sec:rqii}

While RQ1 established that LLM-generated code differs systematically from human code, it also demonstrated that these differences are strictly model-specific and often directionally opposing. In RQ2, we move from statistical and practical significance to predictive utility. We aim to determine whether these distinctive structural signatures provide a sufficient discriminative signal for supervised machine learning models to distinguish between human and LLM authorship reliably.

\subsection{Approach}

To evaluate detection feasibility, we experiment with a diverse suite of eight classifiers representing fundamental learning paradigms:
\begin{itemize}
    \item \textbf{Linear \& Probabilistic:} Logistic Regression (LR) \citep{cox1958regression} and Naive Bayes (NB) \citep{rish2001empirical}.
    \item \textbf{Distance \& Kernel:} K-Nearest Neighbor (KNN) \citep{cover1967nearest} and Support Vector Machine (SVM) \citep{hearst1998support}.
    \item \textbf{Ensemble Trees:} Random Forest (RF) \citep{breiman2001random}, XGBoost (XGB) \citep{chen2016xgboost}, and CatBoost (CB) \citep{prokhorenkova2018catboost}.
    \item \textbf{Neural Networks:} Multi-Layer Perceptron (MLP) \citep{gardner1998artificial}.
\end{itemize}

These algorithms were selected to cover the spectrum from simple interpretable models to complex non-linear estimators, all of which have demonstrated high performance in prior software engineering classification tasks \citep{yang2022predictive, khatoonabadi2024predicting, khatoonabadi2023wasted, idialu2024whodunit, akour2022software, hribar2010software, goyal2022software}.

Crucially, given the findings of RQ1, we do not attempt to train a ``universal" detector. Instead, we train separate binary classifiers for each LLM (Claude 3 Haiku, Claude Haiku 4.5, GPT-3.5, GPT-OSS) at each granularity level (Function and Class), utilizing the metrics extracted in RQ1 as features.

\subsubsection{Feature Selection via AutoSpearman}

High correlation among software metrics can introduce multicollinearity, obscuring model interpretation and inflating standard errors \citep{dormann2013collinearity}. To mitigate this, we employed AutoSpearman \citep{jiarpakdee2018autospearman} on the \emph{training splits} of our dataset. AutoSpearman is an automated, unsupervised feature selection approach designed to minimize collinearity while preserving information.

The process consists of two sequential stages:
\begin{enumerate}
    \item \textbf{Correlation Analysis:} We computed Spearman's rank correlation coefficient ($\rho$) \citep{spearman1987proof} for all feature pairs. For pairs with $|\rho| \geq 0.7$ \citep{kraemer2003measures}, the feature with the highest average correlation to the rest of the dataset was iteratively removed.
    \item \textbf{VIF Analysis:} To address remaining multicollinearity, we calculated the Variance Inflation Factor (VIF) for the surviving features. Features exceeding the threshold of VIF $\geq 5$ \citep{fox2015applied} were removed iteratively, starting with the highest value, until the set stabilized.
\end{enumerate}

AutoSpearman achieved substantial dimensionality reduction across all configurations (66.7\%--76.9\%), yielding compact feature sets (6--12 features) that capture discriminative patterns without redundancy. The selected features are as follows.

\begin{itemize}[leftmargin=*]
    \item Function-level:
    \begin{itemize}
        \item GPT-3.5 and Claude 3 Haiku (6 features): \textit{Blank Lines}, \textit{Comment to Code Ratio}, \textit{Declarative Code Lines}, \textit{Essential Complexity}, \textit{Maximum Nesting}, \textit{Logarithmic Paths}
        \item GPT-OSS and Claude Haiku 4.5 (6 features): \textit{Comment Lines}, \textit{Comment to Code Ratio}, \textit{Declarative Statements}, \textit{Essential Complexity}, \textit{Maximum Nesting}, \textit{Logarithmic Paths}
    \end{itemize}
    \item Class-level:
    \begin{itemize}
        \item GPT-3.5 (11 features): \textit{Average Blank Lines}, \textit{Average Comment Lines}, \textit{Average Essential Complexity}, \textit{Base Classes}, \textit{Comment to Code Ratio}, \textit{Coupled Classes}, \textit{Coupled Classes Modified}, \textit{Derived Classes}, \textit{Instance Methods}, \textit{Instance Variables}, \textit{Maximum Nesting}
        \item GPT-OSS (9 features): \textit{Average Blank Lines}, \textit{Average Essential Complexity}, \textit{Base Classes}, \textit{Comment to Code Ratio}, \textit{Coupled Classes}, \textit{Coupled Classes Modified}, \textit{Derived Classes}, \textit{Instance Methods}, \textit{Instance Variables}
        \item Claude 3 Haiku (12 features): \textit{Average Blank Lines}, \textit{Average Code Lines}, \textit{Average Comment Lines}, \textit{Average Essential Complexity}, \textit{Base Classes}, \textit{Comment to Code Ratio}, \textit{Coupled Classes}, \textit{Coupled Classes Modified}, \textit{Derived Classes}, \textit{Instance Methods}, \textit{Instance Variables}, \textit{Maximum Nesting}
        \item Claude Haiku 4.5 (10 features): \textit{Average Blank Lines}, \textit{Average Comment Lines}, \textit{Average Essential Complexity}, \textit{Base Classes}, \textit{Comment to Code Ratio}, \textit{Coupled Classes}, \textit{Coupled Classes Modified}, \textit{Derived Classes}, \textit{Instance Methods}, \textit{Instance Variables}
    \end{itemize}
\end{itemize}

\subsubsection{Model Training and Selection}

We adopted a rigorous training and validation protocol to ensure the generalizability of our detection models. For each of the eight LLM-granularity configurations, we trained the eight candidate classifiers using the reduced feature sets.

\textbf{Validation Protocol.} We performed 30 repetitions of 10-fold cross-validation \citep{guyon2010model,dietterich1998approximate}. This yields 300 performance measurements per classifier for each configuration, ensuring robust performance estimation and minimizing the variance associated with random data splitting \citep{anguita2012k}.

\textbf{Classifier Selection via Statistical Ranking.} To objectively identify the best classifier, we applied the Scott-Knott ESD test \citep{tantithamthavorn2016empirical,scott1974cluster}. This statistical ranking method employs hierarchical clustering to partition classifiers into distinct, non-overlapping groups based on their AUC-ROC scores. CatBoost consistently ranked in the top-performing Scott-Knott cluster across all eight configurations (see Appendix Section~\ref{appendix:supplementary_ranking}). Consequently, we selected it as the unified classifier for all subsequent evaluations. Its superior performance is attributed to its efficient implementation of ordered boosting, which effectively prevents overfitting on tabular datasets \citep{prokhorenkova2018catboost}.

\subsubsection{Baseline Comparison with Commercial Detectors}

To benchmark our structural approach against the state of practice, we evaluated GPTZero~\citep{adam2026gptzero}, a leading commercial AI detection service. GPTZero provided full research access, allowing us to conduct a comprehensive evaluation across all eight model-granularity configurations. As a general-purpose detector, it outputs a probability score for AI authorship, which we converted to binary classifications using a standard 0.5 threshold. We evaluated GPTZero on the same test sets used for our CatBoost models to ensure a strictly fair comparison.

It is worth mentioning that we also attempted to evaluate Sapling.ai~\citep{saplingContentDetector}. However, initial tests on a small subset of our data (Claude Haiku 3 functions and classes) yielded performance barely distinguishable from random guessing (AUC $\approx$ 0.54). Given this poor predictive utility and the prohibitive cost of scaling the evaluation ($\approx$\$45 per configuration), we did not proceed with further testing of this service.

\subsubsection{Performance Evaluation Metrics}

We evaluated classifier performance using five complementary metrics to provide a comprehensive view of detection capability. In the following definitions, we designate LLM-generated code as the \emph{Positive} class and human-written code as the \emph{Negative} class. Consequently, \emph{TP} (True Positives) refers to LLM-generated code correctly identified as such, \emph{FP} (False Positives) refers to human-authored code incorrectly flagged as LLM-generated, \emph{TN} (True Negatives) refers to correctly identified human-authored code, and \emph{FN} (False Negatives) refers to LLM-generated code missed by the detector.

\begin{itemize}[leftmargin=*]
    \item \textbf{Precision:} Also known as Positive Predictive Value, this metric measures the reliability of the classifier when it predicts the positive class. High precision indicates that when the model flags code as LLM-generated, it is likely correct.
    \begin{equation*}
        Precision = \frac{TP}{TP+FP}
    \end{equation*}

    \item \textbf{Recall:} Also known as Sensitivity, this metric measures the coverage of the classifier. High recall indicates that the model successfully identifies the majority of LLM-generated code samples in the dataset.
    \begin{equation*}
        Recall = \frac{TP}{TP+FN}
    \end{equation*}

    \item \textbf{F1-Score:} This metric is the harmonic mean of Precision and Recall. It provides a single score that balances both concerns, penalizing models that achieve high scores in one metric by sacrificing the other (e.g., a model that flags everything as LLM-generated would have perfect Recall but poor Precision).
    \begin{equation*}
        \textit{F1-Score} = 2 \times \frac{Precision \times Recall}{Precision + Recall}
    \end{equation*}

    \item \textbf{AUC-ROC:} The Area Under the Receiver Operating Characteristic Curve \citep{bradley1997use} measures the classifier's ability to discriminate between classes across all possible decision thresholds. An AUC-ROC of 0.5 indicates random guessing, while 1.0 indicates perfect separation. Unlike F1-Score, AUC-ROC is threshold-independent.

    \item \textbf{Matthews Correlation Coefficient (MCC):} We include MCC \citep{matthews1975comparison} as it provides a balanced measure of the correlation between observed and predicted binary classifications. It considers all four distinct categories of the confusion matrix (TP, TN, FP, FN)  and ranges from -1 (total disagreement) to +1 (perfect prediction), with 0 indicating random prediction.
    \begin{equation*}
        MCC = \frac{TP \times TN - FP \times FN}{\sqrt{(TP+FP)(TP+FN)(TN+FP)(TN+FN)}}
    \end{equation*}
\end{itemize}

\subsubsection{Statistical Evaluation}

We employ two statistical approaches to ensure the reliability of the obtained results:

\textbf{Uncertainty Quantification.} We apply bootstrap resampling \citep{efron1994introduction} with 1,000 iterations to generate robust performance estimates. We report means and 95\% confidence intervals for all metrics.

\textbf{Significance Testing.} We use \textbf{DeLong's test}~\citep{delong1988comparing} to compare AUC-ROC scores between models, which accounts for correlation between ROC curves derived from the same test data. We apply \textbf{Holm-Bonferroni correction}~\citep{holm1979simple} to control the family-wise error rate across pairwise comparisons.

\subsection{Findings}

\subsubsection{Overall Detection Performance}
Tables~\ref{tab:rq2_performance_function} and~\ref{tab:rq2_performance_class} present the detection performance across all eight model-granularity configurations. We compare our structural approach (CatBoost) against the commercial baseline (GPTZero). CatBoost metrics are reported as the Mean [95\% CI] from a 1000-iteration bootstrap evaluation, while GPTZero represents single-run performance on the identical test set.


\begin{table}[h]
\centering
\caption{Detection performance for function-level code across four LLMs. CatBoost metrics reported as Mean [95\% CI]. GPTZero metrics are point estimates. \textbf{Bold} indicates CatBoost outperforms GPTZero.}
\label{tab:rq2_performance_function}
\begin{tabular}{ll|c|c|c|c|}
\cline{3-6}
                                    &          & \textbf{GPT-3.5}                                                              & \textbf{GPT-OSS}                                                              & \textbf{\begin{tabular}[c]{@{}c@{}}Claude 3\\Haiku\end{tabular}}              & \textbf{\begin{tabular}[c]{@{}c@{}}Claude\\Haiku 4.5\end{tabular}}            \\ \hline \hline
\multicolumn{1}{|l|}{\multirow{2}{*}{Precision}} & GPTZero  & 0.71                                                                          & 0.85                                                                          & 0.75                                                                          & 0.78                                                                          \\ \cline{2-6} 
\multicolumn{1}{|l|}{}                           & CatBoost & \textbf{\begin{tabular}[c]{@{}c@{}}0.94\\ {[}0.93, 0.95{]}\end{tabular}}      & \begin{tabular}[c]{@{}c@{}}0.73\\ {[}0.71, 0.75{]}\end{tabular}               & \begin{tabular}[c]{@{}c@{}}0.63\\ {[}0.61, 0.64{]}\end{tabular}               & \begin{tabular}[c]{@{}c@{}}0.66\\ {[}0.64, 0.68{]}\end{tabular}               \\ \hline
\multicolumn{1}{|l|}{\multirow{2}{*}{Recall}}    & GPTZero  & 0.29                                                                           & 0.64                                                                           & \textcolor{blue}{0.34}                                                                           & 0.42                                                                          \\ \cline{2-6} 
\multicolumn{1}{|l|}{}                           & CatBoost & \textbf{\begin{tabular}[c]{@{}c@{}}0.87\\ {[}0.86, 0.88{]}\end{tabular}}      & \textbf{\begin{tabular}[c]{@{}c@{}}0.71\\ {[}0.69, 0.73{]}\end{tabular}}      & \textbf{\begin{tabular}[c]{@{}c@{}}0.64\\ {[}0.63, 0.66{]}\end{tabular}}      & \textbf{\begin{tabular}[c]{@{}c@{}}0.66\\ {[}0.65, 0.68{]}\end{tabular}}      \\ \hline
\multicolumn{1}{|l|}{\multirow{2}{*}{F1-Score}}  & GPTZero  & 0.41                                                                          & 0.73                                                                          & 0.47                                                                          & 0.55                                                                          \\ \cline{2-6} 
\multicolumn{1}{|l|}{}                           & CatBoost & \textbf{\begin{tabular}[c]{@{}c@{}}0.90\\ {[}0.89, 0.91{]}\end{tabular}}      & \begin{tabular}[c]{@{}c@{}}0.72\\ {[}0.71, 0.73{]}\end{tabular}               & \textbf{\begin{tabular}[c]{@{}c@{}}0.64\\ {[}0.62, 0.65{]}\end{tabular}}      & \textbf{\begin{tabular}[c]{@{}c@{}}0.66\\ {[}0.65, 0.68{]}\end{tabular}}      \\ \hline
\multicolumn{1}{|l|}{\multirow{2}{*}{AUC-ROC}}   & GPTZero  & \textcolor{blue}{0.698}                                                                           & \textcolor{blue}{0.862}                                                                           & \textcolor{blue}{0.707}                                                                           & \textcolor{blue}{0.760}                                                                          \\ \cline{2-6} 
\multicolumn{1}{|l|}{}                           & CatBoost & \textbf{\begin{tabular}[c]{@{}c@{}}\textcolor{blue}{0.961}\\ \textcolor{blue}{{[}0.956, 0.965{]}}\end{tabular}}      & \begin{tabular}[c]{@{}c@{}}\textcolor{blue}{0.795}\\ \textcolor{blue}{{[}0.784, 0.807{]}}\end{tabular}               & \begin{tabular}[c]{@{}c@{}}\textcolor{blue}{0.681}\\ \textcolor{blue}{{[}0.667, 0.694{]}}\end{tabular}               & \begin{tabular}[c]{@{}c@{}}\textcolor{blue}{0.714}\\ {[}0.70, 0.73{]}\end{tabular}               \\ \hline
\multicolumn{1}{|l|}{\multirow{2}{*}{MCC}}       & GPTZero  & 0.21                                                                          & 0.54                                                                          & 0.27                                                                          & 0.35                                                                          \\ \cline{2-6} 
\multicolumn{1}{|l|}{}                           & CatBoost & \textbf{\begin{tabular}[c]{@{}c@{}}0.81\\ {[}0.80, 0.83{]}\end{tabular}}      & \begin{tabular}[c]{@{}c@{}}0.45\\ {[}0.43, 0.47{]}\end{tabular}               & \begin{tabular}[c]{@{}c@{}}0.26\\ {[}0.24, 0.28{]}\end{tabular}               & \begin{tabular}[c]{@{}c@{}}0.32\\ {[}0.30, 0.35{]}\end{tabular}               \\ \hline
\end{tabular}
\end{table}

\begin{table}[h]
\centering
\caption{Detection performance for class-level code across four LLMs. CatBoost metrics reported as Mean [95\% CI]. GPTZero metrics are point estimates. \textbf{Bold} indicates CatBoost outperforms GPTZero.}
\label{tab:rq2_performance_class}
\begin{tabular}{ll|c|c|c|c|}
\cline{3-6}
                                    &          & \textbf{GPT-3.5}                                                              & \textbf{GPT-OSS}                                                              & \textbf{\begin{tabular}[c]{@{}c@{}}Claude 3\\Haiku\end{tabular}}              & \textbf{\begin{tabular}[c]{@{}c@{}}Claude\\Haiku 4.5\end{tabular}}            \\ \hline \hline
\multicolumn{1}{|l|}{\multirow{2}{*}{Precision}} & GPTZero  & 0.65                                                                          & 0.86                                                                          & 0.71                                                                          & 0.72                                                                          \\ \cline{2-6} 
\multicolumn{1}{|l|}{}                           & CatBoost & \textbf{\begin{tabular}[c]{@{}c@{}}0.78\\ {[}0.76, 0.79{]}\end{tabular}}      & \begin{tabular}[c]{@{}c@{}}0.72\\ {[}0.70, 0.74{]}\end{tabular}               & \textbf{\begin{tabular}[c]{@{}c@{}}0.72\\ {[}0.70, 0.74{]}\end{tabular}}      & \textbf{\begin{tabular}[c]{@{}c@{}}0.78\\ {[}0.77, 0.80{]}\end{tabular}}      \\ \hline
\multicolumn{1}{|l|}{\multirow{2}{*}{Recall}}    & GPTZero  & 0.15                                                                          & 0.51                                                                          & 0.20                                                                          & 0.21                                                                          \\ \cline{2-6} 
\multicolumn{1}{|l|}{}                           & CatBoost & \textbf{\begin{tabular}[c]{@{}c@{}}0.83\\ {[}0.81, 0.84{]}\end{tabular}}      & \textbf{\begin{tabular}[c]{@{}c@{}}0.75\\ {[}0.73, 0.76{]}\end{tabular}}      & \textbf{\begin{tabular}[c]{@{}c@{}}0.78\\ {[}0.76, 0.79{]}\end{tabular}}      & \textbf{\begin{tabular}[c]{@{}c@{}}0.61\\ {[}0.59, 0.63{]}\end{tabular}}      \\ \hline
\multicolumn{1}{|l|}{\multirow{2}{*}{F1-Score}}  & GPTZero  & 0.25                                                                           & 0.64                                                                           & \textcolor{blue}{0.31}                                                                           & 0.32                                                                          \\ \cline{2-6} 
\multicolumn{1}{|l|}{}                           & CatBoost & \textbf{\begin{tabular}[c]{@{}c@{}}0.80\\ {[}0.79, 0.81{]}\end{tabular}}      & \textbf{\begin{tabular}[c]{@{}c@{}}0.73\\ {[}0.72, 0.75{]}\end{tabular}}      & \textbf{\begin{tabular}[c]{@{}c@{}}0.75\\ {[}0.73, 0.76{]}\end{tabular}}      & \textbf{\begin{tabular}[c]{@{}c@{}}0.68\\ {[}0.67, 0.70{]}\end{tabular}}      \\ \hline
\multicolumn{1}{|l|}{\multirow{2}{*}{AUC-ROC}}   & GPTZero  & \textcolor{blue}{0.526}                                                                           & \textcolor{blue}{0.825}                                                                           & \textcolor{blue}{0.624}                                                                           & \textcolor{blue}{0.658}                                                                          \\ \cline{2-6} 
\multicolumn{1}{|l|}{}                           & CatBoost & \textbf{\begin{tabular}[c]{@{}c@{}}\textcolor{blue}{0.887}\\ {[}0.88, 0.90{]}\end{tabular}}      & \begin{tabular}[c]{@{}c@{}}\textcolor{blue}{0.808}\\ \textcolor{blue}{{[}0.796, 0.821{]}}\end{tabular}               & \textbf{\begin{tabular}[c]{@{}c@{}}\textcolor{blue}{0.830}\\ {[}0.82, 0.84{]}\end{tabular}}      & \textbf{\begin{tabular}[c]{@{}c@{}}\textcolor{blue}{0.806}\\ \textcolor{blue}{{[}0.795, 0.817{]}}\end{tabular}}      \\ \hline
\multicolumn{1}{|l|}{\multirow{2}{*}{MCC}}       & GPTZero  & 0.11                                                                          & 0.47                                                                          & 0.17                                                                          & 0.18                                                                          \\ \cline{2-6} 
\multicolumn{1}{|l|}{}                           & CatBoost & \textbf{\begin{tabular}[c]{@{}c@{}}0.59\\ {[}0.57, 0.61{]}\end{tabular}}      & \begin{tabular}[c]{@{}c@{}}0.46\\ {[}0.43, 0.48{]}\end{tabular}               & \textbf{\begin{tabular}[c]{@{}c@{}}0.47\\ {[}0.45, 0.50{]}\end{tabular}}      & \textbf{\begin{tabular}[c]{@{}c@{}}0.45\\ {[}0.43, 0.48{]}\end{tabular}}      \\ \hline
\end{tabular}
\end{table}


\textcolor{blue}{\textbf{Comparison with Commercial Baselines.} A direct comparison between our model-specific classifiers and the GPTZero commercial baseline (Tables~\ref{tab:rq2_performance_function} and~\ref{tab:rq2_performance_class}) reveals two distinct patterns. On AUC-ROC, the comparison is mixed: GPTZero achieves the higher AUC in four of the eight configurations, namely the three function-level configurations other than GPT-3.5 (by 0.03 to 0.07) and GPT-OSS at the class level (by about 0.02), while our classifiers achieve the higher AUC in the other four (by 0.15 to 0.36), with the largest margins for GPT-3.5 at both granularities. The headline difference between the two approaches lies not in AUC but in recall.} GPTZero achieves a recall between 15\% and 64\% across the eight configurations, while our classifiers achieve 61\% to 87\%. Per-configuration recall deficits for GPTZero range from 7 percentage points (GPT-OSS function) to 68 percentage points (GPT-3.5 class), with a mean deficit of about \textcolor{blue}{39} percentage points. Equivalently, in deployment, GPTZero would miss between 36\% and 85\% of LLM-generated samples, while our classifiers would miss between 13\% and 39\%. The recall deficit is consistent across all eight configurations and reflects a systematic high-precision, low-recall bias in GPTZero rather than an aggregate AUC advantage for our approach.

\textbf{High Detectability of GPT-3.5.} GPT-3.5 is the most detectable model in our study. At the function level, it achieves a near-perfect AUC-ROC of \textcolor{blue}{0.961} (95\% CI [0.956, 0.965]) and an MCC of 0.813, indicating exceptional separability. This represents a substantial performance gap over the next best model at the function level (GPT-OSS, AUC-ROC=\textcolor{blue}{0.795}). Even at the class level, GPT-3.5 remains the top-performing configuration (AUC-ROC=\textcolor{blue}{0.887}), significantly outperforming the next best model at the class level detection by almost 6 percentage points.

\textbf{Universal Detectability.} All models remain detectable well above random chance. Even the most difficult configuration (Claude 3 Haiku at function-level) achieves an AUC-ROC of 0.681 [0.667, 0.694], with 95\% confidence intervals strictly non-overlapping with the random baseline of 0.50.

\subsubsection{Statistical Significance of Performance Differences}
To validate these performance hierarchies, we applied DeLong's test with Holm-Bonferroni correction to all pairwise AUC-ROC comparisons. Table~\ref{tab:rq2_delong} summarizes the results.

\begin{table}[t]
\centering
\caption{DeLong's test results for pairwise AUC-ROC comparisons with Holm-Bonferroni correction. AUC$_A$ and AUC$_B$ are the \textcolor{blue}{bootstrap-mean} AUC-ROC values for the first and second model in each comparison, taken from Tables~\ref{tab:rq2_performance_function} and~\ref{tab:rq2_performance_class}. \textcolor{blue}{$\Delta$ AUC is computed from unrounded AUC values, so it can differ by 0.001 from the difference of the rounded values shown.}}
\label{tab:rq2_delong}
\resizebox{\textwidth}{!}{
\begin{tabular}{l|l|ccc|c}
\toprule
\textbf{Granularity} & \textbf{Comparison (A vs. B)} & \textbf{AUC$_A$} & \textbf{AUC$_B$} & \textbf{$\Delta$ AUC} & \textbf{Significant} \\
\midrule
\multirow{6}{*}{Function}
& GPT-3.5 vs Claude 3 Haiku & \textcolor{blue}{0.961} & \textcolor{blue}{0.681} & \textcolor{blue}{+0.280} & Yes \\
& GPT-3.5 vs Claude Haiku 4.5 & \textcolor{blue}{0.961} & \textcolor{blue}{0.714} & +0.247 & Yes \\
& GPT-3.5 vs GPT-OSS & \textcolor{blue}{0.961} & \textcolor{blue}{0.795} & +0.166 & Yes \\
& GPT-OSS vs Claude Haiku 4.5 & \textcolor{blue}{0.795} & \textcolor{blue}{0.714} & +0.081 & Yes \\
& GPT-OSS vs Claude 3 Haiku & \textcolor{blue}{0.795} & \textcolor{blue}{0.681} & \textcolor{blue}{+0.114} & Yes \\
& Claude Haiku 4.5 vs Claude 3 Haiku & \textcolor{blue}{0.714} & \textcolor{blue}{0.681} & \textcolor{blue}{+0.033} & Yes \\
\midrule
\multirow{6}{*}{Class}
& GPT-3.5 vs Claude Haiku 4.5 & \textcolor{blue}{0.887} & \textcolor{blue}{0.806} & +0.081 & Yes \\
& GPT-3.5 vs GPT-OSS & \textcolor{blue}{0.887} & \textcolor{blue}{0.808} & +0.079 & Yes \\
& GPT-3.5 vs Claude 3 Haiku & \textcolor{blue}{0.887} & 0.830 & \textcolor{blue}{+0.058} & Yes \\
& Claude 3 Haiku vs Claude Haiku 4.5 & 0.830 & \textcolor{blue}{0.806} & \textcolor{blue}{+0.023} & Yes \\
& Claude 3 Haiku vs GPT-OSS & 0.830 & \textcolor{blue}{0.808} & +0.022 & Yes \\
& GPT-OSS vs Claude Haiku 4.5 & \textcolor{blue}{0.808} & \textcolor{blue}{0.806} & +0.002 & No \\
\bottomrule
\end{tabular}}
\end{table}

\textbf{Function-Level Hierarchy.} All six pairwise comparisons are statistically significant. The detectability is clear and distinct: GPT-3.5 $\gg$ GPT-OSS $>$ Claude Haiku 4.5 $>$ Claude 3 Haiku.

\textbf{Class-Level Convergence.} Five of six comparisons remain significant. The exception is GPT-OSS vs. Claude Haiku 4.5 ($\Delta = 0.002, p=0.805$), indicating that the performances of the detectors of these two LLMs are statistically indistinguishable at the class level.

\subsubsection{Granularity Effects}
Table~\ref{tab:rq2_granularity} quantifies the impact of code granularity on detection.

\begin{table}[t]
\centering
\caption{Granularity effect on detection performance. All differences are statistically significant, confirmed by DeLong's test. $\uparrow$ indicates better performance at class-level; $\downarrow$ indicates better performance at function-level.}
\label{tab:rq2_granularity}
\begin{tabular}{lccc}
\toprule
\textbf{Model} & \textbf{Function AUC} & \textbf{Class AUC} & \textbf{$\Delta$} \\
\midrule
Claude 3 Haiku & \textcolor{blue}{0.681} & 0.830 & $\uparrow$ \textcolor{blue}{+0.149} \\
Claude Haiku 4.5 & \textcolor{blue}{0.714} & \textcolor{blue}{0.806} & $\uparrow$ \textcolor{blue}{+0.092} \\
GPT-3.5 & \textcolor{blue}{0.961} & \textcolor{blue}{0.887} & $\downarrow$ \textcolor{blue}{-0.074} \\
GPT-OSS & \textcolor{blue}{0.795} & \textcolor{blue}{0.808} & $\uparrow$ \textcolor{blue}{+0.013} \\
\bottomrule
\end{tabular}
\end{table}

\textbf{Variable Impact.} The effect of granularity is statistically significant for all models but varies in direction. Three models (Claude 3 Haiku, Claude Haiku 4.5, GPT-OSS) show improved detectability at the class level, with Claude 3 Haiku showing the largest jump (\textcolor{blue}{+0.149}). Conversely, GPT-3.5 is the only model that is significantly \emph{harder} to detect at the class level (-0.074).

\subsection{Interpretation}

Our detection analysis establishes three key insights into the practical identification of AI-generated code:

\paragraph{1. The Recall Profile of Structural vs. Textual Detection.}
\textcolor{blue}{GPTZero and our model-specific classifiers differ little in AUC in several configurations, and GPTZero's AUC is in fact the higher of the two in four of the eight, but the two differ substantially in their precision-recall profiles. GPTZero achieves high precision with low recall: across the eight configurations, its recall ranges from 15\% to 64\%, meaning that it misses between 36\% and 85\% of LLM-generated samples even where its AUC is comparable to or higher than ours.} Our model-specific classifiers achieve a recall between 61\% and 87\%, missing between 13\% and 39\% of samples. The difference is not in a generic measure of classifier quality, but in the operating regime each detector occupies: GPTZero is calibrated toward high-confidence positive predictions, which suits adjudicative applications but not screening workflows where missing LLM-generated code is consequential.

\paragraph{2. The ``Function-Class" Detection Trade-off.}
We observe a distinct trade-off between code granularity and detectability.
\begin{itemize}[leftmargin=*]
    \item \textbf{Granularity-Exposed Models (Claude Family, GPT-OSS):} These models are easier to detect at the class level (Claude 3 Haiku: \textcolor{blue}{+0.149}, Claude Haiku 4.5: \textcolor{blue}{+0.092}, GPT-OSS: \textcolor{blue}{+0.013} AUC-ROC). While they effectively mimic human patterns in isolated functions, they struggle with class-level coherence. The added complexity of state management and coupling makes their artificial nature more apparent.
    \item \textbf{Granularity-Masked Model (GPT-3.5):} Conversely, GPT-3.5 becomes harder to detect at the class level (\textcolor{blue}{-0.074} AUC-ROC). Its brevity and lack of comments are clear in functions, but diluted when aggregated into larger class structures.
\end{itemize}

\paragraph{3. Convergence of Advanced Models.}
While GPT-3.5 is easily detected, newer models (Claude Haiku 4.5, GPT-OSS) constitute a significantly harder detection tier. Their class-level performance converges to a similar range ($\approx$ 0.80--0.81 AUC-ROC) and is statistically indistinguishable ($p=0.805$). This implies that advanced LLMs are smoothing out the extreme structural anomalies found in earlier generations, evolving toward a uniform structural profile that remains distinct from human code yet consistent across models.

\begin{tcolorbox}[colback=gray!10, colframe=black, boxrule=0.5pt, arc=2mm]

\textbf{Answer to RQ2:} Structural classifiers distinguish LLM-generated Python code with AUC-ROC between 0.68 and 0.96 across the eight model--granularity configurations. \textcolor{blue}{Compared with the GPTZero commercial detector, the AUC-ROC result is mixed: GPTZero achieves the higher AUC-ROC in four of the eight configurations (by 0.02 to 0.07), while our classifiers achieve the higher AUC-ROC in the other four (by 0.15 to 0.36). The clearer difference is in recall:} our model-specific classifiers achieve recall between 61\% and 87\%, while GPTZero achieves recall between 15\% and 64\%, which indicates that GPTZero is unsuitable as a screening tool for LLM-generated Python code in any configuration we tested. GPT-3.5 is the most detectable model in our study at both granularities, while newer Claude Haiku 4.5 and GPT-OSS converge to a statistically indistinguishable class-level performance.

\end{tcolorbox}

\section{RQ3: \rqiii}
\label{sec:rqiii}

While RQ2 establishes that structural detection is effective, the \emph{mechanisms} driving this performance remain opaque. RQ3 addresses this interpretability gap by deconstructing the ``structural fingerprint'' of each model. We aim to understand: (1) which specific architectural features drive detection for each LLM, (2) whether these discriminative signals are stable across different models, and (3) how granularity shifts the importance of specific metrics. This analysis identifies the unique structural signatures that define each model's generated code.

\subsection{Approach}

\subsubsection{Feature Selection Analysis}

To quantify the stability of discriminative patterns, we analyze the feature sets retained by AutoSpearman (described in RQ2) across configurations. We compute the \emph{Jaccard Similarity Coefficient}~\citep{jaccard1901etude}, which is a metric quantifying the overlap between two sets as the ratio of their intersection to their union, ranging from 0 (no overlap) to 1 (identical sets), between feature sets along two dimensions:
\begin{enumerate}[leftmargin=*]
    \item \textbf{Cross-Model Stability:} Comparing feature sets of different LLMs at the same granularity (e.g., Claude 3 Haiku-generated functions vs. GPT-3.5-generated functions). High similarity here would imply a universal ``LLM fingerprint."
    \item \textbf{Cross-Granularity Stability:} Comparing function-level vs class-level feature sets within the same model. Low similarity here would confirm the ``Inversion Effect" hypothesis, indicating that detection relies on fundamentally different signals at different observation windows.
\end{enumerate}

\subsubsection{Feature Importance via SHAP}
While recursive feature elimination identifies \textit{relevant} features, it offers only a binary signal (kept vs. removed). True interpretability requires quantifying the magnitude and direction of each feature's influence. To achieve this, we employ \emph{SHapley Additive exPlanations (SHAP)} \citep{lundberg2017unified}, a game-theoretic approach that assigns each feature an importance value representing its marginal contribution to the prediction. The choice of SHAP analysis aligns with recent software engineering research, which employs such post-hoc explanation techniques to interpret opaque models and ensure the transparency of automated decision-making~\citep{khatoonabadi2024predicting,idialu2024whodunit,ahmad2025interpretable,haldar2024interpretable}.

We compute SHAP values for every test instance across our eight experimental configurations, aggregating them as the \textit{mean absolute SHAP importance}. This metric goes beyond correlation to pinpoint the specific structural features driving the classifier's decisions for each LLM and granularity.

\subsection{Findings}

\subsubsection{Feature Selection Patterns}

Table~\ref{tab:rq3_feature_overlap} summarizes how feature sets overlap across different configurations. We observe a striking contrast between cross-model and cross-granularity stability. 

We quantify feature stability using the Jaccard similarity coefficient ($J$). For two feature sets $A$ and $B$, this is defined as $J(A,B) = \frac{|A \cap B|}{|A \cup B|}$. While no universal thresholds exist for $J$, we interpret $J > 0.7$ as \emph{strong stability} and $J < 0.2$ as \emph{negligible overlap}. These thresholds follow stability analysis~\citep{nogueira2018stability} principles proposed by Kuncheva~\citep{kuncheva2007stability}, which emphasize distinguishing genuine structural overlap from random intersection. For our feature space, the expected random overlap is $E[J_{random}] \approx 0.11$ (derivation in Appendix Section~\ref{appendix:random_jaccard}).

\begin{itemize}[leftmargin=*]
    \item \textbf{High Cross-Model Overlap:} When comparing different LLMs at the \emph{same} granularity, we find that they select highly consistent features. Cross-model comparisons show high overlap: at the function level, different LLMs share 69\% of selected features on average, while at the class level they share 85.5\%, indicating that structural signals are robust to model architecture changes.
    \item \textbf{Negligible Cross-Granularity Overlap:} Conversely, when comparing the \emph{same} LLM across different granularities, the feature sets are almost entirely disjoint ($J = 0.099$). This value is statistically indistinguishable from random chance, representing an 8.6 times drop in overlapping features compared to cross-model scenarios. This indicates that detection relies on fundamentally different structural signatures depending on the level of abstraction.
\end{itemize}

\begin{table}[h]
\centering
\caption{Feature selection overlap analysis. Stability is high across models but negligible across granularities (indistinguishable from random chance).}
\label{tab:rq3_feature_overlap}
\begin{tabular}{l|l|c|l}
\toprule
\textbf{Comparison Scope} & \textbf{Fixed Factor} & \textbf{Avg. Jaccard} & \textbf{Stability} \\
\midrule
Cross-Model & Class Level & 0.855 & High \\
Cross-Model & Function Level & 0.690 & Moderate \\
Cross-Granularity & Model Architecture & 0.099 & \textbf{Negligible} \\
\bottomrule
\end{tabular}
\end{table}

\subsubsection{Feature Importance Hierarchy (SHAP)}

We used SHAP analysis to identify which features drive the classifier decisions. Table~\ref{tab:rq3_top_features} ranks the top features by their average importance across all configurations.

\textbf{The Universal Feature.} \textit{Comment to Code Ratio} is the only feature selected in all eight configurations. It is also the most important feature overall, with an average SHAP value of 0.777, appearing in the top 3 ranking for 7 of 8 models.

\begin{figure}[t]
\centering
\includegraphics[width=0.9\textwidth]{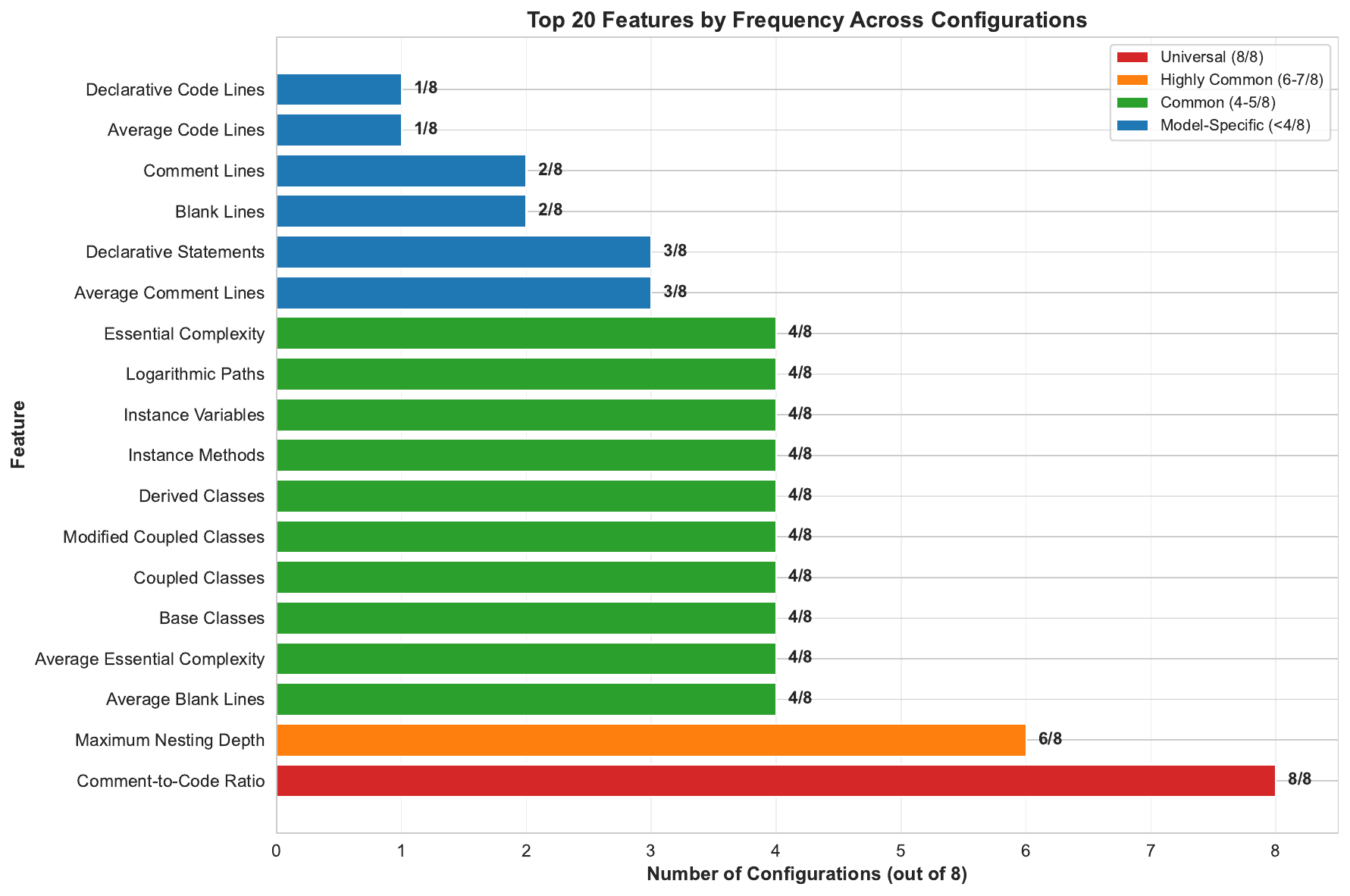}
\caption{Feature frequency across eight configurations. Only \emph{Comment to Code Ratio} (red) appears universally. Most features show model or granularity specificity.}
\label{fig:rq3_feature_frequency}
\end{figure}

\textbf{The Drop-off.} As shown in Figure~\ref{fig:rq3_feature_frequency}, there is a sharp drop in universality after the first metric. Most discriminative patterns are context-dependent, with features like \emph{Average Blank Lines} appearing in only 50\% of configurations despite high importance.

\textbf{Granularity Specialists.} We find that several top-ranked features exhibit strong granularity specificity, appearing exclusively or predominantly at one level of abstraction. At the function level, features like \emph{Declarative Statements} and \emph{Declarative Code Lines} capture algorithmic characteristics reflecting how LLMs structure individual computational units differently from human programmers. At the class level, object-oriented design metrics such as \emph{Coupled Classes} emerge as critical discriminators, capturing structural composition and inter-class relationships that are inherently absent from standalone functions.

\textcolor{blue}{\paragraph{Category-Level Feature Importance.} The individual-feature analysis above raises a natural question: does the granularity effect manifest as a shift in \emph{which category} of metric dominates, not just which specific feature is selected? We answer this by aggregating each configuration's full set of selected-feature SHAP importances by the three metric categories introduced in Section~\ref{sec:metric-motivation} (comment-related stylometry, complexity, and structural-count stylometry), extending structural-count to include Instance Methods and Instance Variables and complexity to include the class-level coupling and inheritance metrics (Base Classes, Derived Classes, Coupled Classes, Coupled Classes Modified), consistent with the stylometry/complexity split already used in Appendix Table~\ref{tab:metrics_definitions}. Table~\ref{tab:rq3_category_shap} reports the results.}

\textcolor{blue}{Two patterns are directional and hold without exception across all four models. First, the comment category's share of total SHAP importance drops from function to class level in every model: from 29.5\% to 19.9\% (Claude 3 Haiku), 37.2\% to 26.9\% (Claude Haiku 4.5), 64.5\% to 29.8\% (GPT-3.5), and 48.5\% to 23.0\% (GPT-OSS). Averaged across models, comment-category share falls from 44.9\% at the function level to 24.9\% at the class level. Second, the structural-count category's share rises from function to class level in every model, from 41.9\% to 45.6\%, 41.1\% to 58.6\%, 16.5\% to 26.2\%, and 10.2\% to 34.8\% respectively, averaging 27.4\% at the function level and 41.3\% at the class level. The complexity category, by contrast, does not move monotonically: it increases for three of the four models but decreases for Claude Haiku 4.5 (21.7\% to 14.5\%), making it the least granularity-consistent of the three categories.}

\textcolor{blue}{This pattern also holds at the level of individual rank-1 features: at the function level, three of the four models have a comment-category feature as the single most important predictor; at the class level, only one does, with the top feature displaced by a structural-count metric (Claude 3 Haiku's \emph{Average Code Lines}, Claude Haiku 4.5's \emph{Average Blank Lines}) or a complexity metric (GPT-3.5's \emph{Maximum Nesting Depth}). Together with the near-zero cross-granularity Jaccard overlap reported below, this gives a more specific account of the granularity effect than ``different features are selected'': the comment-density signal that dominates function-level detection is systematically displaced, not merely supplemented, by structural verbosity and coupling signals at the class level, most consistently by the class-specific metrics (Instance Methods, Instance Variables, Coupled Classes) that have no function-level analogue.}

\begin{table}[h]
\centering
\caption{\textcolor{blue}{Category-level share of total SHAP importance per configuration. Comment = Comment Lines, Average Comment Lines, Comment to Code Ratio. Complexity = cyclomatic-family metrics, Essential Complexity, Maximum Nesting Depth, Logarithmic Paths, and (class level) Base/Derived/Coupled Classes. Structural-count = line/statement counts and (class level) Instance Methods/Variables.}}
\label{tab:rq3_category_shap}
\textcolor{blue}{
\begin{tabular}{llccc}
\toprule
\textbf{Model} & \textbf{Granularity} & \textbf{Comment} & \textbf{Complexity} & \textbf{Structural-count} \\
\midrule
Claude 3 Haiku & Function & 29.5\% & 28.6\% & 41.9\% \\
Claude Haiku 4.5 & Function & 37.2\% & 21.7\% & 41.1\% \\
GPT-3.5 & Function & 64.5\% & 18.9\% & 16.5\% \\
GPT-OSS & Function & 48.5\% & 41.3\% & 10.2\% \\
\midrule
Claude 3 Haiku & Class & 19.9\% & 34.5\% & 45.6\% \\
Claude Haiku 4.5 & Class & 26.9\% & 14.5\% & 58.6\% \\
GPT-3.5 & Class & 29.8\% & 44.1\% & 26.2\% \\
GPT-OSS & Class & 23.0\% & 42.2\% & 34.8\% \\
\bottomrule
\end{tabular}}
\end{table}

\begin{table}[h]
\centering
\caption{Top features by average SHAP importance across configurations. ``Frequency" indicates how many configurations the feature appears.}
\label{tab:rq3_top_features}
\begin{tabular}{l|c|c|c|c}
\toprule
\textbf{Feature} & \textbf{Frequency} & \textbf{Avg. SHAP} & \textbf{Max. SHAP} & \textbf{Top 3} \\
\midrule
Comment to Code Ratio & 8/8 & 0.777 & 3.795 & 7 \\
Average Blank Lines & 4/8 & 0.703 & 1.807 & 2 \\
Declarative Code Lines & 1/8 & 0.631 & 0.631 & 1 \\
Average Code Lines & 1/8 & 0.603 & 0.603 & 1 \\
Average Comment Lines & 3/8 & 0.594 & 0.740 & 3 \\
Maximum Nesting Depth & 6/8 & 0.401 & 1.034 & 2 \\
\bottomrule
\end{tabular}
\end{table}

\subsubsection{Configuration-Specific Variation}

Tables~\ref{tab:rq3_top3_function} and~\ref{tab:rq3_top3_class} show the top 3 features for each specific model. While \textbf{Comment to Code Ratio} is dominant, its magnitude and rank vary significantly across configurations, revealing distinct markers for different models.

\begin{table}[t]
\centering
\caption{Top 3 features by SHAP importance for function-level detection.}
\label{tab:rq3_top3_function}
\scriptsize
\begin{tabular}{ll|c|c|c|c|}
\cline{3-6}
                                    &        & \textbf{GPT-3.5}                                                              & \textbf{GPT-OSS}                                                              & \textbf{\begin{tabular}[c]{@{}c@{}}Claude 3\\Haiku\end{tabular}}              & \textbf{\begin{tabular}[c]{@{}c@{}}Claude\\Haiku 4.5\end{tabular}}            \\ \hline \hline
\multicolumn{1}{|l|}{\multirow{3}{*}{Rank 1}} & Feature & \textbf{\begin{tabular}[c]{@{}c@{}}Comment-to-\\Code Ratio\end{tabular}}      & \begin{tabular}[c]{@{}c@{}}Comment\\Lines\end{tabular}                        & \textbf{\begin{tabular}[c]{@{}c@{}}Comment-to-\\Code Ratio\end{tabular}}      & \begin{tabular}[c]{@{}c@{}}Declarative\\Statements\end{tabular}               \\ \cline{2-6} 
\multicolumn{1}{|l|}{}                        & SHAP   & 3.80                                                                          & 0.57                                                                          & 0.34                                                                          & 0.62                                                                          \\ \hline
\multicolumn{1}{|l|}{\multirow{3}{*}{Rank 2}} & Feature & \begin{tabular}[c]{@{}c@{}}Maximum\\Nesting\end{tabular}                      & \begin{tabular}[c]{@{}c@{}}Essential\\Complexity\end{tabular}                 & \begin{tabular}[c]{@{}c@{}}Blank\\Lines\end{tabular}                          & \begin{tabular}[c]{@{}c@{}}Comment\\Lines\end{tabular}                        \\ \cline{2-6} 
\multicolumn{1}{|l|}{}                        & SHAP   & 0.70                                                                          & 0.42                                                                          & 0.29                                                                          & 0.38                                                                          \\ \hline
\multicolumn{1}{|l|}{\multirow{3}{*}{Rank 3}} & Feature & \begin{tabular}[c]{@{}c@{}}Declarative\\Code Lines\end{tabular}               & \textbf{\begin{tabular}[c]{@{}c@{}}Comment-to-\\Code Ratio\end{tabular}}      & \begin{tabular}[c]{@{}c@{}}Declarative\\Statements\end{tabular}               & \textbf{\begin{tabular}[c]{@{}c@{}}Comment-to-\\Code Ratio\end{tabular}}      \\ \cline{2-6} 
\multicolumn{1}{|l|}{}                        & SHAP   & 0.63                                                                          & 0.32                                                                          & 0.19                                                                          & 0.18                                                                          \\ \hline
\end{tabular}
\end{table}

\begin{table}[t]
\centering
\caption{Top 3 features by SHAP importance for class-level detection.}
\label{tab:rq3_top3_class}
\scriptsize
\begin{tabular}{ll|c|c|c|c|}
\cline{3-6}
                                    &        & \textbf{GPT-3.5}                                                              & \textbf{GPT-OSS}                                                              & \textbf{\begin{tabular}[c]{@{}c@{}}Claude 3\\Haiku\end{tabular}}              & \textbf{\begin{tabular}[c]{@{}c@{}}Claude\\Haiku 4.5\end{tabular}}            \\ \hline \hline
\multicolumn{1}{|l|}{\multirow{3}{*}{Rank 1}} & Feature & \begin{tabular}[c]{@{}c@{}}Maximum\\Nesting\end{tabular}                      & \textbf{\begin{tabular}[c]{@{}c@{}}Comment-to-\\Code Ratio\end{tabular}}      & \begin{tabular}[c]{@{}c@{}}Average\\Code Lines\end{tabular}                   & \begin{tabular}[c]{@{}c@{}}Average\\Blank Lines\end{tabular}                  \\ \cline{2-6} 
\multicolumn{1}{|l|}{}                        & SHAP   & 1.03                                                                          & 0.55                                                                          & 0.60                                                                          & 1.81                                                                          \\ \hline
\multicolumn{1}{|l|}{\multirow{3}{*}{Rank 2}} & Feature & \begin{tabular}[c]{@{}c@{}}Average\\Comment Lines\end{tabular}                & \begin{tabular}[c]{@{}c@{}}Modified\\Classes\end{tabular}                     & \begin{tabular}[c]{@{}c@{}}Modified\\Classes\end{tabular}                     & \begin{tabular}[c]{@{}c@{}}Average\\Comment Lines\end{tabular}                \\ \cline{2-6} 
\multicolumn{1}{|l|}{}                        & SHAP   & 0.74                                                                          & 0.47                                                                          & 0.34                                                                          & 0.73                                                                          \\ \hline
\multicolumn{1}{|l|}{\multirow{3}{*}{Rank 3}} & Feature & \textbf{\begin{tabular}[c]{@{}c@{}}Comment-to-\\Code Ratio\end{tabular}}      & \begin{tabular}[c]{@{}c@{}}Average\\Blank Lines\end{tabular}                  & \begin{tabular}[c]{@{}c@{}}Average\\Comment Lines\end{tabular}                & \textbf{\begin{tabular}[c]{@{}c@{}}Comment-to-\\Code Ratio\end{tabular}}      \\ \cline{2-6} 
\multicolumn{1}{|l|}{}                        & SHAP   & 0.54                                                                          & 0.36                                                                          & 0.32                                                                          & 0.28                                                                          \\ \hline
\end{tabular}
\end{table}

\begin{itemize}[leftmargin=*]
    \item \textbf{The Outlier (GPT-3.5 Function):} \emph{Comment to Code Ratio} has an extreme SHAP value of \textbf{3.795}, which is 4.9 times higher than the global average. This single feature can effectively solve the detection task for GPT-3.5.
    \item \textbf{The Human-Mimic (Claude Haiku 4.5):} This model successfully suppresses the ``Comment Ratio" signal. It is the only model where \emph{Comment to Code Ratio} falls to 3rd place at both function and class levels. Instead, detection relies on subtle structural markers like \emph{Declarative Statements} (0.62) at the function level and \emph{Average Blank Lines} (1.81) at the class level.
    \item \textbf{The Verbose Coder (Claude 3 Haiku Class):} Uniquely, this configuration is primarily detected by \emph{Average Code Lines} (0.60), confirming RQ1 findings that Claude 3 Haiku's class structures differ significantly in length from human norms.
    \item \textbf{The Absolute Commenter (GPT-OSS):} While most models are detected by the \emph{ratio} of comments, GPT-OSS function-level detection is driven more by the \emph{absolute count} of \emph{Comment Lines} (0.57). This suggests that while it may get the ratio right, the sheer volume of documentation it generates remains statistically distinct.
\end{itemize}

\begin{figure}[t]
\centering
\begin{subfigure}[b]{0.48\textwidth}
    \includegraphics[width=\textwidth]{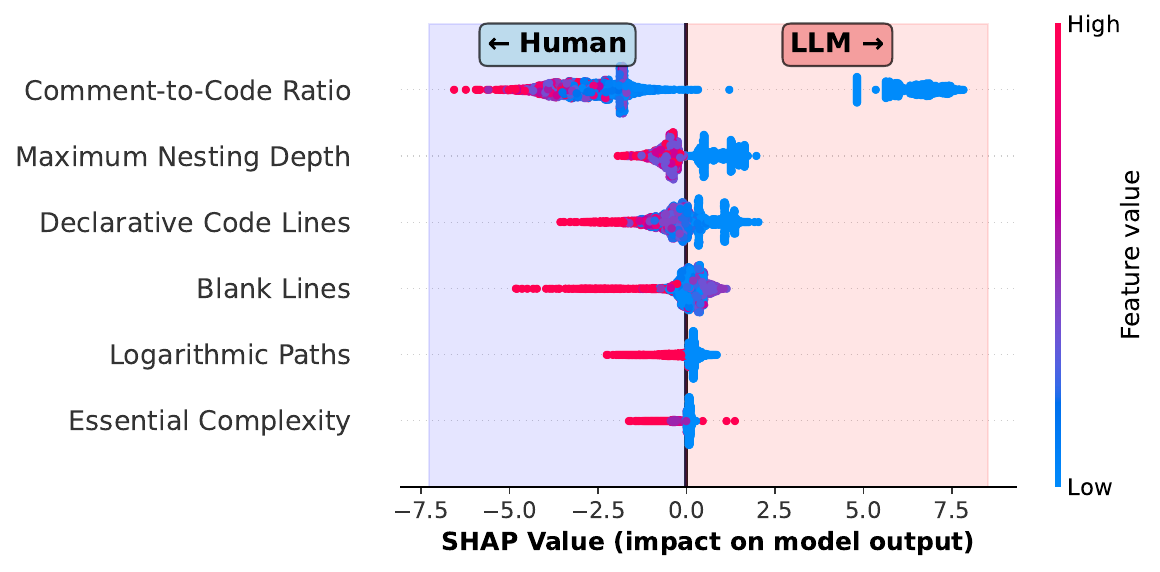}
    \caption{GPT-3.5 Function-level}
\end{subfigure}
\begin{subfigure}[b]{0.48\textwidth}
    \includegraphics[width=\textwidth]{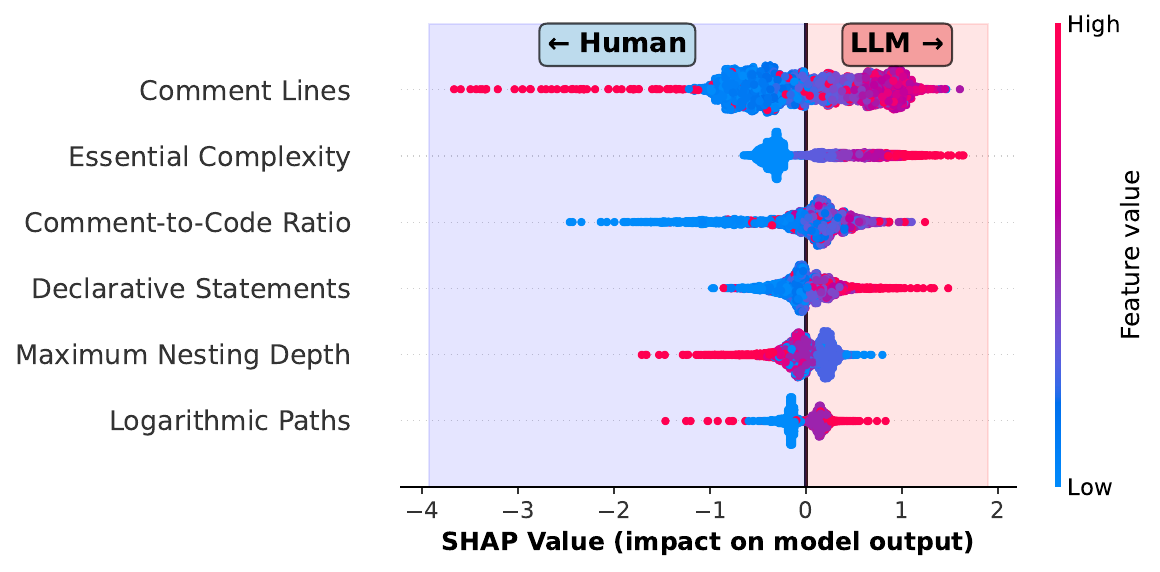}
    \caption{GPT-OSS Function-level}
\end{subfigure}

\begin{subfigure}[b]{0.48\textwidth}
    \includegraphics[width=\textwidth]{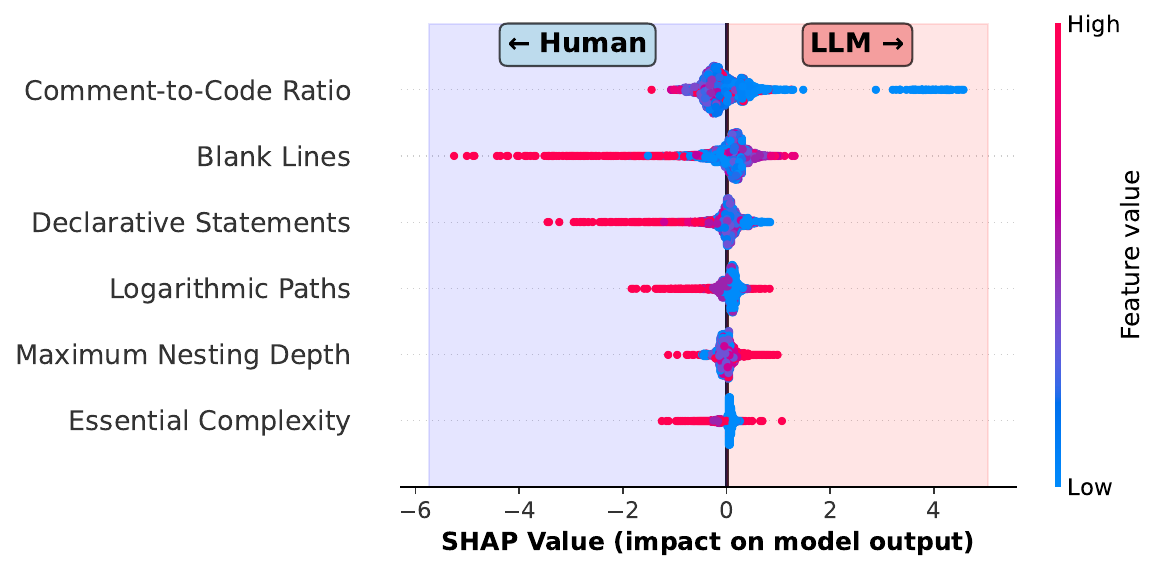}
    \caption{Claude 3 Haiku Function-level}
\end{subfigure}
\begin{subfigure}[b]{0.48\textwidth}
    \includegraphics[width=\textwidth]{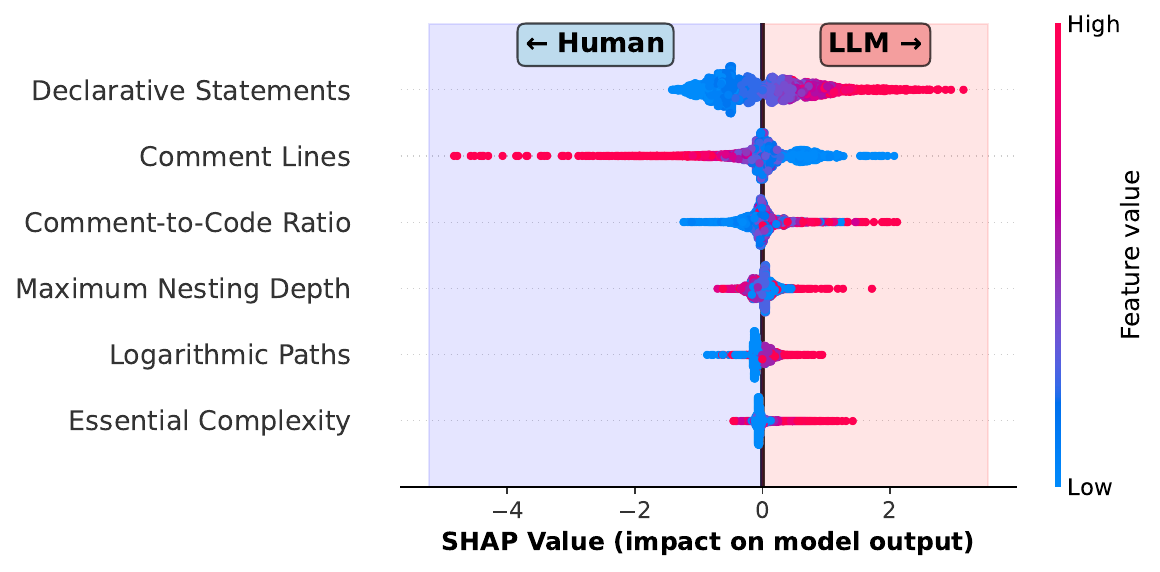}
    \caption{Claude Haiku 4.5 Function-level}
\end{subfigure}

\caption{SHAP beeswarm plots for \textbf{function-level} detection across all four LLMs. Each point is a test instance; the x-axis position shows impact on prediction (left pushes toward ``Human'', right toward ``LLM'').}
\label{fig:rq3_shap_function}
\end{figure}

\begin{figure}[t]
\centering
\begin{subfigure}[b]{0.48\textwidth}
    \includegraphics[width=\textwidth]{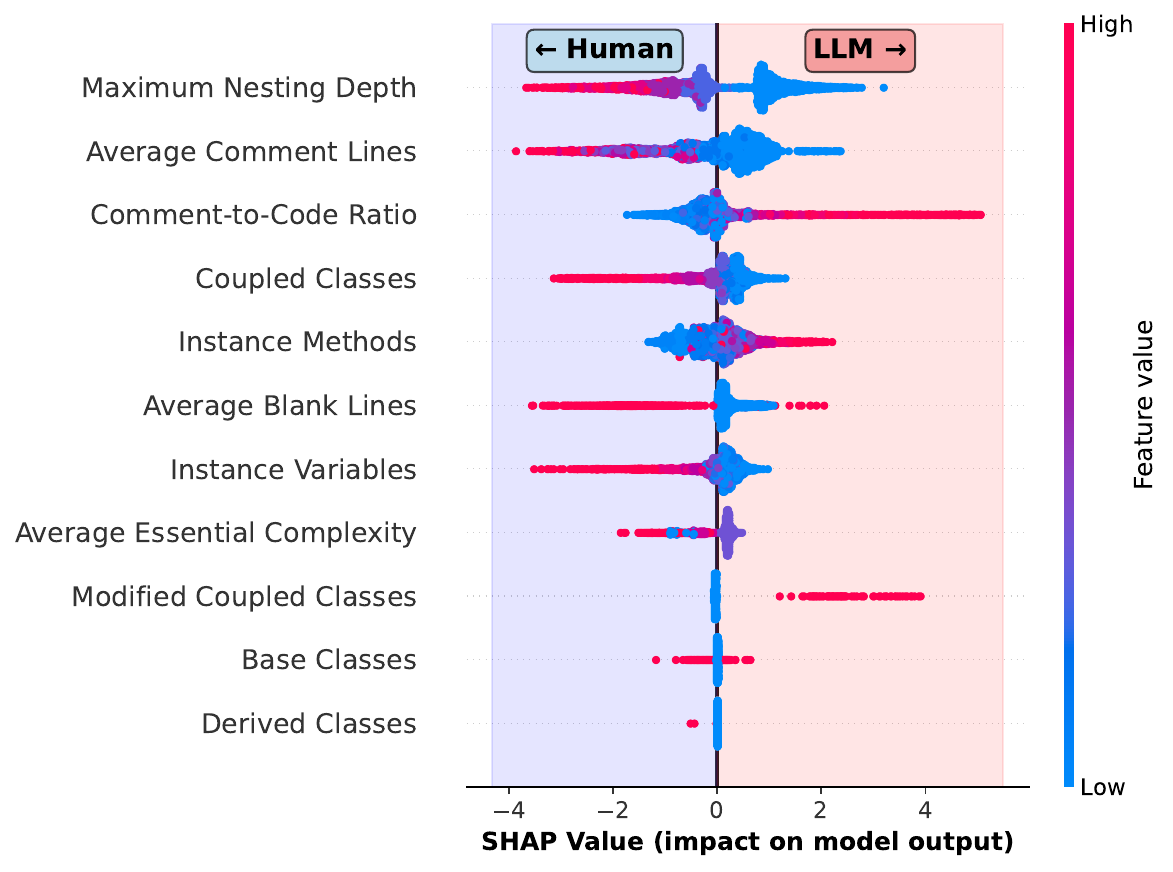}
    \caption{GPT-3.5 Class-level}
\end{subfigure}
\begin{subfigure}[b]{0.48\textwidth}
    \includegraphics[width=\textwidth]{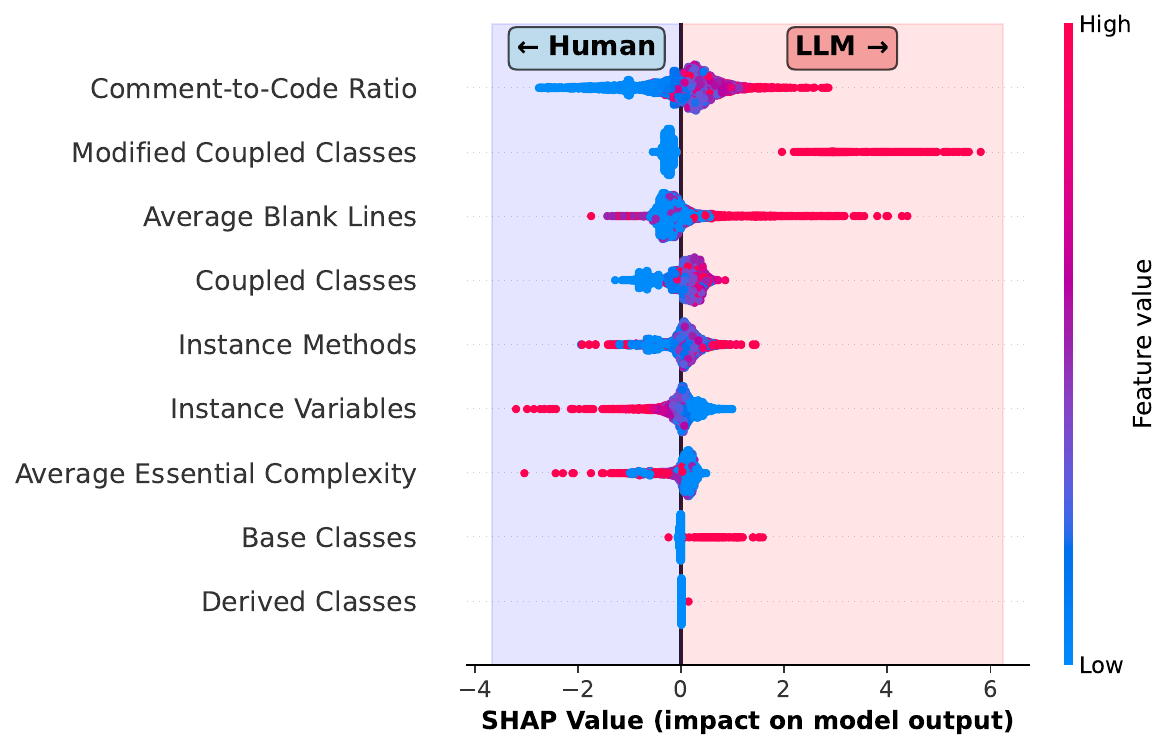}
    \caption{GPT-OSS Class-level}
\end{subfigure}

\begin{subfigure}[b]{0.48\textwidth}
    \includegraphics[width=\textwidth]{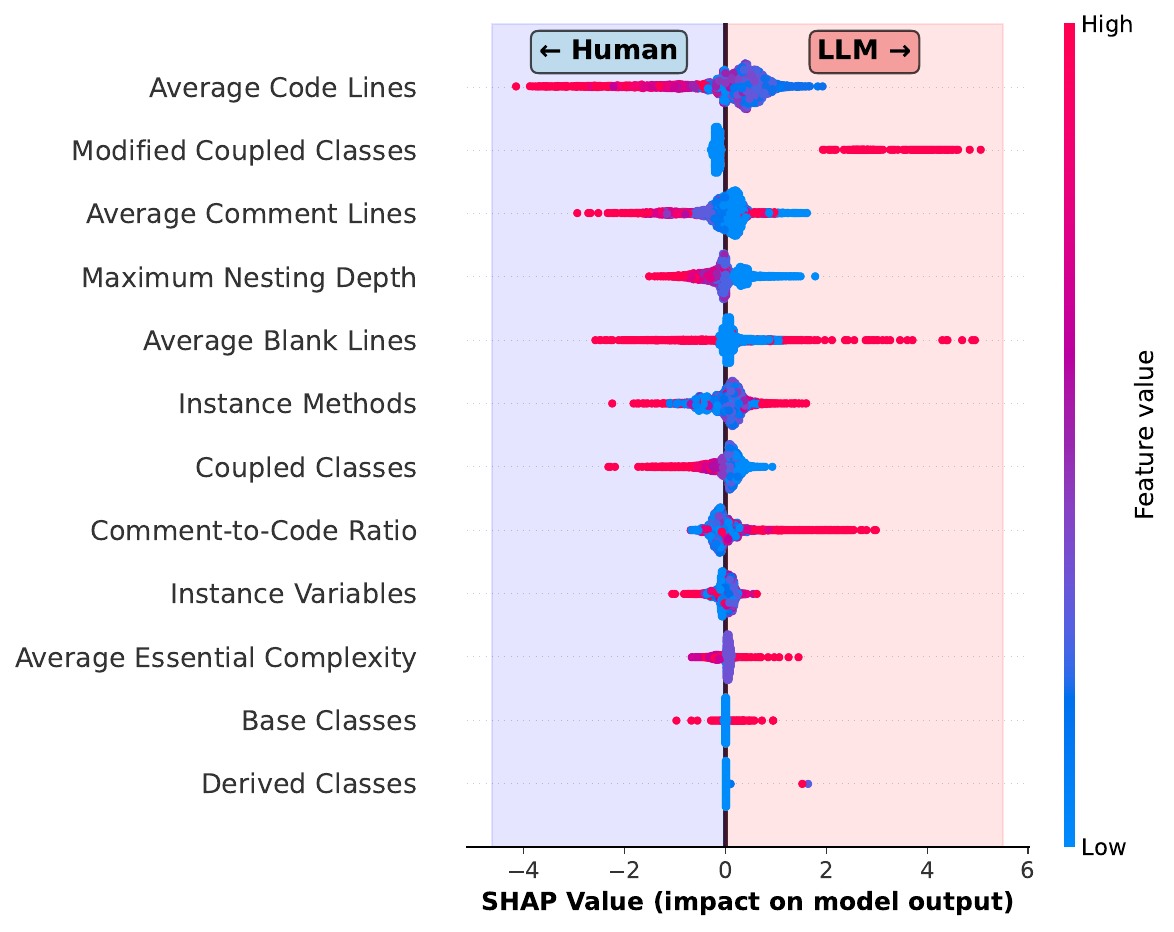}
    \caption{Claude 3 Haiku Class-level}
\end{subfigure}
\begin{subfigure}[b]{0.48\textwidth}
    \includegraphics[width=\textwidth]{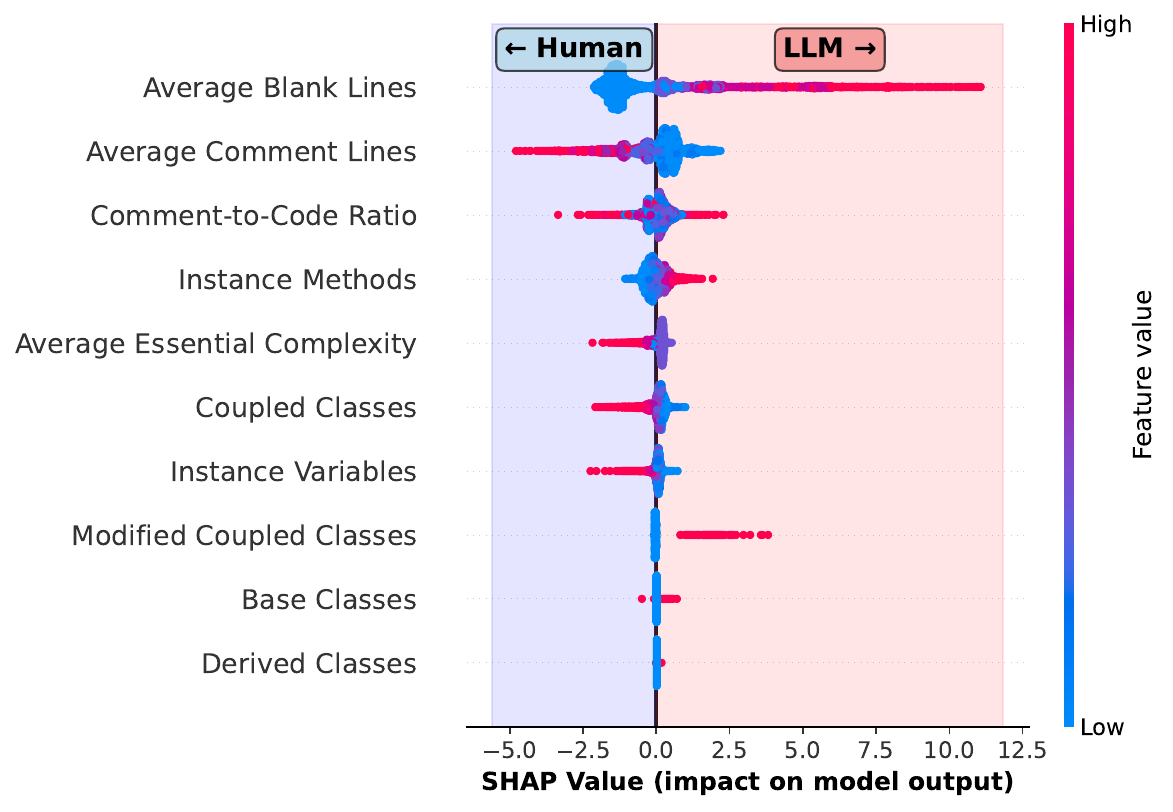}
    \caption{Claude Haiku 4.5 Class-level}
\end{subfigure}

\caption{SHAP beeswarm plots for \textbf{class-level} detection across all four LLMs. Compared to the function-level plots in Figure~\ref{fig:rq3_shap_function}, feature importance is more distributed.}
\label{fig:rq3_shap_class}
\end{figure}

Figures~\ref{fig:rq3_shap_function} and~\ref{fig:rq3_shap_class} visualize these distinctions via SHAP beeswarm plots. In the function-level comparisons (Figure~\ref{fig:rq3_shap_function}), observe the \emph{Comment to Code Ratio} feature at the top of the y-axis. For GPT-3.5, the ``tail" of this feature extends far to the right, indicating a massive contribution toward predicting ``LLM" authorship. In contrast, the same feature for Claude 3 Haiku shows a compressed distribution, reflecting its much lower discriminative power. At the class level (Figure~\ref{fig:rq3_shap_class}), the hierarchy shifts: models like Claude Haiku 4.5 show a more distributed feature importance, with architectural metrics like \emph{Average Blank Lines} overtaking commenting patterns.
 
\subsection{Interpretation}

Our feature analysis clarifies the mechanisms behind the detection results in RQ2.

\paragraph{1. Granularity Overrides Architecture.}
The most critical insight is that code granularity affects the structural signature more strongly than the model architecture itself, as evident by the Jaccard coefficient values. This explains the ``Inversion Effect" observed in RQ2. Our study shows that detection is not about finding a single ``LLM fingerprint"; it is about identifying two distinct signatures:
\begin{itemize}
    \item \textbf{The Micro-Signature (Function):} Relies on \emph{implementation details} like statement counts, comment density, and nesting depths.
    \item \textbf{The Macro-Signature (Class):} Relies on \emph{implementation details} as well as \emph{organizational structure} like whitespace management and coupling.
\end{itemize}
This proves that ``universality" in detection is impractical unless the detector is granularity-aware.

\paragraph{2. The ``Comment Gap" and Human-Likeness.}
The high-frequency presence of \textit{Comment to Code Ratio} indicates that all LLMs consistently ``over-comment" relative to humans. However, the \emph{magnitude} of this feature acts as a proxy for ``human-likeness":
\begin{itemize}[leftmargin=*]
    \item \textbf{GPT-3.5 (SHAP 3.80):} The massive importance value confirms it is an outlier. Its commenting behaviour is so distinct from humans that it becomes a clear indicator.
    \item \textbf{Claude Haiku 4.5 (SHAP 0.18):} The low importance value suggests its commenting style has converged significantly toward human norms, making it much harder to detect via this metric alone.
\end{itemize}

\paragraph{3. Discriminative Markers are Context-Dependent.}
We find that specific structural markers only emerge in specific contexts. For example, \textbf{Maximum Nesting Depth} (complexity) is a top predictor for GPT-3.5 but not as strong a predictor for Claude models. This supports the ``Directional Split" from RQ1: GPT-3.5 is detected by its \emph{simplicity} (low nesting), whereas other models are identified by alternative structural anomalies. This heterogeneity confirms that effective detection requires model-specific feature weighting rather than a one-size-fits-all metric threshold.

\begin{tcolorbox}[colback=gray!10, colframe=black, boxrule=0.5pt, arc=2mm]
\textbf{Answer to RQ3:} Granularity drives feature selection more than model architecture, resulting in disjoint structural signatures between function and class levels ($J < 0.1$). While \textit{Comment to Code Ratio} is the sole universal discriminator, its predictive weight varies, necessitating detection strategies tailored to the specific level of code abstraction.
\end{tcolorbox}

%

\section{RQ4: \rqiv}
\label{sec:rqiv}

The previous three research questions established that LLM-generated code differs structurally from human-written code (RQ1), that structural classifiers can exploit these differences for detection (RQ2), and that the discriminative power of these classifiers concentrates on a small set of features dominated by Comment to Code Ratio and related comment-density metrics (RQ3). This concentration of discriminative power motivates a natural follow-up question: if a developer adopting LLM-generated code knows that comment density is the dominant detection signal, can they evade the classifier by manipulating that single class of features? In this section, we test this question empirically.

\subsection{Approach}
\label{sec:rqiv_approach}

\textbf{Threat Model.} We consider a developer who has already generated code from an LLM and wishes to pass it off as human-written. The developer modifies the generated code by removing all comments before submitting it for review. The developer does not retrain or fine-tune the LLM and does not change the generation prompt; the modification is purely post-hoc. The detector encounters the modified code without warning.

We target comment removal specifically because RQ3 identified Comment to Code Ratio as the dominant discriminator in most model--granularity configurations, with other comment-related features (Comment Lines, Average Comment Lines) ranking among the top three discriminators for several others. A targeted attack on the most discriminative class of features is the strongest test of how brittle detection is when its dominant signal is removed.

We frame the attack as a post-hoc transformation rather than as regeneration under a different prompt (e.g., ``do not add comments''). Two reasons motivate this choice. First, post-hoc removal is the lower-cost attack: it requires no API access and no additional inference. Second, regenerating under a different prompt confounds the analysis. A re-prompted LLM may change not only its commenting behaviour but also its variable naming, structural patterns, and verbosity, making it impossible to attribute the resulting change in detection performance to comment removal specifically. The post-hoc approach isolates the variable cleanly.

\textcolor{blue}{This choice is also consistent with established practice in the closely related detection literature. None of the empirical studies we are aware of test comment or gamifiable-feature evasion via regeneration under a modified prompt; where they test analogous robustness questions, they consistently do so post-hoc. Suh et al.~\citep{suh2024empirical} construct comment-free, uniform-variable-name, and uniform-method-name code variants by editing the abstract syntax tree of already-generated code with Tree-Sitter, rather than re-prompting the generating model. Idialu et al.~\citep{idialu2024whodunit} train a classifier variant that excludes gameable features (e.g., empty lines, whitespace) from the feature space entirely, reporting that it still achieves an F1-score and AUC-ROC of 0.89, against 0.91 for the full feature set. Nguyen et al.~\citep{nguyen2024gptsniffer} report that removing comments as a preprocessing step improves their classifier's accuracy. Yang et al.~\citep{yang2023zero} test a related but distinct ``revision attack'' in which a separate model masks and refills random lines of already-generated code, again post-hoc rather than through re-prompting. We are not aware of any published work in this space, including Ye et al.~\citep{ye2025uncovering}, Xu et al.~\citep{xu2025distinguishing}, Park et al.~\citep{park2025detection}, Xu et al.~\citep{xu2024one}, and Shi et al.~\citep{shi2025between}, that tests detector robustness via a prompt-level instruction to the generating model.}

\textbf{Operationalization.} Our trained classifiers consume \textit{Understand}-extracted feature vectors, not raw code. Removing all comments from a code sample affects only three features in the \textit{Understand} catalogue: \textit{Comment Lines}, \textit{Average Comment Lines}, and \textit{Comment to Code Ratio}. Per \textit{Understand}'s documentation, \textit{Code Lines} counts any line containing executable code, regardless of whether the line also contains a comment, so comment removal leaves \textit{Code Lines} invariant.\footnote{See \url{https://blog.scitools.com/numerical-evidence/}. A line such as \textit{x = 1\# initialize} counts toward both \textit{Code Lines} and \textit{Comment Lines}; stripping the comment leaves \textit{Code Lines} unchanged.}

\textbf{Evaluation Protocol.} For each of the eight model-granularity configurations, we evaluate the trained classifier on the full held-out test set (5{,}794 function-level samples and 4{,}766 class-level samples per configuration). We compute baseline metrics on the unmodified test set and adversarial metrics on the same test set after the comment-feature zeroing transformation has been applied to LLM-labelled rows. We report mean, and 95\% bootstrap confidence intervals from 1{,}000 resamples, matching the protocol used elsewhere in this paper.

\subsection{Findings}
\label{sec:rqiv_findings}

Table~\ref{tab:adversarial_auc} reports baseline and adversarial AUC-ROC for all eight configurations, with 95\% confidence intervals and the resulting delta. Comment stripping has an asymmetric effect across configurations: three classifiers become substantially \emph{more} accurate under the transformation, four are essentially unaffected, and one, GPT-OSS at the class level, collapses to near-random detection performance.

\begin{table}[t]
\centering
\caption{Detection performance under comment stripping across all eight model-granularity configurations. Each row is evaluated on the full held-out test set with no retraining. Values are mean [95\% CI] from a 1{,}000-iteration bootstrap. The $\Delta$ column reports the change in AUC mean from baseline to adversarial. The single configuration where the adversarial attack succeeds (GPT-OSS, class level) is highlighted.}
\label{tab:adversarial_auc}
\resizebox{\textwidth}{!}{
\begin{tabular}{llccc}
\toprule
\textbf{Model} & \textbf{Granularity} & \textbf{Baseline AUC} & \textbf{Adversarial AUC} & \textbf{$\Delta$ AUC} \\
\midrule
Claude 3 Haiku   & Function & 0.681 [\textcolor{blue}{0.667, 0.694}] & 0.985 [0.981, 0.988] & +0.304 \\
Claude 3 Haiku   & Class    & \textcolor{blue}{0.830} [0.818, 0.841] & 0.858 [0.848, 0.868] & \textcolor{blue}{+0.028} \\
Claude Haiku 4.5 & Function & \textcolor{blue}{0.714} [0.700, 0.726] & 0.994 [0.993, 0.996] & \textcolor{blue}{+0.280} \\
Claude Haiku 4.5 & Class    & 0.806 [0.795, 0.817] & 0.788 [0.775, 0.801] & --0.018 \\
GPT-3.5          & Function & 0.961 [0.956, 0.965] & 1.000 [0.999, 1.000] & +0.039 \\
GPT-3.5          & Class    & 0.887 [0.879, 0.896] & 0.903 [0.894, 0.911] & +0.015 \\
GPT-OSS          & Function & 0.795 [0.784, 0.807] & 0.850 [0.840, 0.860] & +0.055 \\
\textbf{GPT-OSS} & \textbf{Class} & \textbf{0.808 [0.796, 0.821]} & \textbf{0.490 [0.473, 0.506]} & \textbf{--0.319} \\
\bottomrule
\end{tabular}}
\end{table}

\textbf{The GPT-OSS class-level collapse.} For GPT-OSS class-level detection, adversarial AUC reaches 0.490 with the upper 95\% confidence bound at 0.506, making it statistically indistinguishable from random guessing. The collapse is not limited to AUC: recall drops from 0.746 to 0.312 (the classifier now misses 69\% of LLM-generated samples versus 25\% at baseline), and Matthews correlation coefficient drops from 0.458 to 0.025, indicating that the predictions and the true labels become essentially uncorrelated. This is the only configuration in our study where the adversarial attack matches the catastrophic-collapse mental model implicit in the comment-stripping threat.

\textbf{Function-level amplification.} For three function-level configurations, comment stripping makes the classifiers substantially \emph{more} accurate. Claude 3 Haiku function-level AUC rises from 0.681 to 0.985 ($\Delta = +0.304$); Claude Haiku 4.5 function-level AUC rises from \textcolor{blue}{0.714} to 0.994 ($\Delta = \textcolor{blue}{+0.280}$); and the GPT-3.5 function-level classifier shows a smaller delta ($\Delta = +0.039$) because baseline detection was already near-ceiling at AUC 0.961. The counterintuitive direction of these changes---adversarial transformation \emph{improving} detection---requires a feature-attribution explanation we provide in Section~\ref{sec:rqiv_interpretation}.

\textbf{Largely-unaffected configurations.} The four remaining configurations (Claude 3 Haiku class, Claude Haiku 4.5 class, GPT-3.5 class, GPT-OSS function) show small positive or near-zero deltas, and detection performance does not change meaningfully under comment stripping. The fact that detection survives the perturbation in these cases suggests that the classifiers in question rely on a broader set of structural features than comments alone.

\subsection{Interpretation}
\label{sec:rqiv_interpretation}

The asymmetric pattern documented in Section~\ref{sec:rqiv_findings} calls for explanation rather than only description: out of eight model-granularity configurations, three classifiers became substantially more accurate under comment stripping, four were essentially unaffected, and one collapsed to near-random performance. Similar to RQ3 (Section~\ref{sec:rqiii}), we use SHAP-based feature attribution to explain this asymmetry. The mechanism that governs an individual configuration's response to comment stripping turns out to be the joint product of two factors: (i) which feature dominates the trained classifier's predictions, as captured by SHAP magnitudes, and (ii) the direction of the LLM-versus-human difference for that feature, as established in Section~\ref{sec:rqi}. Three qualitatively distinct mechanisms emerge from this joint structure, and they together explain the asymmetric outcome.

We compute mean-absolute SHAP values for each trained classifier on both the baseline and the adversarial feature matrices, using CatBoost's native SHAP implementation. Tables~\ref{tab:adv_shap_gpt_oss_class} and~\ref{tab:adv_shap_claude_haiku45_function} report the top three features by baseline SHAP importance for the two most representative configurations and their behaviour under the adversarial transformation. The remaining six configurations are reported in Appendix~\ref{appendix:adversarial_shap}.

\begin{table}[t]
\centering
\small
\caption{Top three features by baseline SHAP importance for the GPT-OSS class-level classifier, with their SHAP values under the adversarial (comment-stripping) transformation. The dominant feature (Comment to Code Ratio) increases in SHAP magnitude by 90\% under the transformation, but the direction of its influence becomes systematically wrong, driving the AUC collapse from 0.808 to 0.490.}
\label{tab:adv_shap_gpt_oss_class}
\resizebox{\textwidth}{!}{
\begin{tabular}{clcccc}
\toprule
\textbf{Rank} & \textbf{Feature} & \textbf{Baseline SHAP} & \textbf{Adversarial SHAP} & \textbf{$\Delta$} & \textbf{$\Delta$\%} \\
\midrule
1 & Comment to Code Ratio & 0.55 & 1.04 & +0.49 & +90\% \\
2 & Coupled Classes (Modified) & 0.47 & 0.41 & -0.06 & -13\% \\
3 & Average Blank Lines & 0.36 & 0.58 & +0.22 & +62\% \\
\bottomrule
\end{tabular}}
\end{table}

\begin{table}[t]
\centering
\small
\caption{Top three features by baseline SHAP importance for the Claude Haiku 4.5 function-level classifier, with their SHAP values under the adversarial (comment-stripping) transformation. Comment-related features (ranks 2 and 3) become dominant under the transformation, pushing the AUC from \textcolor{blue}{0.714} to 0.994.}
\label{tab:adv_shap_claude_haiku45_function}
\resizebox{\textwidth}{!}{
\begin{tabular}{clcccc}
\toprule
\textbf{Rank} & \textbf{Feature} & \textbf{Baseline SHAP} & \textbf{Adversarial SHAP} & \textbf{$\Delta$} & \textbf{$\Delta$\%} \\
\midrule
1 & Declarative Statements & 0.62 & 0.54 & -0.08 & -13\% \\
2 & Comment Lines & 0.38 & 1.04 & +0.66 & +172\% \\
3 & Comment to Code Ratio & 0.18 & 0.57 & +0.39 & +220\% \\
\bottomrule
\end{tabular}}
\end{table}

\textbf{Mechanism 1: Direction-flip collapse.} For the GPT-OSS class-level classifier (Table~\ref{tab:adv_shap_gpt_oss_class}), Comment to Code Ratio is the highest-importance feature in the baseline ranking, with mean $|$SHAP$|$~$=$~0.55. RQ1 established that GPT-OSS at the class level produces code with \emph{more} comments than human authors (see Table~\ref{tab:gptoss_class}), so the classifier learned the decision rule ``high comment ratio implies LLM-generated.'' Under the adversarial transformation, LLM samples are forced to \textit{Comment to Code Ratio}~$=$~$0$, which the classifier reads as the human-like end of the spectrum. The feature's SHAP magnitude actually grows after the transformation, to 1.04: the classifier relies on the feature \emph{more}, not less. But the direction in which it relies on the feature has become systematically wrong, and the classifier becomes confidently incorrect. Because no other feature in the GPT-OSS class-level selection carries comparable importance, no corrective signal emerges, and detection performance collapses to chance.

\textbf{Mechanism 2: Alignment amplification.} For three function-level configurations (Claude 3 Haiku, Claude Haiku 4.5, and GPT-3.5), RQ1 established that the LLMs produce code with \emph{fewer} or comparable numbers of comments to human authors (see Table~\ref{tab:claude3_func}, Table~\ref{tab:claude45_func}, and Table~\ref{tab:gpt35_func}). The classifiers learned the decision rule ``low comment values imply LLM-generated.'' Under the adversarial transformation, LLM samples are forced to the extreme low end of the comment-feature scale, amplifying rather than weakening the classifier's correct signal. Table~\ref{tab:adv_shap_claude_haiku45_function} shows this clearly for the Claude Haiku 4.5 function-level: Comment Lines grows in mean $|$SHAP$|$ from 0.38 (baseline) to 1.04 (adversarial), and Comment to Code Ratio from 0.18 to 0.57. Both features become near-deterministic indicators of LLM origin once they take the constant value zero across all LLM samples. The decision boundary that the classifier learned at training time happens to align with the adversary's transformation, and accuracy increases as a consequence.

\textbf{Mechanism 3: Robustness through redundancy.} For the four configurations where comment stripping has a small or near-zero effect on AUC (Claude 3 Haiku class, Claude Haiku 4.5 class, GPT-3.5 class, GPT-OSS function), the dominant SHAP features are not entirely comment-related (see Figures~\ref{fig:rq3_shap_class} and ~\ref{fig:rq3_shap_function}). The Claude Haiku 4.5 class-level classifier, for instance, is dominated by Average Blank Lines (mean $|$SHAP$|$~$=$~1.81), with comment features secondary; the GPT-3.5 class-level classifier is dominated by Maximum Nesting ($|$SHAP$|$~$=$~1.03). These classifiers rely on a portfolio of structural features rather than on comments alone, and the discriminative power of their non-comment features is unaffected by the adversarial transformation.

\textbf{The catastrophic-collapse case is the unique intersection.} The three mechanisms above are not equally common, and their joint structure explains why catastrophic detection failure occurs in exactly one of the eight configurations. Mechanism~1 (collapse) requires a classifier that relies dominantly on a comment-related feature \emph{and} an LLM-human directional configuration where stripping moves LLM samples toward the human distribution. Mechanism~2 (amplification) is the same dominance condition combined with the opposite directional configuration. Mechanism~3 (redundancy) covers all configurations where the dominance condition does not hold. Of the eight configurations we tested, only GPT-OSS class-level satisfies both conditions for Mechanism~1 simultaneously; the other configurations fall into Mechanism~2 (three configurations) or Mechanism~3 (four configurations). The uniqueness of the collapse outcome is therefore a structural property of the specific intersection of feature-importance distribution and feature-direction direction, not a coincidence.

\textbf{Implications for detector design.} Two design implications follow from this mechanism analysis. First, classifiers whose decisions are dominated by a single high-SHAP feature are intrinsically more fragile under adversarial perturbation: even when that single feature is the strongest individual discriminator at training time, its dominance becomes a vulnerability if the adversary knows the feature and can manipulate its value. The GPT-OSS class-level case demonstrates that this vulnerability can be exploited with a transformation as low-cost as removing comments. Detector training procedures that explicitly penalize single-feature dominance, for example, by adding regularization that flattens the SHAP-importance distribution across features, may improve adversarial robustness at some cost to baseline accuracy. We do not evaluate this tradeoff here, but the mechanism analysis suggests it is worth investigating. Second, the asymmetry across configurations means that ensembles over model-specific classifiers may be more robust than any single classifier in the ensemble, since the configurations that collapse under one adversarial transformation are not necessarily the same as those that would collapse under others.

\begin{tcolorbox}[colback=gray!10, colframe=black, boxrule=0.5pt, arc=2mm]
\textbf{Answer to RQ4:} Adversarial robustness to comment stripping is highly asymmetric across the eight model--granularity configurations. Seven of the eight classifiers are either unaffected or made \emph{more} accurate by the transformation, while one (GPT-OSS at the class level) collapses from AUC 0.808 to AUC 0.490. The function-level Claude 3 Haiku and Claude Haiku 4.5 classifiers, by contrast, become substantially more accurate under the same transformation, with AUC rising from 0.681 to 0.985 and from \textcolor{blue}{0.714} to 0.994, respectively. The asymmetry indicates that adversarial vulnerability is a configuration-specific property that does not align with baseline detection performance.
\end{tcolorbox}

\section{Discussion}
\label{sec:discussion}

Our comparative analysis of LLM-generated code detection reveals critical insights that challenge prevailing assumptions in existing literature and highlight the complexity of building robust detection systems in practice.

\subsection{The GPT-3.5 Detection Paradox}

A systematic review of the literature reveals a striking bias: the vast majority of LLM code detection research focuses predominantly on GPT-3.5. Studies by Xu et al.~\cite{xu2025distinguishing}, Nguyen et al.~\cite{nguyen2024gptsniffer}, Shi et al.~\cite{shi2025between}, and Yang et al.~\cite{yang2023zero} have consistently reported high detection performance for GPT-3.5, with AUC-ROC scores frequently exceeding 0.90. These results have established an implicit narrative that LLM-generated code detection is a largely solved problem.

Our findings demonstrate that while GPT-3.5 exhibits exceptional detectability (AUC-ROC 0.961 at function-level; 0.887 at class-level), this characteristic is \emph{an anomaly rather than a baseline}. Contemporary models like Claude 3 Haiku, Claude Haiku 4.5, and GPT-OSS show substantially lower detectability, creating a performance gap of up to \textcolor{blue}{28.0} percentage points. The field's heavy reliance on GPT-3.5 creates a misleading impression of generalizability, when in reality, its high detectability stems from unique stylometric signatures that newer models have effectively shed.

\subsection{The Fallacy of Cross-LLM Evaluation}

A concerning pattern in existing literature~\citep{suh2024empirical} involves evaluating detectors trained on one LLM by testing them on others, framing the resulting degradation as a limitation of the detector. Our findings reveal a fundamental flaw in this paradigm: \emph{different LLMs imprint fundamentally different structural distributions}. While feature sets may overlap, the magnitude of their discriminative signals varies dramatically.

Consequently, when a GPT-3.5-trained detector fails on Claude-generated code, it is not a failure of the model, but a change in the task definition. Each LLM requires a detector calibrated to its specific structural fingerprint. Cross-LLM evaluation without retraining measures robustness to distribution shift, not intrinsic detector quality.

\subsection{Generalization to Post-Cutoff Repositories}
\label{sec:temporal-gap}

Beyond cross-model differences, our trained classifiers exhibit substantial brittleness when applied to repositories that lie outside the training-data distribution of both our classifiers and the LLMs that generated the code. We evaluated all eight trained models on a smaller secondary dataset assembled exclusively from GitHub repositories created after December~31, 2024 ($n=846$ functions, $555$ classes). This cutoff date was chosen to lie after the training cutoffs of all four LLMs in our study, so that no code in this secondary dataset could have been seen during LLM training. The results show substantial degradation: AUC-ROC drops by an average of 26.3 percentage points (range: 14.7--39.7~pp) relative to the main test set.

All eight configurations fall outside their main-test-set 95\% confidence intervals on the post-cutoff data. In several cases, performance collapses to levels indistinguishable from random guessing (e.g., Claude 3 Haiku class-level: AUC 0.432; Claude Haiku 4.5 function-level: AUC 0.372). GPT-3.5 remains the most detectable model relative to the others on the post-cutoff data, but its absolute detection performance also suffers a significant drop. The breadth of the effect across configurations indicates that the brittleness is not specific to one LLM, one granularity, or one detector.

\paragraph{Two non-exclusive interpretations of the degradation.} \textcolor{blue}{Because the training corpora of the four LLMs are not publicly disclosed, we cannot verify whether they overlap with our data, and the observed degradation is therefore consistent with two non-exclusive mechanisms.} The first interpretation is \emph{temporal generalization brittleness}: LLM generation behaviour is evolving rapidly, both because of underlying model improvements and because of fine-tuning shifts over time, and classifiers trained on earlier outputs fail to recognize newer ones. Under this interpretation, the post-cutoff drop reflects a genuine distribution shift in LLM behaviour. The second interpretation is \emph{data contamination}: LLMs may have been trained on portions of the CodeSearchNet corpus from which our main dataset is drawn, in which case the high baseline performance on the main test set is partially an artifact of LLMs reproducing memorized human-written patterns that the structural classifiers happen to discriminate against the LLMs' generated output. Under this interpretation, the post-cutoff drop reflects the disappearance of an artifactual boost that was present on the main test set but cannot be present on a corpus the LLMs have never seen. Both mechanisms would produce the asymmetry we observe between the main and post-cutoff test sets, and both are individually consistent with the magnitude and directional pattern of the drop. \textcolor{blue}{The drop may also be partly driven by differences in data characteristics between the post-cutoff and main datasets (e.g., code complexity and length), which would affect detection performance independently of either mechanism.} \textcolor{blue}{Disentangling them would require access to the training corpus composition of the four LLMs we studied.} We therefore report the magnitude of the effect without claiming a specific causal mechanism.

\paragraph{Implications for detection methodology.} Whichever mechanism dominates, the practical implication is the same: classifiers trained on standard, GitHub-derived datasets cannot be assumed to generalize to repositories outside the training distribution of either the classifier or the underlying LLM. If the dominant mechanism is temporal drift, detection systems must be retrained or updated continuously to keep pace with LLM evolution. If the dominant mechanism is contamination, baseline performance on standard benchmarks is systematically optimistic and contamination-free evaluation is necessary for accurate measurement of detector capability. In practice, both mechanisms are likely to operate simultaneously, and detection deployments should treat post-cutoff or contamination-free evaluation as a separate, more demanding performance regime rather than as a stress test of the main result. We return to the contamination interpretation as a threat to construct validity in Section~\ref{sec:threats}.

\textcolor{blue}{\paragraph{Why a similarity-based memorization probe would not resolve this.} A natural proxy for the contamination interpretation is to measure the similarity between each LLM's generated code and the matched human-written reference for the same docstring, on the reasoning that a memorized example should be reproduced near-verbatim. We considered and rejected this approach because similarity alone cannot distinguish memorization from convergent, independent generation: many of the functions and classes in our dataset admit only a small number of idiomatic implementations (e.g., a getter, a canonical sort wrapper, a standard string-formatting routine), and an LLM that has never seen a given example can converge on a near-identical implementation purely because the solution space is small. This confound is worse than it first appears at the granularity level: we would expect function-level generations to show higher raw similarity to their human references than class-level generations regardless of memorization, both because functions are shorter and more likely to admit a single idiomatic form, and because CodeSearchNet~\citep{husain2019codesearchnet} is itself function-centric, so a naive reading of any function-versus-class similarity gap is equally consistent with ``LLMs memorized more function-level examples'' and ``function-level problems are more often trivial regardless of training exposure.'' Without independent knowledge of each model's training-data composition, a similarity-based probe would produce a number that admits both an innocent and a contamination-consistent explanation, and reporting it would risk implying a resolution the data does not support. We therefore do not include such an analysis, and continue to report the post-cutoff degradation as informative about the combined effect of temporal drift and contamination without attempting to further decompose it.}

\subsection{Granularity-Dependent Detection Patterns}

We identify a critical, previously undocumented interaction between model architecture and code granularity. While newer models (Claude 3 Haiku, Claude Haiku 4.5, GPT-OSS) exhibit superior detectability at the class level, GPT-3.5 displays an inverted profile, performing significantly better at the function level. 

This divergence suggests a fundamental shift in generation mechanics: GPT-3.5's structural fingerprint is most pronounced in \emph{local algorithmic logic} (function-level), whereas newer models leave clearer traces in their \emph{macro-level organizational structure} (class-level).

The negligible cross-granularity feature overlap confirms this dichotomy. Since granularity impacts feature selection 8.6 times more strongly than model architecture, function and class detection are effectively distinct tasks requiring disjoint feature spaces. Consequently, detection systems must be specialized by abstraction level; a detector optimized for local logic will fail to capture the organizational artifacts that define modern LLM-generated code.

\textcolor{blue}{\paragraph{Use cases and technical challenges specific to each granularity.} As motivated in the Introduction, the two granularities serve different workflows: pull-request and commit review is scoped to individual functions or methods, where the question is whether a specific piece of behaviour was machine-written, whereas architectural reviews and codebase audits assess whole classes, where the question is whether the design of an abstraction was machine-generated. Beyond requiring disjoint feature spaces, these two settings pose different technical challenges for building a robust detector. Function-level detection operates over a comparatively narrow feature space (18 metrics; Section~\ref{sec:dataset}) and, per the category-level analysis in Section~\ref{sec:rqiii}, relies disproportionately on a single category of signal: comment-related features account for 44.9\% of total SHAP importance at the function level on average, compared to 24.9\% at the class level. A detector concentrated on one category is correspondingly more exposed to targeted manipulation of that category, though as Section~\ref{sec:rqiv} shows, the practical consequence of such manipulation is model-specific rather than uniformly harmful: comment stripping degrades the one configuration (GPT-OSS, class level) whose dominant signal is over-commenting relative to humans, while it improves detection for the three function-level models whose dominant signal is under-commenting, since removing comments pushes them further from, not closer to, the human baseline.}

\textcolor{blue}{Class-level detection, by contrast, faces a larger and more heterogeneous unit: a class contains more text (several methods together with attributes and docstrings), more internal complexity (the control flow of every method contributes to a single sample), and relations among its members and to other classes. It accordingly operates over a substantially larger feature space (39 metrics) that includes quantities with no function-level analogue: coupling and inheritance metrics (Coupled Classes, Base Classes, Derived Classes), which describe relationships between classes rather than properties of a single unit, and aggregate statistics (Average, Sum, and Maximum variants of per-method metrics) that require a design decision about how to summarize behaviour across a class's methods. \textcolor{blue}{Because our standalone-class design (Section~\ref{sec:dataset}) excludes classes that inherit from, or are inherited by, other classes, these relationship metrics capture only what each class exhibits in isolation, and} our results do not speak to the added complexity of detection in codebases where classes participate in non-trivial inheritance structures, itself a distinct technical challenge for class-level detection in practice that we return to in Section~\ref{sec:limitations}.}

\subsection{Universal Features with Model-Specific Magnitudes}

Our SHAP analysis uncovers a critical nuance in feature universality: while specific features are consistently selected, their predictive power varies wildly. The \emph{Comment to Code Ratio} emerges as the sole universal discriminator, appearing in all eight configurations and ranking in the top-3 for seven of them. This suggests it captures a fundamental stylometric property distinguishing LLM generation from human craft, irrespective of the model architecture.

However, the \emph{magnitude} of this signal is highly volatile. For GPT-3.5, the SHAP importance of \emph{Comment to Code Ratio} (3.795 at function-level) is an order of magnitude higher (10 times--21 times) than in any other configuration. This disproportionate reliance explains the ``GPT-3.5 trap": detectors trained on GPT-3.5 achieve exceptional performance by overfitting to an atypically loud signal that effectively vanishes in newer models. While the consistent presence of commenting patterns offers a theoretical basis for universal detection, practical implementation requires model-specific calibration. We cannot rely on the ``strong" signals of early models to carry over to modern architectures; instead, detectors must be sensitive to the far subtler expressions of these features in state-of-the-art LLMs.

\subsection{Detectability and Code Quality}
\label{sec:detectability-quality}

The structural metrics that distinguish LLM-generated code from human-written code in our analysis are not arbitrary features; several of them are well-established proxies for code quality in the software engineering literature. Cyclomatic Complexity, in particular, \textcolor{blue}{was originally proposed as a basis for managing test-path coverage and modularization~\citep{mccabe1976complexity}, and subsequent empirical work has documented a project-dependent, rather than universal, association with post-release defect-proneness~\citep{nagappan2006mining}.} Maximum Nesting Depth, Comment to Code Ratio, and lines-of-code measures are similarly recognized inputs to standard maintainability assessments and bug-prediction models. The connection between our detection signals and code quality is therefore not a separate empirical relationship that we need to establish in this paper; it is a direct consequence of the fact that the most discriminative metrics in our analysis are precisely the metrics that the software engineering community has long treated as code-quality proxies. Our findings can be read as showing that LLM-generated Python code differs from human-written code along structural dimensions known to correlate with maintainability and defect-proneness\textcolor{blue}{, albeit with associations whose strength varies by project, matching the broader pattern of context-dependence documented for these metrics}.

The directional findings from RQ1 give this connection concrete substance. Claude 3 Haiku and GPT-3.5 generate code that is shorter and sparsely documented relative to human-written counterparts, while Claude Haiku 4.5 and GPT-OSS generate code that is longer, more nested, and (in the case of GPT-OSS at the class level) more heavily commented. Each directional shift moves the generated code along a dimension that has practical maintainability implications. We caution that our study measures structural signatures rather than quality outcomes such as observed defect rates or developer-perceived maintainability, and we do not claim that LLM-generated code is uniformly more or less maintainable than human-written code. Such a claim would require a follow-up study on defect histories and maintenance records of LLM-attributed code, which we identify as future work. Within the scope of our measurements, however, the systematic structural shifts we observe imply that organizations adopting LLM-generated code at scale should expect their codebase quality profile to shift along these established metric axes, and that the direction of the shift will depend on which LLM is used. The detection signals we identify are, in this sense, simultaneously signals about the structural quality of the resulting code.

\subsection{Where Structural Detection Beats GPTZero, and Where It Does Not}

The headline difference between our structural classifiers and the GPTZero commercial baseline is not aggregate AUC but recall. Across the eight configurations, GPTZero achieves recall between 15\% and 64\%, while our model-specific classifiers achieve 61\% to 87\%. GPTZero's high precision combined with low recall reflects a systematic skew toward the ``human'' label: it labels 76--85\% of all samples as human-written and misses a substantial majority of LLM-generated code. This calibration suits adjudicative use cases where a flag is consequential, but is unsafe for screening workflows where a false negative passes LLM-generated code through review undetected.

\textcolor{blue}{On AUC, the picture is mixed. GPTZero's AUC is higher in four of the eight configurations: the three function-level configurations other than GPT-3.5 (by 0.03 to 0.07) and GPT-OSS at the class level (by about 0.02). Our structural classifiers' AUC is higher in the other four, by 0.15 to 0.36, with the largest margins for GPT-3.5 at both granularities. The recall deficit, however, holds uniformly across all eight configurations and is the consistent failure mode of textual-perplexity detection on code.} The directionally correct conclusion is therefore not that structural features dominate textual features in general, but that structural features dominate in the recall regime that screening workflows require, and that GPTZero's calibration choice is the reason.

\subsection{Structural Interpretability versus Neural Accuracy}
\label{sec:interpretability_tradeoff}

\textcolor{blue}{The comparison above is against GPTZero, a general-purpose commercial detector not built for source code. A more demanding comparison is against code-specific neural detectors trained directly for this task. CodeGPTSensor~\citep{xu2025distinguishing}, which applies contrastive learning over a UniXcoder semantic encoder, and GPTSniffer~\citep{nguyen2024gptsniffer}, a CodeBERT-based classifier, both report substantially higher AUC-ROC than any configuration in our study when evaluated on their respective training distributions. We do not dispute this: within a single-model evaluation, learned semantic representations capture discriminative signal that a compact set of 6--12 interpretable structural metrics does not.}

\textcolor{blue}{We do not read this gap as evidence of a general, inherent accuracy-interpretability trade-off. Rudin~\citep{rudin2019stop} has argued persuasively that for many structured prediction problems, well-engineered interpretable models can match black-box accuracy, and that the belief in an unavoidable trade-off is often a failure to try hard enough on the interpretable side rather than a mathematical necessity. Our claim is narrower and empirical: within the specific, deliberately compact feature space used here, we do not close the accuracy gap to neural approaches on their own single-model benchmarks. This feature space is compact precisely because it must remain small enough to support the SHAP-based explanatory analysis in RQ3 and the metric-level adversarial analysis in RQ4. Whether a richer but still-interpretable feature space could close the accuracy gap is a question for future work, not one this study resolves.}

\textcolor{blue}{What our results do speak to is whether the cross-model generalization failure we report in RQ2 and RQ3 is an artifact specific to structural features, or a property of the detection problem itself. This generalization failure means that no structural feature or feature magnitude transfers cleanly across all four LLMs. Suh et al.~\citep{suh2024empirical} evaluated GPTSniffer, the same CodeBERT-based detector cited above, across ChatGPT, GPT-4, Gemini Pro, and Starcoder2, and found that while it achieved strong performance on its training distribution, it showed limited generalization across generators. A semantic, code-specific detector therefore exhibits the same qualitative pattern that we document for structural metrics: strong within-distribution accuracy, but weak cross-model transfer. This does not prove that richer features cannot solve cross-model generalization, but it weakens the assumption that they trivially would; the difficulty may lie in the diversity of what different LLMs actually produce, not in which family of features is used to detect it.}

\textcolor{blue}{Given this, we scope our own conclusions accordingly: our claim that no universal structural signature exists is a claim about the interpretable-metric space we study, not a claim about detection in general; and our claim above that structural features outperform text-based detection is specifically a claim relative to GPTZero's calibration failure, not relative to specialized neural detectors, against which we do not currently have a retrained, cross-model comparison.}

\subsection{Deployment Guidance}
\label{sec:deployment_guidance}

Although the principal contribution of this work is the mechanistic characterization of how LLM-generated Python code differs structurally from human-written code, our results have practical implications for organizations that consider deploying structural detectors as part of a code-review workflow. In this subsection, we address three practitioner-facing questions: (1) what the AUC values we report mean operationally, (2) how to handle the realistic deployment scenario in which the LLM behind a code sample is unknown, and (3) what the brittleness findings from Sections~\ref{sec:rqiv} and~\ref{sec:temporal-gap} imply for deployment over time.

\textbf{Interpreting AUC values operationally.} AUC measures the probability that a randomly chosen LLM-generated sample receives a higher classifier score than a randomly chosen human-written sample. An AUC of 0.96 (GPT-3.5 function-level) means the classifier ranks the two correctly 96\% of the time; an AUC of \textcolor{blue}{0.68--0.83} (other configurations) means the ranking is correct \textcolor{blue}{68--83}\% of the time, leaving a non-trivial fraction of samples on the wrong side of any chosen threshold. For deployment, what matters is not the AUC itself but the precision and recall at the operating threshold a team chooses. Tables~\ref{tab:rq2_performance_function} and~\ref{tab:rq2_performance_class} report these at the default 0.5 threshold; in practice, teams should choose the threshold based on the asymmetry of false-positive and false-negative costs in their workflow. A code-review screening tool, for example, may prefer a low threshold that catches most LLM-generated samples (high recall) even at the cost of additional human review of false positives. An academic-integrity proceeding may prefer a high threshold that flags only high-confidence cases (high precision), even at the cost of missing some LLM-generated submissions. The classifier outputs we report support either operating regime; the choice is application-specific.

\textbf{Handling the unknown-LLM scenario via ensembles.} The realistic deployment scenario is that the team does not know which LLM (if any) produced a given code sample, while our trained classifiers are model-specific and do not generalize well to LLMs other than the one they were trained on. The natural deployment strategy is therefore not to choose one classifier but to deploy an ensemble of all available model-specific classifiers, running them in parallel and combining their outputs into a single decision.

\textcolor{blue}{Combining the classifiers' outputs, however, is less straightforward than it may first appear. AutoSpearman selects a different feature subset for each model-granularity configuration (Section~\ref{sec:rqii}), and the eight classifiers differ substantially in intrinsic separability (AUC-ROC ranging from 0.681 to 0.961); as a result, a given predicted-probability value does not necessarily carry the same meaning across classifiers without additional calibration (e.g., Platt scaling or isotonic regression fit on a held-out set), which we do not undertake in this study. This specifically limits combination rules that compare or aggregate probability magnitudes across classifiers, such as taking the maximum or the average predicted probability across the ensemble: a classifier that tends to produce more extreme scores would dominate such a rule independent of whether it is more accurate. A majority-vote rule, which counts each classifier's own binary decision at its own already-reported threshold rather than combining probability magnitudes directly, sidesteps this calibration requirement and is accordingly the more defensible of the available options to propose without further validation. More sophisticated combination strategies, such as a meta-classifier trained on the per-LLM detector outputs, would require labelled multi-LLM data that is typically not available outside research settings, and calibrating and empirically validating any of these combination rules against a mixed-origin evaluation set is left to future work. We present the ensemble strategy itself, rather than any specific combination rule, as the practical takeaway for deployment: relying on a single model-specific classifier is unsafe when the generating LLM is unknown, and a majority-vote ensemble across all available classifiers is the more conservative starting point among the options we considered.}

The adversarial-robustness analysis (Section~\ref{sec:rqiv}) provides additional support for the ensemble approach in general: the single configuration where comment stripping collapses detection performance (GPT-OSS class-level) is one of eight, and the remaining seven configurations continue to discriminate correctly under the same adversarial transformation\textcolor{blue}{, so a majority-vote ensemble would not be dominated by this one failure mode}.

\textbf{Limits of the ensemble approach.} An ensemble built from the four LLMs in this study covers four LLMs; it does not cover the (rapidly growing) population of contemporary code-generating models. New LLMs released after our study would not be detected by any classifier in the ensemble unless the ensemble is retrained with samples from those models. We do not claim that the ensemble is a complete solution to the unknown-LLM problem; rather, we claim that it is the right operational strategy given the cross-LLM generalization findings of this paper, and that the strategy is extensible (additional classifiers can be added as additional LLMs become important). The most consequential limitation of the ensemble approach is that the cross-model generalization findings imply detector deployments must be \emph{maintained}: as new LLMs are released, new model-specific classifiers must be trained and added to the ensemble. This is operationally non-trivial but is the unavoidable consequence of the lack of cross-LLM generalization in interpretable structural detectors.

\textbf{Maintaining detectors over time.} The brittleness to post-cutoff repositories (Section~\ref{sec:temporal-gap}) implies that even within a single LLM, the structural fingerprint of generated code shifts over time. Whether this shift reflects genuine model evolution or whether it reflects contamination effects that disappear on contamination-free data, the practical implication is the same: classifiers trained on data from a given time window cannot be assumed to generalize forward indefinitely. Deployment-grade detection systems must adopt a continuous-evaluation regime in which classifier performance is periodically reassessed against fresh contamination-free samples, and classifiers retrained when performance degrades below an acceptable threshold. We acknowledge that this is a substantial operational burden, but it appears unavoidable given the rate at which contemporary LLMs are evolving.

\textbf{Why general-purpose detectors do not substitute for domain-specific ones.} A natural question for practitioners is whether off-the-shelf commercial detection tools, such as GPTZero, offer a less burdensome alternative to maintaining a domain-specific ensemble. Our results indicate that they do not: GPTZero's recall on our test sets ranges from 15\% to 64\% across configurations (versus 61\% to 87\% for our model-specific detectors), meaning that a deployment of GPTZero would miss between 36\% and 85\% of LLM-generated samples. The failure mode is systematic rather than random: GPTZero exhibits a strong bias toward the ``human'' label, which produces high precision but very low recall and renders the tool unsafe as a screening instrument in workflows where false negatives are consequential. Closer inspection of which configurations GPTZero handles best suggests that its training distribution is dominated by natural-language text and earlier-generation LLM outputs, with limited adaptation to the syntactically-constrained domain of code. We argue that practical detection systems must therefore be \emph{domain-specific}, built upon structural software metrics or other code-aware representations rather than adapted from general-purpose text-detection tools. Within this domain-specific regime, the calibration challenge identified throughout this paper---that universal signals such as Comment to Code Ratio carry vastly different magnitudes across LLMs---suggests that future architectures should move toward calibration-aware systems that can adapt signal interpretations to the LLM at hand, rather than treating the structural fingerprint as model-invariant.

\section{Limitations}
\label{sec:limitations}

While our study provides comprehensive insights into LLM-generated code detection, several limitations warrant discussion.

Firstly, our study deliberately focuses on structural software metrics (complexity, coupling, nesting depth, commenting ratios) rather than text-based features (token sequences, embeddings, language model representations). This design choice reflects our goal: not only to detect LLM-generated code but to \emph{explain} why code exhibits LLM characteristics through interpretable structural signatures. Text-based deep learning approaches, such as CodeGPTSensor~\cite{xu2025distinguishing} and GPTSniffer~\cite{nguyen2024gptsniffer}, while potentially achieving higher detection accuracy, operate as black boxes that provide limited insight into which code characteristics drive detection. This paradigm difference between structural analysis and textual analysis means we cannot directly compare with these academic baselines without training them from scratch on our dataset, which would require substantial computational resources (estimated 100+ GPU-hours for CodeGPTSensor alone across eight configurations) and conflate fundamentally different detection approaches. \textcolor{blue}{Our evaluation of GPTZero, a general-purpose commercial detector requiring no training, shows that structural features yield substantially higher recall than generic textual-perplexity detection on code, although the AUC comparison is mixed (Section~\ref{sec:rqii}). Against code-specific neural detectors purpose-built for this task, the comparison is different, and we do not claim an accuracy advantage there: as discussed in Section~\ref{sec:interpretability_tradeoff}, published results for CodeGPTSensor and GPTSniffer report higher raw accuracy than our structural classifiers achieve. Our contribution is a compact, interpretable feature space that supports the explanatory (RQ3) and adversarial-robustness (RQ4) analyses that black-box neural detectors do not support, not a claim of superior accuracy over specialized neural approaches.}

Future work directly comparing structural feature-based detection with text-based deep learning approaches on identical datasets would provide valuable insights into the relative strengths of these complementary paradigms. Our work establishes structural metrics as a viable and interpretable approach for detection.

Secondly, the deliberate inclusion of only standalone code artifacts is a limitation of this study. Our analysis focuses on standalone functions and classes that can be generated independently without extensive contextual dependencies. We excluded code requiring substantial surrounding context (methods dependent on complex class hierarchies, functions requiring project-specific libraries or state) for practical reasons: providing sufficient context would require significantly longer prompts, dramatically increasing API costs while potentially exceeding token limits, and incomplete context increases the likelihood of receiving non-functional code from LLMs.

However, real software systems contain both standalone and context-dependent code. Detection characteristics of highly coupled code integrated into existing projects might differ from our observed patterns, as such code may be constrained by project-specific conventions, architectural patterns, or API usage that reduce distinctive LLM signatures. This represents a tradeoff between experimental control and ecological validity. Our focus on standalone artifacts provides clean cross-model comparisons while reflecting common scenarios (utility functions, self-contained classes, code snippets), but future work examining detection within complete projects with complex dependencies would complement our findings.

Thirdly, we acknowledge a limitation of our adversarial analysis. We evaluate only one adversarial transformation — complete comment removal. We chose this attack because RQ3 identified comments as the dominant discriminator. After all, it is the lowest-cost evasion available to a developer, and it cleanly tests the prominence of comment features in the trained classifiers. More sophisticated attacks (rewriting comments to mimic human style rather than removing them, structural perturbations that affect non-comment features, or re-prompting the LLM with project-specific style examples) are out of scope for this study and remain an important direction for future work. We do not claim that classifiers that survive comment stripping are robust to all adversarial transformations.

Finally, our findings are conditional on the Python scope. Whether the granularity effects, model-specific signatures, and feature-importance patterns we observe also hold in JavaScript, TypeScript, Go, Rust, or other languages with comparably independent function and class constructs is an empirical question we leave for future work.

\section{Threats to Validity}
\label{sec:threats}

We discuss potential threats to the validity of our study and steps taken to mitigate them.

\subsection*{Internal Validity}

Our feature set comprises 18 function-level and 39 class-level structural metrics. While other potentially discriminative features might exist, we employed AutoSpearman feature selection to identify statistically significant features in a non-parametric manner, reducing dependence on initial feature choices. The consistency of \emph{Comment to Code Ratio} as a universal discriminator across all configurations provides evidence that our feature set captures fundamental distinguishing characteristics.

\subsection*{External Validity}

First, our reliance on the CodeSearchNet Python dataset may limit generalizability to proprietary codebases or languages with distinct paradigms (e.g., functional or low-level systems). However, we focused on fundamental structural metrics, such as cyclomatic complexity and nesting depth, that act as language-agnostic proxies for code logic. The consistency of these structural patterns across four distinct LLMs suggests they reflect inherent properties of neural code generation rather than dataset-specific artifacts.

Second, given the rapid evolution of LLMs, the specific detectability scores for current models may eventually drift. However, our core contribution lies in identifying the \emph{mechanisms} of detection: specifically, that newer models are shedding the distinct stylometric signals of GPT-3.5 and that detection success is heavily granularity-dependent. These architectural trends—particularly the divergence between local and macro-level consistency—provide a robust framework for evaluating future models, regardless of specific version updates.

\subsection*{Construct Validity}

First, our study focuses exclusively on \emph{authorship attribution}, not code quality. We do not claim that detectability correlates with correctness, security, or maintainability. Instead, our research addresses the need for transparency in code provenance, with direct implications for academic integrity and copyright compliance.

Second, we utilize SHAP analysis to quantify feature importance. It is crucial to note that SHAP measures \emph{predictive contribution}, not causality. High SHAP values indicate a strong association with LLM-generated code but do not necessarily imply a causal mechanism. We interpret these findings conservatively, prioritizing patterns that remain consistent across configurations (e.g., the robustness of the \emph{Comment to Code Ratio}) over isolated signals.

Third, to ensure statistical robustness, we employed a rigorous validation protocol: 30 repetitions of 10-fold cross-validation, bootstrap confidence intervals ($n=1000$), and DeLong's test with Holm-Bonferroni correction for pairwise comparisons. While multiple comparisons across eight configurations inherently increase the risk of Type I errors, the magnitude of the observed differences (e.g., the dominance of GPT-3.5) and their structural consistency suggest these are genuine phenomena rather than statistical artifacts.

Finally, we acknowledge a threat related to the training data of the LLMs we evaluate. \textcolor{blue}{Our function-level dataset is drawn from CodeSearchNet, and our class-level dataset is drawn from the same underlying GitHub repositories. Both are publicly available and may overlap with the training corpora of the four LLMs we study.} None of the four model providers discloses training-corpus composition at the level needed to verify this overlap directly, but the possibility cannot be ruled out. If such overlap exists, the structural patterns we observe in LLM-generated code may partly reflect memorization of CodeSearchNet examples rather than the LLMs' intrinsic generation tendencies, which would inflate detection performance on the main test set. We addressed this concern with a separate evaluation on a contamination-free secondary dataset constructed from post-cutoff repositories that share zero overlap with CodeSearchNet (Section~\ref{sec:temporal-gap}); the substantial detection performance drop we observe on this contamination-free set is consistent with the contamination hypothesis but is also consistent with temporal generalization brittleness (LLM behaviour evolving, independent of any contamination effect)\textcolor{blue}{, and it may also reflect differences in data characteristics between the two datasets (e.g., code complexity and length), independent of contamination. We cannot disambiguate these explanations with publicly available information, and we report this multi-factor interpretation as a residual threat to construct validity.} \textcolor{blue}{A natural alternative would be to directly measure the similarity between each LLM's output and its matched human reference, but as discussed in Section~\ref{sec:temporal-gap}, such a measure is confounded by problem triviality and by the function-centric composition of CodeSearchNet, and would not reliably distinguish memorization from convergent independent generation.}

\section{Related Work}
\label{sec:related_work}

In this section, we situate our work within the broader landscape of code detection and related software engineering research.

\subsection{LLM-Generated Code Detection}

Detection of LLM-generated code has emerged as a critical research area, driven by concerns regarding academic integrity, attribution, and software quality assurance.

\subsubsection{Feature-Based Detection Approaches}

Idialu et al.~\cite{idialu2024whodunit} trained gradient boosting classifiers to detect GPT-4-generated code at the function level using programming competition problems. Their work identified stylometric features as primary discriminators between human and LLM-generated code. However, their study focused exclusively on competitive programming tasks and did not examine class-level detection or compare multiple LLMs.

Shi et al.~\cite{shi2025between} proposed DetectCodeGPT, a perturbation-based technique inspired by the \emph{naturalness hypothesis} of code~\cite{rahman2019natural,hindle2012naturalness}. Their approach analyzes differences between machine and human-written code by perturbing stylistic tokens such as whitespace and newlines. Despite its innovation, their evaluation centred primarily on GPT-3.5, limiting insights into cross-model generalization.

Xu et al.~\cite{xu2024one} investigated the efficacy of perplexity-based methods across C, C++, and Python. Comparing perplexity approaches against feature-based and pre-training methods, they found that perplexity performs well for C/C++ but exhibits variable performance across languages and difficulty levels. Although their work examined GPT-4o, Gemini-1.0, and Llama-3.1, it did not explore stylometric feature importance or class-level detection patterns.

Park et al.~\cite{park2025detection} developed LPcodedec to detect LLM-paraphrased code using coding style features. Their work addresses the critical scenario of plagiarism via LLM-based paraphrasing, introducing features related to naming consistency, structure, and readability. While their focus on paraphrase detection complements direct generation detection, their analysis centers on identifying the specific paraphrasing model rather than examining fundamental differences in detectability across architectures.

\subsubsection{Deep Learning and Pre-trained Model Approaches}

Nguyen et al.~\cite{nguyen2024gptsniffer} proposed GPTSniffer, a CodeBERT-based classifier for detecting ChatGPT-generated code. They reported high accuracy and demonstrated that GPTSniffer outperforms general-purpose detectors (e.g., GPTZero) on source code. However, as a neural approach, it provides limited interpretability regarding the specific characteristics driving detection. Similarly, Oedingen et al.~\cite{oedingen2024chatgpt} demonstrated that embedding-based approaches combined with supervised learning can achieve 98\% accuracy in detecting ChatGPT-generated Python code, though such models remain constrained to their training distributions.

Xu et al.~\cite{xu2025distinguishing} developed CodeGPTSensor, using contrastive learning with a UniXcoder-based semantic encoder to distinguish ChatGPT from human-written code. They curated the HMCorp dataset, containing 550k pairs of human and ChatGPT Python/Java code. While effective for ChatGPT, the approach was designed specifically for that model, leaving its transferability to other LLMs untested. A follow-up work~\cite{yin2025detecting} introduced CodeGPTSensor+, which employs adversarial training to improve robustness, yet similarly focuses on ChatGPT-generated content.

Xu and Sheng~\cite{xu2025codevision} proposed CodeVision, utilizing 2D token probability maps combined with vision models (ResNet and ViT) to detect generated code. Their approach preserves spatial code structures and demonstrates robustness across languages. Evaluation on GPT-3.5 and GPT-4 showed strong performance, but the work lacks examination of other contemporary models or feature-level explanations.

\subsubsection{Zero-Shot and Training-Free Methods}

Yang et al.~\cite{yang2023zero} adapted DetectGPT for code by using surrogate white-box models for probability estimation. They demonstrated effectiveness on text-davinci-003, GPT-3.5, and GPT-4, suggesting that smaller models (e.g., PolyCoder-160M) can serve as universal code detectors. However, their evaluation relied primarily on competitive programming datasets and did not extend to the architectures of contemporary models.

Ye et al.~\cite{ye2025uncovering} developed a zero-shot detector based on code rewriting similarity, observing that differences between LLM-rewritten and original code are smaller when the original is synthetic. Using self-supervised contrastive learning, they improved upon existing zero-shot detectors on APPS and MBPP benchmarks. This approach provides an orthogonal perspective to feature-based methods, but was evaluated primarily on GPT-3.5 Python code.

Ashkenazi et al.~\cite{ashkenazi2025zero} proposed Approximated Task Conditioning (ATC), observing that conditioning probability distributions on the original task prompt reveals notable differences between human and machine code. Their method achieves strong results across Python, C++, and Java; however, the evaluation focused on within-language performance rather than systematic cross-model differences.

\subsection{Comparative and Multi-Model Studies}

While most research focuses on single LLMs, a small number of recent works have begun examining detection across multiple models. These studies highlight the gap our work addresses.

Demirok and Kutlu~\cite{demirok2025aigcodeset} introduced AIGCodeSet, a dataset of code generated by models including CodeLlama-34B, Codestral-22B, and Gemini 1.5 Flash. Their focus was on error correction (generating code to fix bugs) rather than de novo generation. While they acknowledge model diversity, their analysis does not systematically compare detectability patterns across models or investigate granularity effects.

Suh et al.~\cite{suh2024empirical} conducted an empirical study evaluating existing detectors on code from ChatGPT, GPT-4, Gemini Pro, and Starcoder2. They found that natural language detectors perform poorly on code and that even code-specific tools like GPTSniffer show limited generalization. Their work provides valuable insights into tool limitations but focuses on evaluating off-the-shelf detectors rather than investigating the fundamental structural differences that drive detectability. \textcolor{blue}{This generalization failure for a semantic, code-specific detector is consistent with the cross-model instability we document for structural features (Section~\ref{sec:rqii}, Section~\ref{sec:rqiii}), and informs the scoped interpretation of that instability we offer in Section~\ref{sec:interpretability_tradeoff}: the difficulty of cross-model detection may be a property of the underlying problem rather than a limitation specific to any one family of features.}

\subsection{LLM-Generated Text Detection}

Text detection informs our methodological approach but faces different constraints. Beresneva~\cite{beresneva2016computer} surveyed early computer-authored text detection, focusing on statistical methods for machine translation. Jawahar et al.~\cite{jawahar2020automatic} provided a comprehensive survey of detection for sophisticated LLMs like GPT-2. Tang et al.~\cite{tang2024science} categorized approaches into black-box and white-box detection, highlighting watermarking as a promising direction. Yang et al.~\cite{yang2023survey} and Wu et al.~\cite{wu2023survey} identified zero-shot and training-based detection as the dominant paradigms.

Code presents unique challenges compared to text due to its rigid syntactic structure, lower entropy, and functional constraints. Our work builds on these text-detection foundations while addressing the specific requirements of software engineering contexts.

\subsection{Related Applications of LLMs in Software Engineering}

Beyond detection, LLMs have been applied to numerous SE tasks. Abedu et al.~\cite{abedu2024llm} studied challenges in using chatbots for repository mining. Kang et al.~\cite{kang2023large} investigated LLMs for bug reproduction and program repair. Wang et al.~\cite{wang2023codet5+} proposed CodeT5+ to support tasks like natural-language-to-code generation. Other applications include automated code review~\cite{liu2020retrieval,li2022automating}, comment generation~\cite{li2022auger}, and summarization~\cite{ahmed2022few}. This broader body of work demonstrates the pervasive integration of LLMs into development workflows, reinforcing the need for robust detection capabilities.

\subsection*{Our Contributions Relative to Existing Work}

Our work addresses critical gaps through a systematic multi-model comparison using interpretable structural software metrics rather than black-box features. We provide the first cross-granularity investigation, revealing that function and class detection rely on fundamentally different signatures, with granularity effects dominating model differences. Through rigorous statistical methodology and SHAP analysis, we identify universal discriminative features while explaining why detectors trained on one model fail on others due to dramatic magnitude variations in shared features. We reassess the cross-LLM evaluation paradigm where detectors are tested across models without retraining, demonstrating that this measures distribution shift rather than detector quality. Unlike competitive programming studies, we analyze real-world functions and classes from open-source projects, directly addressing practical mixed-authorship scenarios.

\section{Conclusion}
\label{sec:conclusion}

We conducted a systematic comparative analysis of LLM-generated code detection across four contemporary models and two granularities using structural software metrics. Our results reveal that GPT-3.5 is an anomaly: its exceptional detectability is fundamentally unrepresentative of contemporary models (Claude 3 Haiku, Claude Haiku 4.5, GPT-OSS), which demonstrate substantially lower detectability. This performance gap proves that detection findings derived solely from GPT-3.5 do not generalize to the broader landscape of modern LLMs.

Mechanistically, we discovered that detection is driven more by granularity than by model architecture. Feature overlap between function and class levels is negligible, indicating that classifiers rely on distinct structural signatures depending on the abstraction level. Furthermore, while we identified universal discriminators like the \emph{Comment to Code Ratio}, their predictive magnitude varies drastically across models. This explains the failure of cross-LLM transfer: detectors optimized for the loud signals of GPT-3.5 fail to perceive the subtler signatures of newer architectures.

\textcolor{blue}{Practically, our structural classifiers achieve substantially higher recall than commercial general-purpose detectors like GPTZero, although the comparison on AUC-ROC is mixed.} However, as LLMs continue to evolve, the field must move beyond monolithic, single-model evaluations. Future detection frameworks must be granularity-aware and calibrated to the specific structural fingerprints of diverse architectures, rather than relying on the fading artifacts of early-generation models.

\section*{Acknowledgments}

We thank GPTZero for their generous provision of API credits. This support, which distinguishes them from other commercial vendors who declined research access, was instrumental in conducting the comprehensive baseline evaluation.

\section*{Declarations}

\subsection*{Funding}
This research was supported by the \textbf{Natural Sciences and Engineering Research Council of Canada (NSERC)} under Grant Number: \textit{CREATE 555406-2021}.

\subsection*{Conflict of Interest}
\textbf{Emad Shihab} has ongoing collaborations with EMSE Editorial Board members: \emph{Foutse Khomh}, \emph{Sarah Nadi}, \emph{Mei Nagappan}, and \emph{Alexander Serebrenik}.

\subsection*{Ethical Approval}
This article does not contain any studies with human participants or animals performed by any of the authors.

\subsection*{Informed Consent}
Not applicable.

\subsection*{Author Contributions}
\textbf{Musfiqur Rahman:} Conceptualization, Methodology, Implementation, Formal analysis, Investigation, Data curation, Writing - original draft, Writing - review \& editing, Visualization. \textbf{SayedHassan Khatoonabadi:} Methodology, Writing - review, Validation. \textbf{Ahmad Abdellatif:} Conceptualization, Writing - review. \textbf{Emad Shihab:} Conceptualization, Writing - review, Supervision, Project administration, Funding acquisition.

\phantomsection
\subsection*{Data Availability Statement} \label{sec:replication-package}
The datasets generated and analyzed during the current study, along with all code and analysis scripts, are publicly available in our replication package at: \url{https://github.com/mrsumitbd/LLM-generated-code-detection\_Replication-Package}.

\subsection*{Clinical Trial Number}
Not applicable.

\bibliographystyle{spbasic}      

\bibliography{references}   

\appendix

\section{Appendix}

\subsection{Derivation of Expected Random Feature Overlap}
\label{appendix:random_jaccard}

To determine whether the observed cross-granularity feature overlap is significant, we calculate the expected Jaccard similarity ($E[J_{random}]$) for two feature subsets selected uniformly at random from the total feature space.

Let $N$ be the total size of the feature space (the union of all available features). Based on our class-level analysis, $N = 39$.

Let $k_{class}$ be the number of features selected by the class-level model. We observe a maximum selection size of $k_{class} = 12$.

Let $k_{func}$ be the number of features selected by the function-level model. We observe a constant selection size of $k_{func} = 6$.

First, we calculate the expected size of the intersection ($E[|I|]$) between these two random subsets.

\begin{equation}
    E[|I|] = \frac{k_{class} \times k_{func}}{N}
\end{equation}

Substituting our values:

\begin{equation}
    E[|I|] = \frac{12 \times 6}{39} = \frac{72}{39} \approx 1.846
\end{equation}

Next, we approximate the expected Jaccard index. The Jaccard index $J$ is defined as the ratio of the intersection to the union:

\begin{equation}
    J = \frac{|A \cap B|}{|A \cup B|} = \frac{|I|}{k_{class} + k_{func} - |I|}
\end{equation}

Using the first-order approximation $E[f(x)] \approx f(E[x])$:

\begin{equation}
    E[J_{random}] \approx \frac{E[|I|]}{k_{class} + k_{func} - E[|I|]}
\end{equation}

Substituting the expected intersection derived above:

\begin{equation}
    E[J_{random}] \approx \frac{1.846}{12 + 6 - 1.846} = \frac{1.846}{16.154} \approx 0.114
\end{equation}

Thus, if the models were selecting features purely at random, we would expect a Jaccard overlap of approximately $0.11$. Our observed cross-granularity overlap of $0.099$ falls below this random expectation, confirming that the feature sets are effectively disjoint.

\subsection{Definition of Software Metrics Studied in this Paper}
\label{appendix:supplementary_materials}

\begin{table}[h]
\centering
\caption{Software metrics used for LLM-generated code detection. Metrics are extracted using SciTools Understand and cover code stylometry and complexity dimensions at both function and class granularities.}
\label{tab:metrics_definitions}
\resizebox{\textwidth}{!}{
\begin{tabular}{lp{10cm}}
\toprule
\textbf{Metric} & \textbf{Definition} \\
\midrule
\multicolumn{2}{l}{\textbf{\textit{Code Stylometry Metrics}}} \\
All Methods & Total number of methods in a class, including inherited methods \\
Average Blank Lines & Mean number of blank lines per method in a class \\
Average Code Lines & Mean number of code lines per method in a class \\
Average Comment Lines & Mean number of comment lines per method in a class \\
Average Lines & Mean total number of lines per method in a class \\
Blank Lines & Number of lines containing only whitespace \\
Code Lines & Number of lines containing executable or declarative code \\
Comment Lines & Number of lines containing comments or documentation \\
Comment to Code Ratio & Ratio of comment lines to code lines \\
Declarative Code Lines & Number of lines containing variable or constant declarations \\
Declarative Statements & Number of declaration statements (variables, constants, imports) \\
Executable Code Lines & Number of lines containing executable statements \\
Executable Statements & Number of executable statements (assignments, function calls, control flow) \\
Instance Methods & Number of instance methods defined in a class \\
Instance Variables & Number of instance variables (attributes) defined in a class \\
Lines & Total number of lines including code, comments, and blank lines \\
Methods & Number of methods defined in a class (excluding inherited methods) \\
Statements & Total number of statements (declarative + executable) \\
\midrule
\multicolumn{2}{l}{\textbf{\textit{Code Complexity Metrics}}} \\
Average Cyclomatic Complexity & Mean cyclomatic complexity across all methods in a class \\
Average Essential Complexity & Mean essential complexity across all methods in a class \\
Average Modified Cyclomatic Complexity & Mean modified cyclomatic complexity across all methods in a class \\
Average Strict Cyclomatic Complexity & Mean strict cyclomatic complexity across all methods in a class \\
Average Strict Modified Cyclomatic Complexity & Mean strict modified cyclomatic complexity across all methods in a class \\
Base Classes & Number of direct base classes (immediate parent classes) \\
Coupled Classes & Number of classes that are coupled to this class through method calls or attribute access \\
Coupled Classes Modified & Modified count of coupled classes including indirect dependencies \\
Cyclomatic Complexity & Number of linearly independent paths through code (McCabe's metric) \\
Derived Classes & Number of classes that directly inherit from this class \\
Essential Complexity & Measure of unstructured control flow; lower values indicate more structured code \\
Logarithmic Paths & Logarithm (base 2) of the number of unique execution paths \\
Maximum Cyclomatic Complexity & Highest cyclomatic complexity among all methods in a class \\
Maximum Essential Complexity & Highest essential complexity among all methods in a class \\
Maximum Inheritance Tree & Maximum depth of the inheritance hierarchy \\
Maximum Modified Cyclomatic Complexity & Highest modified cyclomatic complexity among all methods in a class \\
Maximum Nesting Depth & Maximum depth of nested control structures (if, while, for, etc.) \\
Maximum Strict Cyclomatic Complexity & Highest strict cyclomatic complexity among all methods in a class \\
Maximum Strict Modified Cyclomatic Complexity & Highest strict modified cyclomatic complexity among all methods in a class \\
Modified Cyclomatic Complexity & Variant of cyclomatic complexity treating multiple conditions in a single decision as one \\
Paths & Total number of unique execution paths through the code \\
Strict Cyclomatic Complexity & Cyclomatic complexity calculated with stricter counting of decision points \\
Strict Modified Cyclomatic Complexity & Modified cyclomatic complexity with stricter decision point counting \\
Sum Cyclomatic Complexity & Sum of cyclomatic complexity across all methods in a class \\
Sum Essential Complexity & Sum of essential complexity across all methods in a class \\
Sum Modified Cyclomatic Complexity & Sum of modified cyclomatic complexity across all methods in a class \\
Sum Strict Cyclomatic Complexity & Sum of strict cyclomatic complexity across all methods in a class \\
Sum Strict Modified Cyclomatic Complexity & Sum of strict modified cyclomatic complexity across all methods in a class \\
\bottomrule
\end{tabular}}
\end{table}

\subsection{Degrees and Directions of Differences}

Tables \ref{tab:claude3_func}, \ref{tab:claude45_func}, \ref{tab:gpt35_func}, \ref{tab:gptoss_func}, \ref{tab:claude3_class}, \ref{tab:claude45_class}, \ref{tab:gpt35_class}, \ref{tab:gptoss_class} represent the differences in all extracted features between LLM-generated code and human-written code. The `Effect' column represents the degree of difference. In the `Direction' column, $\uparrow$ means LLM-generated code has values greater than human-written code, and $\downarrow$ means the opposite.


\begin{table}[htbp]
    \centering
    \begin{minipage}[t]{0.49\textwidth}
        \centering
        \caption{Claude 3 Haiku (Function)}
        \label{tab:claude3_func}
        \resizebox{\linewidth}{!}{%
            \begin{tabular}{|l|cccc|}
            \toprule
            \textbf{Feature} & \textbf{$p$-val} & \textbf{Cliff's $\delta$} & \textbf{Effect} & \textbf{Direction} \\
            \midrule
            Lines & $1.13e^{-208}$ & -0.209 & Small & $\downarrow$ \\
            Blank Lines & $8.15e^{-64}$ & -0.112 & Negl. & $\downarrow$ \\
            Code Lines & $6.49e^{-96}$ & -0.141 & Negl. & $\downarrow$ \\
            Decl. Code Lines & $2.45e^{-20}$ & -0.062 & Negl. & $\downarrow$ \\
            Exec. Code Lines & $7.00e^{-110}$ & -0.151 & Small & $\downarrow$ \\
            Comment Lines & $< 10^{-300}$ & -0.254 & Small & $\downarrow$ \\
            Paths & $2.28e^{-36}$ & -0.084 & Negl. & $\downarrow$ \\
            Paths Log(x) & $9.19e^{-46}$ & -0.082 & Negl. & $\downarrow$ \\
            Statements & $6.94e^{-62}$ & -0.112 & Negl. & $\downarrow$ \\
            Decl. Stmts & $3.24e^{-28}$ & -0.074 & Negl. & $\downarrow$ \\
            Exec. Stmts & $6.34e^{-74}$ & -0.123 & Negl. & $\downarrow$ \\
            Cyclomatic & $4.89e^{-39}$ & -0.087 & Negl. & $\downarrow$ \\
            Mod. Cyclomatic & $4.89e^{-39}$ & -0.087 & Negl. & $\downarrow$ \\
            Strict Cyclomatic & $1.54e^{-52}$ & -0.102 & Negl. & $\downarrow$ \\
            Str. Mod. Cyclo & $1.54e^{-52}$ & -0.102 & Negl. & $\downarrow$ \\
            Essential & $5.39e^{-30}$ & -0.057 & Negl. & $\downarrow$ \\
            Max Nesting & $3.81e^{-08}$ & -0.036 & Negl. & $\downarrow$ \\
            Comm/Code Ratio & $3.23e^{-92}$ & -0.138 & Negl. & $\downarrow$ \\
            \bottomrule
            \end{tabular}%
        }
    \end{minipage}
    \hfill 
    \begin{minipage}[t]{0.49\textwidth}
        \centering
        \caption{\textcolor{blue}{Claude} Haiku 4.5 (Function)}
        \label{tab:claude45_func}
        \resizebox{\linewidth}{!}{%
            \begin{tabular}{|l|cccc|}
            \toprule
            \textbf{Feature} & \textbf{$p$-val} & \textbf{Cliff's $\delta$} & \textbf{Effect} & \textbf{Direction} \\
            \midrule
            Lines & $1.73e^{-207}$ & 0.209 & Small & $\uparrow$ \\
            Blank Lines & $< 10^{-300}$ & 0.375 & Med & $\uparrow$ \\
            Code Lines & $2.39e^{-257}$ & 0.232 & Small & $\uparrow$ \\
            Decl. Code Lines & $< 10^{-300}$ & 0.292 & Small & $\uparrow$ \\
            Exec. Code Lines & $4.79e^{-173}$ & 0.190 & Small & $\uparrow$ \\
            Comment Lines & $4.69e^{-02}$ & -0.013 & Negl. & $\downarrow$ \\
            Paths & $4.90e^{-106}$ & 0.147 & Negl. & $\uparrow$ \\
            Paths Log(x) & $2.13e^{-113}$ & 0.138 & Negl. & $\uparrow$ \\
            Statements & $7.41e^{-260}$ & 0.233 & Small & $\uparrow$ \\
            Decl. Stmts & $< 10^{-300}$ & 0.303 & Small & $\uparrow$ \\
            Exec. Stmts & $3.76e^{-154}$ & 0.179 & Small & $\uparrow$ \\
            Cyclomatic & $6.73e^{-125}$ & 0.159 & Small & $\uparrow$ \\
            Mod. Cyclomatic & $6.73e^{-125}$ & 0.159 & Small & $\uparrow$ \\
            Str. Cyclomatic & $1.80e^{-107}$ & 0.148 & Small & $\uparrow$ \\
            Str. Mod. Cyclo & $1.80e^{-107}$ & 0.148 & Small & $\uparrow$ \\
            Essential & $2.69e^{-46}$ & 0.078 & Negl. & $\uparrow$ \\
            Max Nesting & $6.32e^{-84}$ & 0.128 & Negl. & $\uparrow$ \\
            Comm/Code Ratio & $5.97e^{-178}$ & -0.193 & Small & $\downarrow$ \\
            \bottomrule
            \end{tabular}%
        }
    \end{minipage}
\end{table}

\begin{table}[htbp]
    \centering
    \begin{minipage}[t]{0.49\textwidth}
        \centering
        \caption{GPT-3.5 (Function)}
        \label{tab:gpt35_func}
        \resizebox{\linewidth}{!}{%
            \begin{tabular}{|l|cccc|}
            \toprule
            \textbf{Feature} & \textbf{$p$-val} & \textbf{$\delta$} & \textbf{Effect} & \textbf{Direction} \\
            \midrule
            Lines & $< 10^{-300}$ & -0.685 & Large & $\downarrow$ \\
            Blank Lines & $< 10^{-300}$ & -0.416 & Med & $\downarrow$ \\
            Code Lines & $< 10^{-300}$ & -0.493 & Large & $\downarrow$ \\
            Decl. Code Lines & $< 10^{-300}$ & -0.363 & Med & $\downarrow$ \\
            Exec. Code Lines & $< 10^{-300}$ & -0.506 & Large & $\downarrow$ \\
            Comment Lines & $< 10^{-300}$ & -0.811 & Large & $\downarrow$ \\
            Paths & $< 10^{-300}$ & -0.350 & Med & $\downarrow$ \\
            Paths Log(x) & $< 10^{-300}$ & -0.234 & Small & $\downarrow$ \\
            Statements & $< 10^{-300}$ & -0.463 & Med & $\downarrow$ \\
            Decl. Stmts & $< 10^{-300}$ & -0.373 & Med & $\downarrow$ \\
            Exec. Stmts & $< 10^{-300}$ & -0.467 & Med & $\downarrow$ \\
            Cyclomatic & $< 10^{-300}$ & -0.355 & Med & $\downarrow$ \\
            Mod. Cyclomatic & $< 10^{-300}$ & -0.355 & Med & $\downarrow$ \\
            Str. Cyclomatic & $< 10^{-300}$ & -0.367 & Med & $\downarrow$ \\
            Str. Mod. Cyclo & $< 10^{-300}$ & -0.367 & Med & $\downarrow$ \\
            Essential & $1.33e^{-204}$ & -0.142 & Negl. & $\downarrow$ \\
            Max Nesting & $< 10^{-300}$ & -0.323 & Small & $\downarrow$ \\
            Comm/Code Ratio & $< 10^{-300}$ & -0.557 & Large & $\downarrow$ \\
            \bottomrule
            \end{tabular}%
        }
    \end{minipage}
    \hfill
    \begin{minipage}[t]{0.49\textwidth}
        \centering
        \caption{GPT-OSS (Function)}
        \label{tab:gptoss_func}
        \resizebox{\linewidth}{!}{%
            \begin{tabular}{|l|cccc|}
            \toprule
            \textbf{Feature} & \textbf{$p$-val} & \textbf{$\delta$} & \textbf{Effect} & \textbf{Direction} \\
            \midrule
            Lines & $< 10^{-300}$ & 0.406 & Med & $\uparrow$ \\
            Blank Lines & $< 10^{-300}$ & 0.362 & Med & $\uparrow$ \\
            Code Lines & $< 10^{-300}$ & 0.316 & Small & $\uparrow$ \\
            Decl. Code Lines & $< 10^{-300}$ & 0.273 & Small & $\uparrow$ \\
            Exec. Code Lines & $< 10^{-300}$ & 0.312 & Small & $\uparrow$ \\
            Comment Lines & $< 10^{-300}$ & 0.454 & Med & $\uparrow$ \\
            Paths & $< 10^{-300}$ & 0.318 & Small & $\uparrow$ \\
            Paths Log(x) & $< 10^{-300}$ & 0.282 & Small & $\uparrow$ \\
            Statements & $< 10^{-300}$ & 0.345 & Med & $\uparrow$ \\
            Decl. Stmts & $< 10^{-300}$ & 0.262 & Small & $\uparrow$ \\
            Exec. Stmts & $< 10^{-300}$ & 0.339 & Med & $\uparrow$ \\
            Cyclomatic & $< 10^{-300}$ & 0.307 & Small & $\uparrow$ \\
            Mod. Cyclomatic & $< 10^{-300}$ & 0.307 & Small & $\uparrow$ \\
            Str. Cyclomatic & $< 10^{-300}$ & 0.299 & Small & $\uparrow$ \\
            Str. Mod. Cyclo & $< 10^{-300}$ & 0.299 & Small & $\uparrow$ \\
            Essential & $< 10^{-300}$ & 0.295 & Small & $\uparrow$ \\
            Max Nesting & $1.09e^{-209}$ & 0.202 & Small & $\uparrow$ \\
            Comm/Code Ratio & $2.19e^{-219}$ & 0.214 & Small & $\uparrow$ \\
            \bottomrule
            \end{tabular}%
        }
    \end{minipage}
\end{table}

\begin{table}[htbp]
    \centering
    \begin{minipage}[t]{0.49\textwidth}
        \centering
        \caption{Claude 3 Haiku (Class)}
        \label{tab:claude3_class}
        \resizebox{\linewidth}{!}{%
            \begin{tabular}{|l|cccc|}
            \toprule
            \textbf{Feature} & \textbf{$p$-val} & \textbf{$\delta$} & \textbf{Eff} & \textbf{Direction} \\
            \midrule
            Avg Lines & $< 10^{-300}$ & -0.298 & Sm & $\downarrow$ \\
            Avg Blank & $1.32e^{-170}$ & -0.127 & Neg & $\downarrow$ \\
            Avg Code & $< 10^{-300}$ & -0.314 & Sm & $\downarrow$ \\
            Avg Comment & $1.14e^{-59}$ & -0.113 & Neg & $\downarrow$ \\
            Avg Cyclo & $< 10^{-300}$ & -0.315 & Sm & $\downarrow$ \\
            Avg Mod Cyclo & $< 10^{-300}$ & -0.315 & Sm & $\downarrow$ \\
            Avg Essent & $3.46e^{-164}$ & -0.163 & Sm & $\downarrow$ \\
            Base Classes & $8.83e^{-02}$ & 0.004 & Neg & $\uparrow$ \\
            Coupled & $2.54e^{-124}$ & -0.174 & Sm & $\downarrow$ \\
            Derived & $3.88e^{-02}$ & 0.001 & Neg & $\uparrow$ \\
            Inst Methods & $3.74e^{-13}$ & 0.054 & Neg & $\uparrow$ \\
            Inst Vars & $3.49e^{-04}$ & -0.026 & Neg & $\downarrow$ \\
            Methods & $1.81e^{-15}$ & 0.059 & Neg & $\uparrow$ \\
            Lines & $7.19e^{-103}$ & -0.161 & Sm & $\downarrow$ \\
            Code Lines & $4.95e^{-127}$ & -0.179 & Sm & $\downarrow$ \\
            Comment Lines & $3.32e^{-58}$ & -0.120 & Neg & $\downarrow$ \\
            Statements & $3.95e^{-189}$ & -0.219 & Sm & $\downarrow$ \\
            Max Cyclo & $< 10^{-300}$ & -0.306 & Sm & $\downarrow$ \\
            Max Nesting & $< 10^{-300}$ & -0.246 & Sm & $\downarrow$ \\
            Comm/Code & $9.44e^{-04}$ & -0.025 & Neg & $\downarrow$ \\
            Sum Cyclo & $7.54e^{-80}$ & -0.141 & Neg & $\downarrow$ \\
            Sum Essential & $5.67e^{-11}$ & -0.049 & Neg & $\downarrow$ \\
            \bottomrule
            \end{tabular}%
        }
    \end{minipage}
    \hfill
    \begin{minipage}[t]{0.49\textwidth}
        \centering
        \caption{\textcolor{blue}{Claude} Haiku 4.5 (Class)}
        \label{tab:claude45_class}
        \resizebox{\linewidth}{!}{%
            \begin{tabular}{|l|cccc|}
            \toprule
            \textbf{Feature} & \textbf{$p$-val} & \textbf{$\delta$} & \textbf{Eff} & \textbf{Dir} \\
            \midrule
            Avg Lines & $2.04e^{-20}$ & 0.069 & Neg & $\uparrow$ \\
            Avg Blank & $< 10^{-300}$ & 0.302 & Sm & $\uparrow$ \\
            Avg Code & $1.03e^{-07}$ & 0.040 & Neg & $\uparrow$ \\
            Avg Comment & $4.09e^{-13}$ & 0.052 & Neg & $\uparrow$ \\
            Avg Cyclo & 0.959 & 0.000 & Neg & $\uparrow$ \\
            Avg Mod Cyclo & 0.943 & 0.001 & Neg & $\uparrow$ \\
            Avg Essent & $5.31e^{-03}$ & -0.018 & Neg & $\downarrow$ \\
            Base Classes & 0.607 & 0.001 & Neg & $\uparrow$ \\
            Coupled & $1.20e^{-03}$ & -0.024 & Neg & $\downarrow$ \\
            Derived & 0.405 & 0.000 & Neg & $\uparrow$ \\
            Inst Methods & $7.10e^{-21}$ & 0.070 & Neg & $\uparrow$ \\
            Inst Vars & $4.30e^{-06}$ & 0.034 & Neg & $\uparrow$ \\
            Methods & $1.35e^{-24}$ & 0.076 & Neg & $\uparrow$ \\
            Lines & $1.11e^{-28}$ & 0.083 & Neg & $\uparrow$ \\
            Code Lines & $7.97e^{-18}$ & 0.064 & Neg & $\uparrow$ \\
            Comment Lines & $2.24e^{-06}$ & 0.035 & Neg & $\uparrow$ \\
            Statements & 0.375 & 0.007 & Neg & $\uparrow$ \\
            Max Cyclo & 0.853 & 0.001 & Neg & $\uparrow$ \\
            Max Nesting & $6.98e^{-04}$ & 0.025 & Neg & $\uparrow$ \\
            Comm/Code & 0.866 & 0.001 & Neg & $\uparrow$ \\
            Sum Cyclo & $3.48e^{-13}$ & 0.054 & Neg & $\uparrow$ \\
            Sum Essential & $2.74e^{-10}$ & 0.047 & Neg & $\uparrow$ \\
            \bottomrule
            \end{tabular}%
        }
    \end{minipage}
\end{table}

\begin{table}[htbp]
    \centering
    \begin{minipage}[t]{0.49\textwidth}
        \centering
        \caption{GPT-3.5 (Class)}
        \label{tab:gpt35_class}
        \resizebox{\linewidth}{!}{%
            \begin{tabular}{|l|cccc|}
            \toprule
            \textbf{Feature} & \textbf{$p$-val} & \textbf{$\delta$} & \textbf{Eff} & \textbf{Direction} \\
            \midrule
            Avg Lines & $< 10^{-300}$ & -0.514 & Lg & $\downarrow$ \\
            Avg Blank & $< 10^{-300}$ & -0.185 & Sm & $\downarrow$ \\
            Avg Code & $< 10^{-300}$ & -0.567 & Lg & $\downarrow$ \\
            Avg Comment & $2.44e^{-176}$ & -0.192 & Sm & $\downarrow$ \\
            Avg Cyclo & $< 10^{-300}$ & -0.478 & Lg & $\downarrow$ \\
            Avg Mod Cyclo & $< 10^{-300}$ & -0.478 & Lg & $\downarrow$ \\
            Avg Essent & $< 10^{-300}$ & -0.226 & Sm & $\downarrow$ \\
            Base Classes & 0.365 & 0.002 & Neg & $\uparrow$ \\
            Coupled & $< 10^{-300}$ & -0.324 & Sm & $\downarrow$ \\
            Derived & 0.763 & 0.000 & Neg & $\uparrow$ \\
            Inst Methods & $6.41e^{-04}$ & 0.025 & Neg & $\uparrow$ \\
            Inst Vars & $3.08e^{-140}$ & -0.184 & Sm & $\downarrow$ \\
            Methods & $5.15e^{-04}$ & 0.026 & Neg & $\uparrow$ \\
            Lines & $< 10^{-300}$ & -0.328 & Sm & $\downarrow$ \\
            Code Lines & $< 10^{-300}$ & -0.392 & Med & $\downarrow$ \\
            Comment Lines & $2.45e^{-215}$ & -0.232 & Sm & $\downarrow$ \\
            Statements & $< 10^{-300}$ & -0.410 & Med & $\downarrow$ \\
            Max Cyclo & $< 10^{-300}$ & -0.522 & Lg & $\downarrow$ \\
            Max Nesting & $< 10^{-300}$ & -0.527 & Lg & $\downarrow$ \\
            Comm/Code & $3.19e^{-04}$ & -0.027 & Neg & $\downarrow$ \\
            Sum Cyclo & $2.10e^{-265}$ & -0.260 & Sm & $\downarrow$ \\
            Sum Essential & $1.54e^{-51}$ & -0.113 & Neg & $\downarrow$ \\
            \bottomrule
            \end{tabular}%
        }
    \end{minipage}
    \hfill
    \begin{minipage}[t]{0.49\textwidth}
        \centering
        \caption{GPT-OSS (Class)}
        \label{tab:gptoss_class}
        \resizebox{\linewidth}{!}{%
            \begin{tabular}{|l|cccc|}
            \toprule
            \textbf{Feature} & \textbf{$p$-val} & \textbf{$\delta$} & \textbf{Eff} & \textbf{Direction} \\
            \midrule
            Avg Lines & $1.32e^{-215}$ & 0.234 & Sm & $\uparrow$ \\
            Avg Blank & $1.44e^{-267}$ & 0.212 & Sm & $\uparrow$ \\
            Avg Code & $1.35e^{-43}$ & 0.103 & Neg & $\uparrow$ \\
            Avg Comment & $< 10^{-300}$ & 0.421 & Med & $\uparrow$ \\
            Avg Cyclo & $1.22e^{-04}$ & 0.028 & Neg & $\uparrow$ \\
            Avg Mod Cyclo & $1.12e^{-04}$ & 0.028 & Neg & $\uparrow$ \\
            Avg Essent & $1.76e^{-51}$ & 0.098 & Neg & $\uparrow$ \\
            Base Classes & $2.15e^{-05}$ & 0.010 & Neg & $\uparrow$ \\
            Coupled & $9.94e^{-214}$ & 0.231 & Sm & $\uparrow$ \\
            Derived & 0.763 & 0.000 & Neg & $\uparrow$ \\
            Inst Methods & $3.51e^{-51}$ & 0.112 & Neg & $\uparrow$ \\
            Inst Vars & 0.506 & 0.005 & Neg & $\uparrow$ \\
            Methods & $8.37e^{-70}$ & 0.132 & Neg & $\uparrow$ \\
            Lines & $7.68e^{-204}$ & 0.228 & Sm & $\uparrow$ \\
            Code Lines & $2.12e^{-72}$ & 0.135 & Neg & $\uparrow$ \\
            Comment Lines & $< 10^{-300}$ & 0.413 & Med & $\uparrow$ \\
            Statements & $2.69e^{-07}$ & 0.038 & Neg & $\uparrow$ \\
            Max Cyclo & $6.07e^{-12}$ & 0.051 & Neg & $\uparrow$ \\
            Max Nesting & $9.02e^{-10}$ & 0.045 & Neg & $\uparrow$ \\
            Comm/Code & $< 10^{-300}$ & 0.358 & Med & $\uparrow$ \\
            Sum Cyclo & $1.03e^{-41}$ & 0.101 & Neg & $\uparrow$ \\
            Sum Essential & $8.02e^{-87}$ & 0.148 & Sm & $\uparrow$ \\
            \bottomrule
            \end{tabular}%
        }
    \end{minipage}
\end{table}

\subsection{Rankings of Detection Models} \label{appendix:supplementary_ranking}
CB: CatBoost, XGB: XGBoost, also known as Extreme Gradient Boosting, MLP: Multi-Layer Perceptron, RF: Random Forest, KNN: K-Nearest Neighbours, SVM: Support Vector Machine, LR: Logistic Regression, NB: Naive Bayes

\begin{figure*}[h]
\centering
\begin{subfigure}[h]{0.48\textwidth}
    \includegraphics[width=\textwidth]{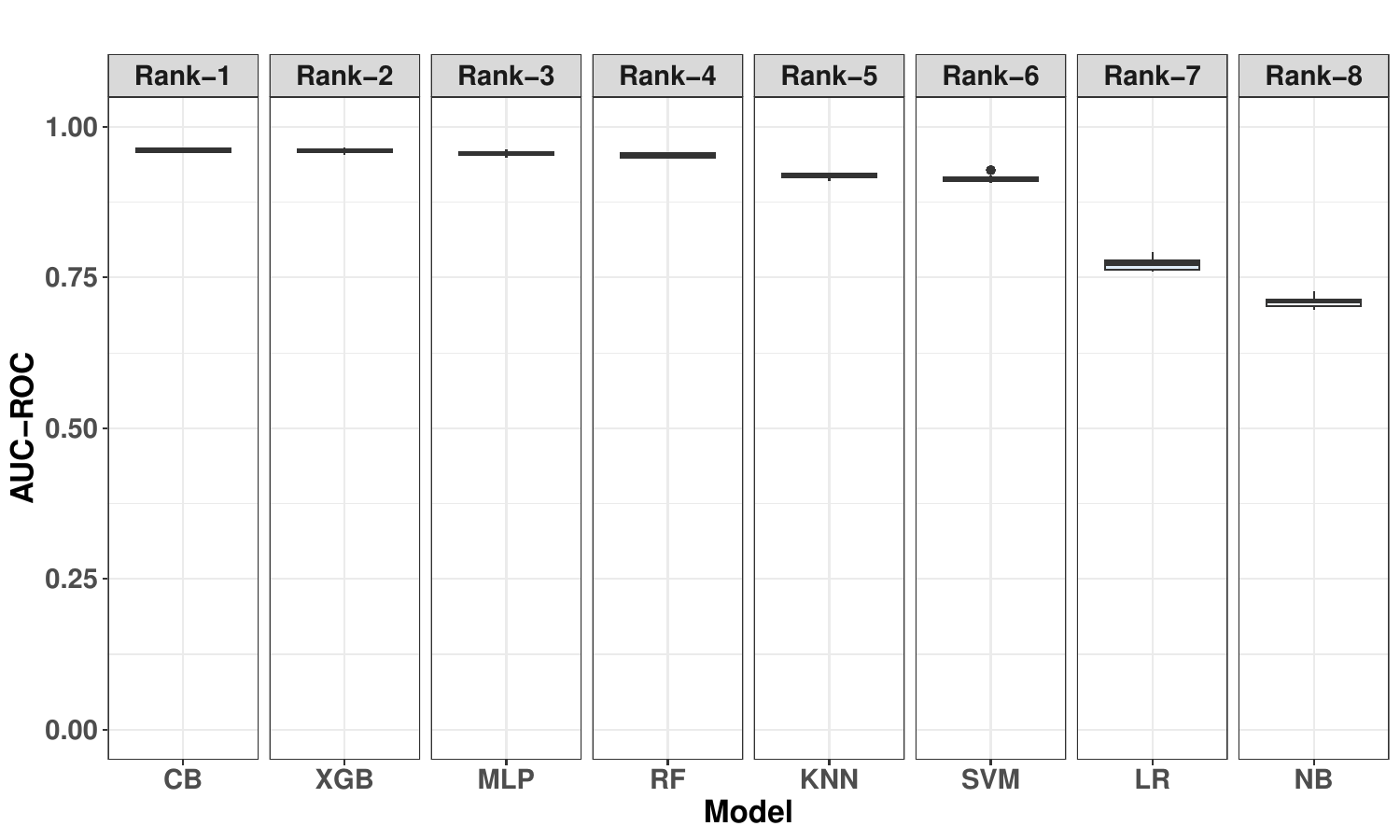}
    \caption{GPT-3.5 Function-level}
\end{subfigure}
\hfill
\begin{subfigure}[h]{0.48\textwidth}
    \includegraphics[width=\textwidth]{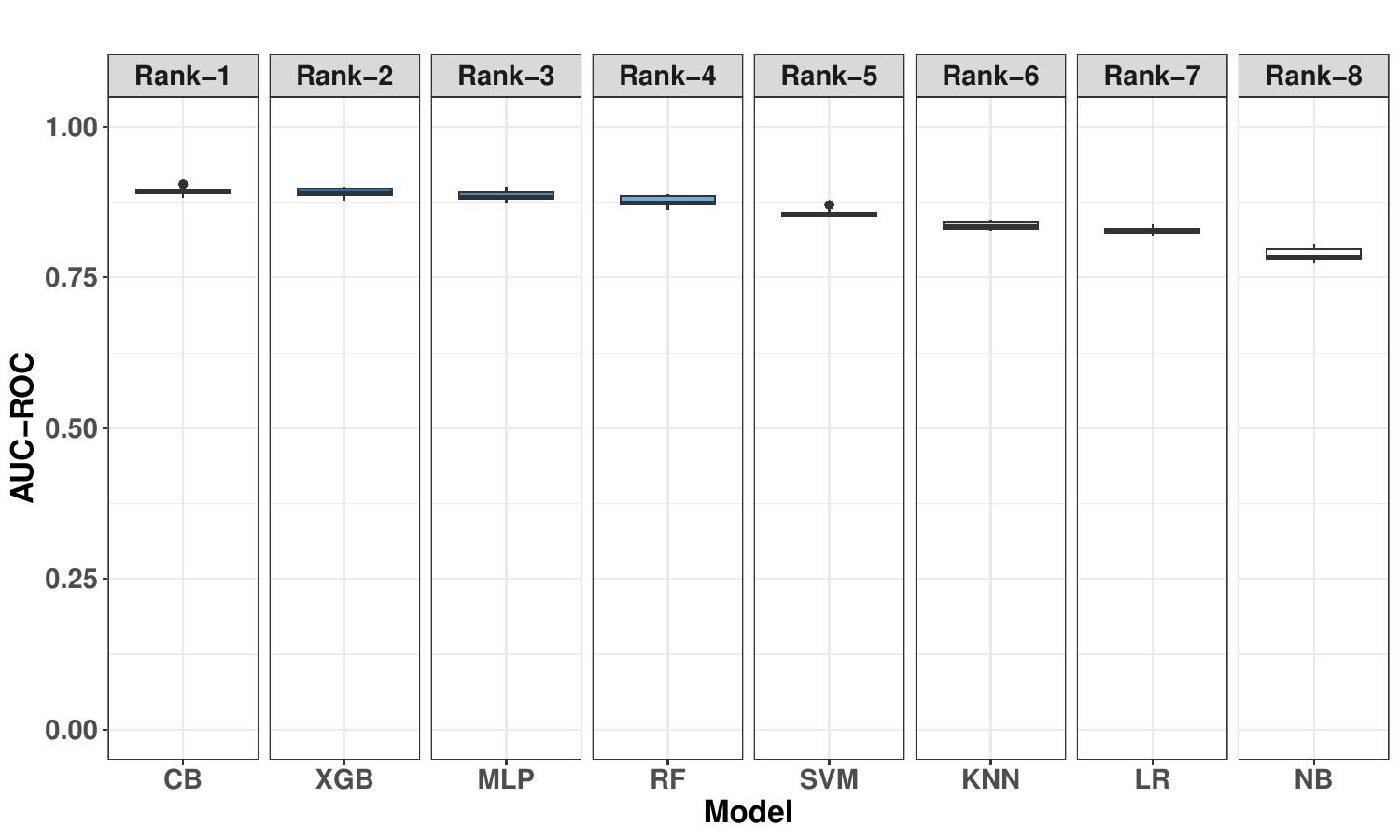}
    \caption{GPT-3.5 Class-level}
\end{subfigure}

\begin{subfigure}[h]{0.48\textwidth}
    \includegraphics[width=\textwidth]{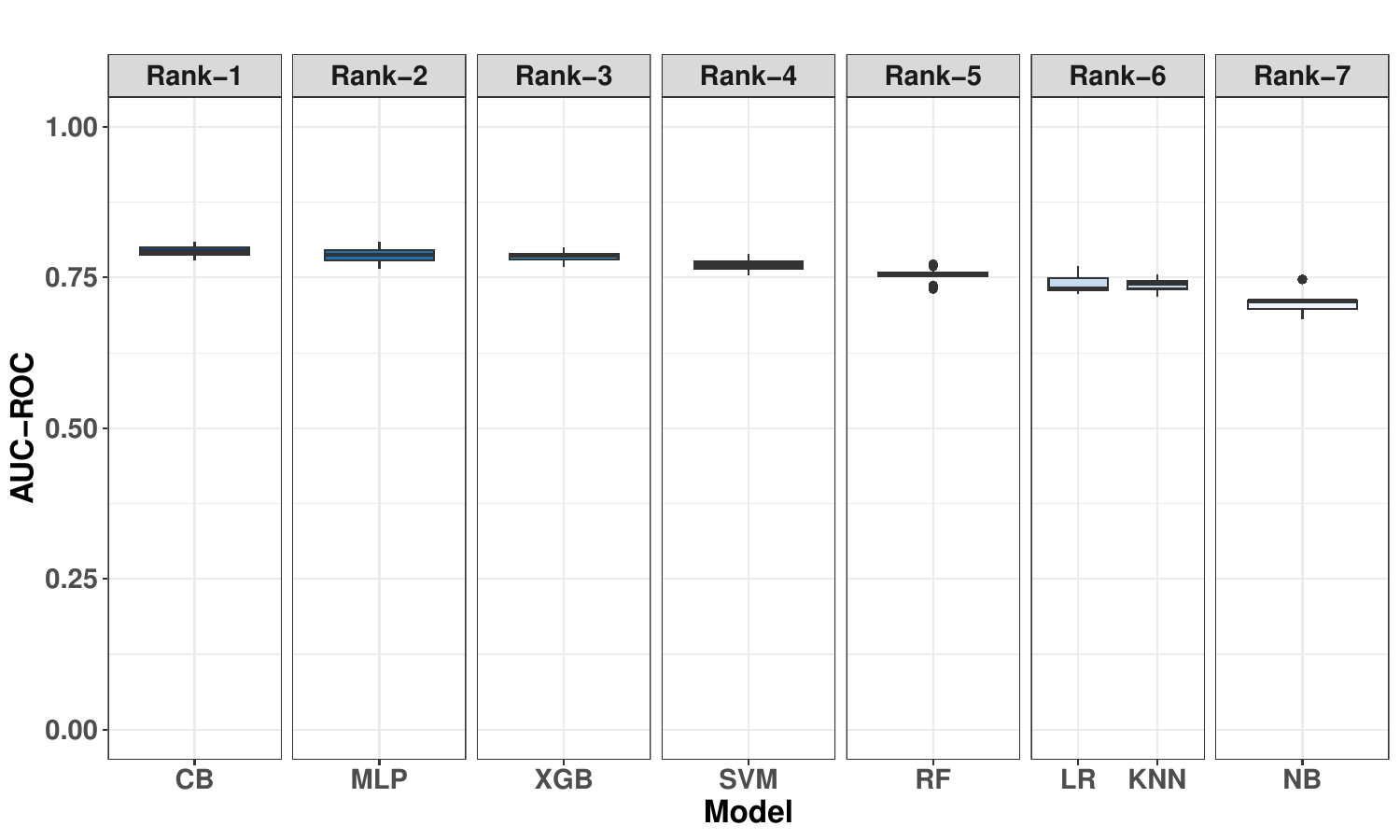}
    \caption{GPT-OSS Function-level}
\end{subfigure}
\hfill
\begin{subfigure}[h]{0.48\textwidth}
    \includegraphics[width=\textwidth]{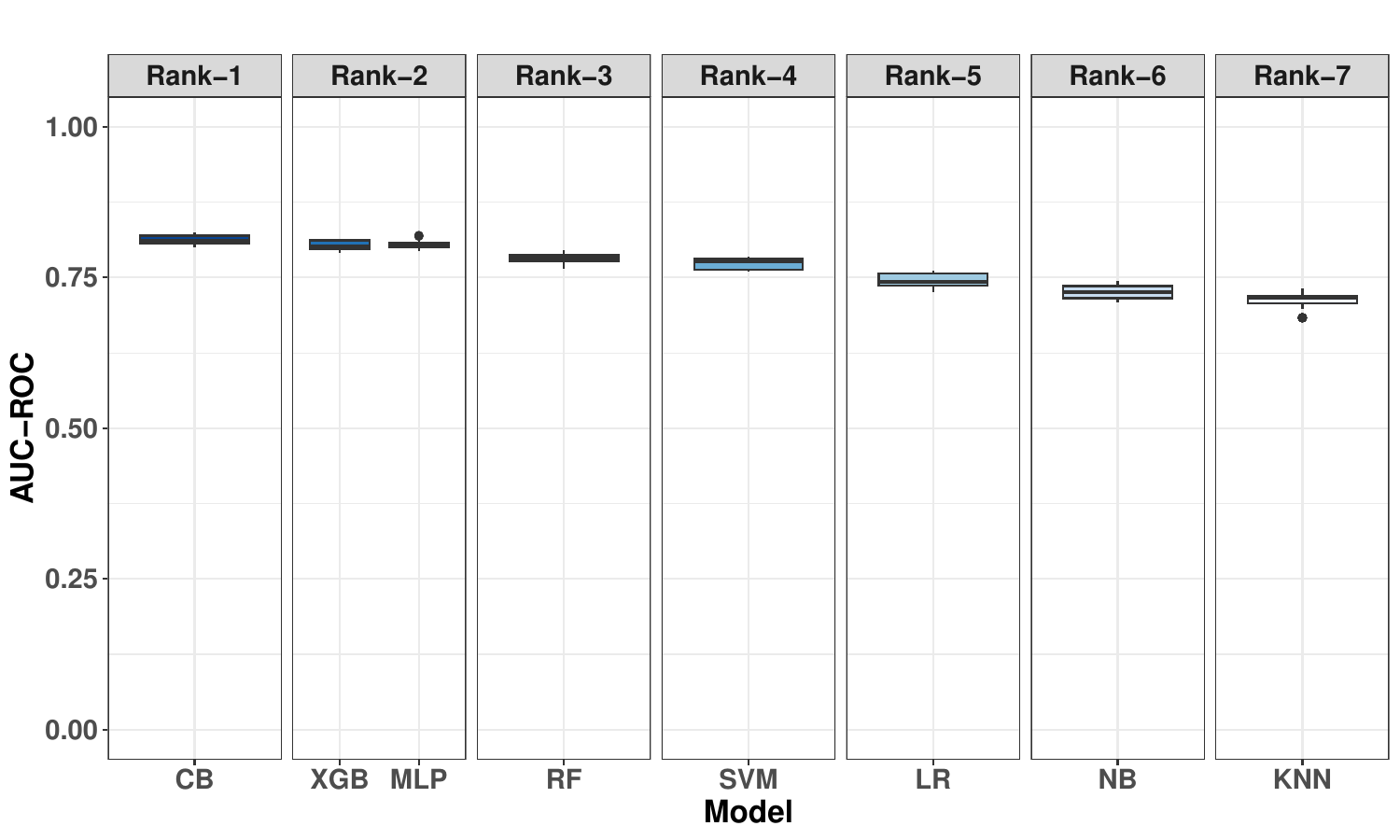}
    \caption{GPT-OSS Class-level}
\end{subfigure}

\begin{subfigure}[h]{0.48\textwidth}
    \includegraphics[width=\textwidth]{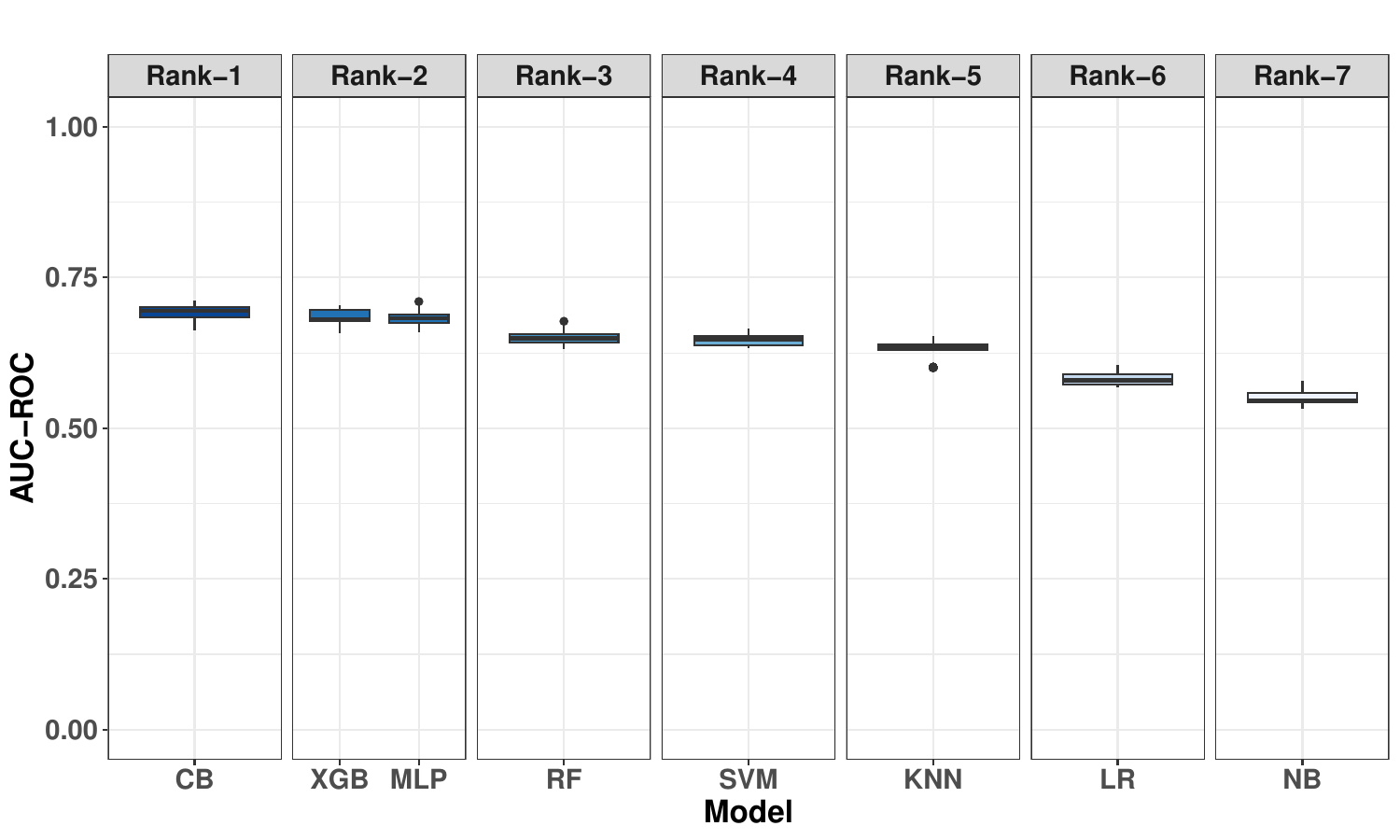}
    \caption{Claude 3 Haiku Function-level}
\end{subfigure}
\begin{subfigure}[h]{0.48\textwidth}
    \includegraphics[width=\textwidth]{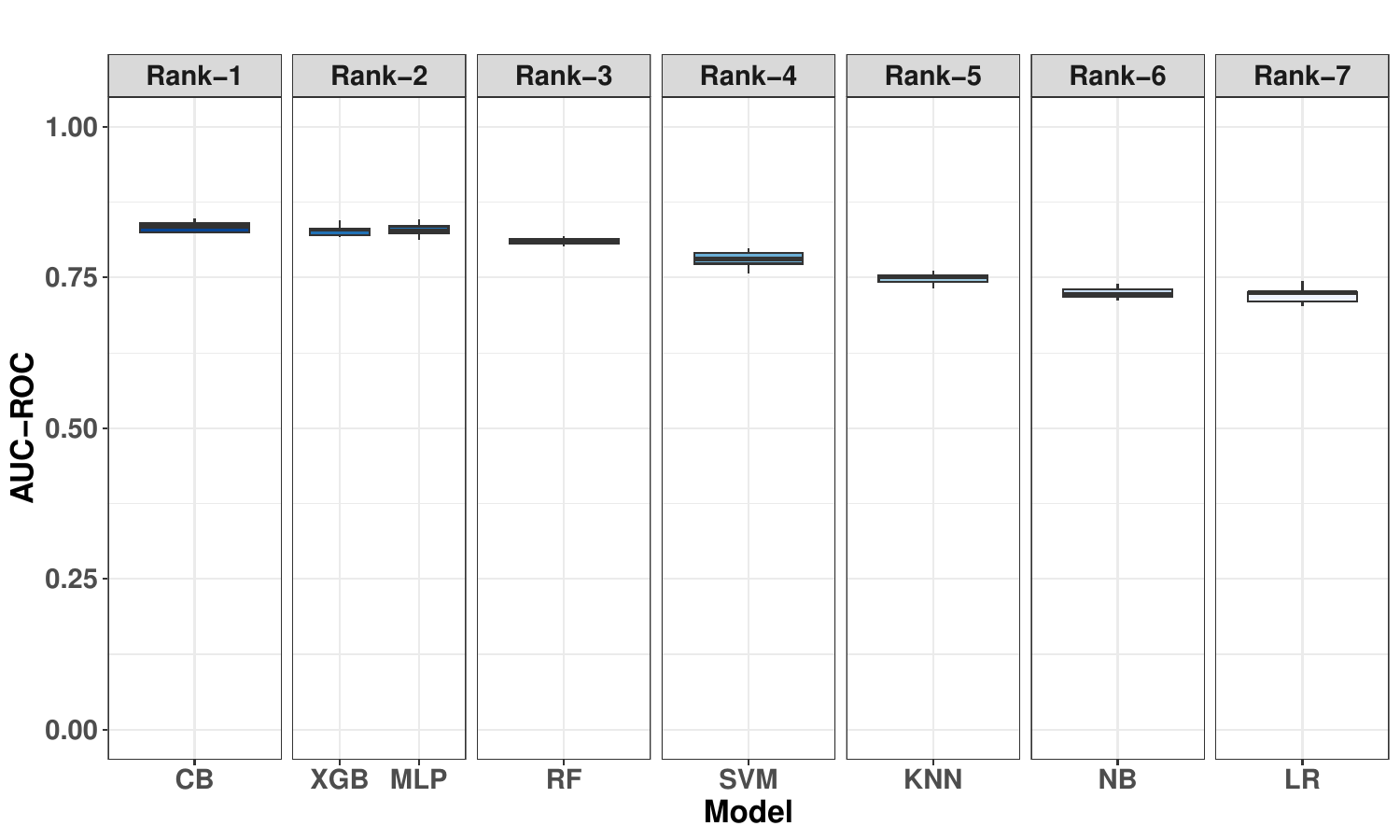}
    \caption{Claude 3 Haiku Class-level}
\end{subfigure}

\begin{subfigure}[h]{0.48\textwidth}
    \includegraphics[width=\textwidth]{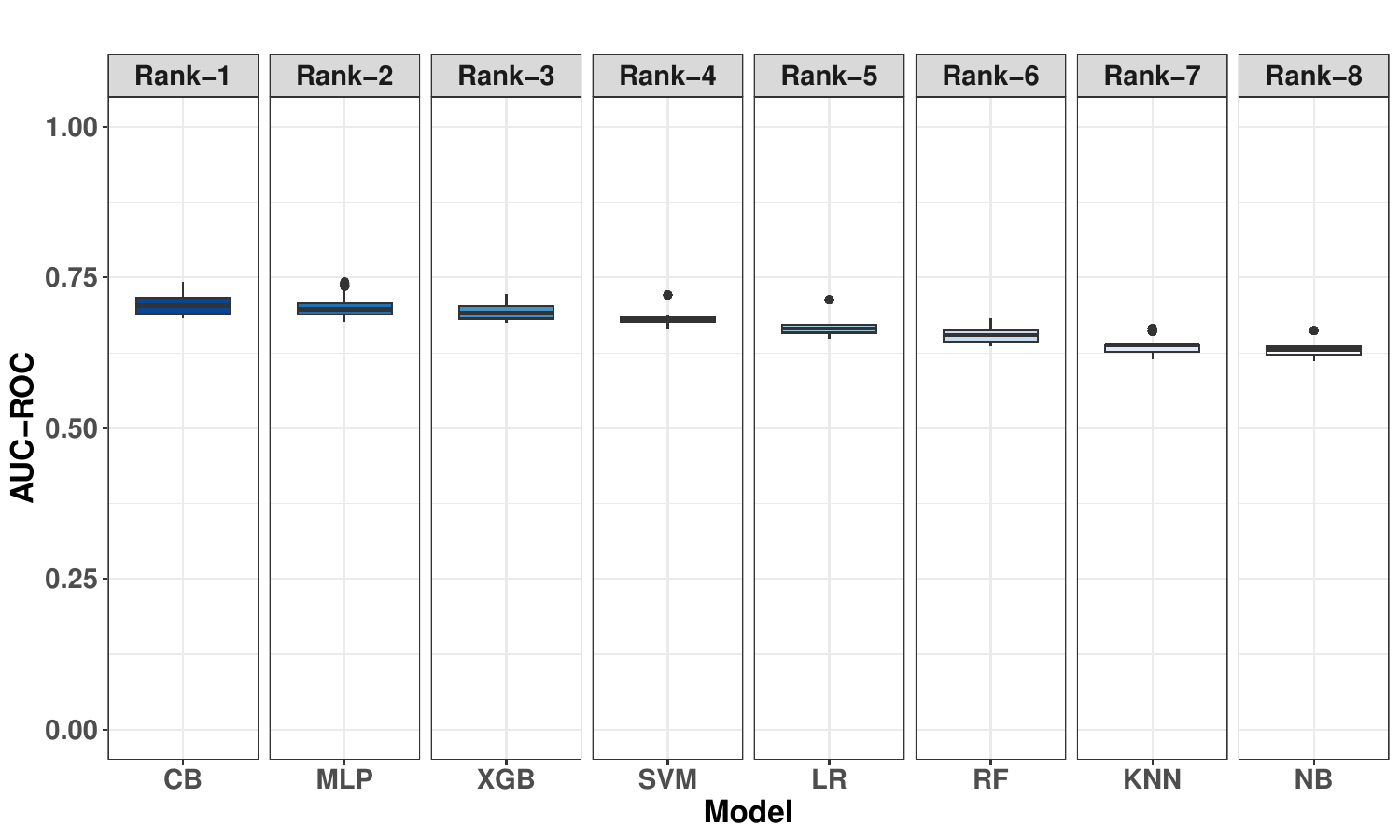}
    \caption{Claude Haiku 4.5 Function-level}
\end{subfigure}
\hfill
\begin{subfigure}[h]{0.48\textwidth}
    \includegraphics[width=\textwidth]{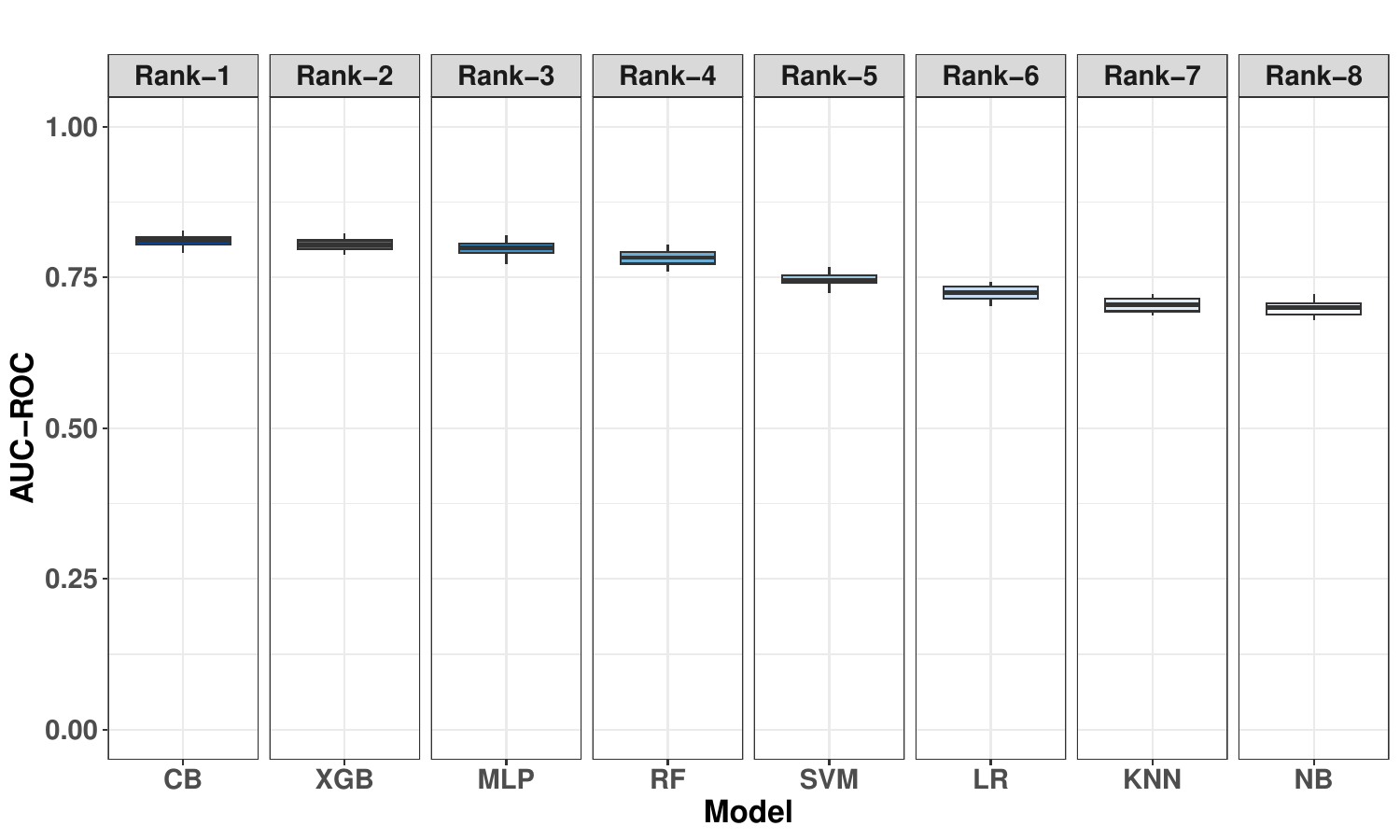}
    \caption{Claude Haiku 4.5 Class-level}
\end{subfigure}

\caption{Rankings based on ScottKnott ESD tests for each LLM-granularity configuration.}
\label{fig:rq3_shap_all}
\end{figure*}

\clearpage

\subsection{Detection Performance Comparison across Models}

\begin{table}[h]
\centering
\caption{Pairwise comparison of detection performance (AUC-ROC) across models. Statistical significance was determined using the DeLong test with Holm-Bonferroni correction.}
\label{tab:pairwise_auc_comparison}
\resizebox{\textwidth}{!}{%
\begin{tabular}{lllcccccc}
\toprule
\textbf{Granularity} & \textbf{Model A} & \textbf{Model B} & \textbf{AUC A} & \textbf{AUC B} & \textbf{$\Delta$ AUC} & \textbf{$Z$} & \textbf{$p$-value} & \textbf{Significant} \\
\midrule
\multirow{6}{*}{Function} 
 & GPT-3.5 & GPT-OSS & 0.961 & 0.795 & 0.166 & 25.52 & $< 10^{-5}$ & Yes \\
 & GPT-3.5 & \textcolor{blue}{Claude} Haiku 4.5 & 0.961 & 0.714 & 0.247 & 34.99 & $< 10^{-5}$ & Yes \\
 & GPT-3.5 & Claude 3 Haiku & 0.961 & \textcolor{blue}{0.681} & \textcolor{blue}{0.280} & 40.34 & $< 10^{-5}$ & Yes \\
 & GPT-OSS & \textcolor{blue}{Claude} Haiku 4.5 & 0.795 & 0.714 & 0.081 & 10.16 & $< 10^{-5}$ & Yes \\
 & GPT-OSS & Claude 3 Haiku & 0.795 & \textcolor{blue}{0.681} & \textcolor{blue}{0.114} & 11.60 & $< 10^{-5}$ & Yes \\
 & \textcolor{blue}{Claude} Haiku 4.5 & Claude 3 Haiku & 0.714 & \textcolor{blue}{0.681} & \textcolor{blue}{0.033} & 3.53 & $4.11\times 10^{-4}$ & Yes \\
\midrule
\multirow{6}{*}{Class} 
 & GPT-3.5 & Claude 3 Haiku & 0.887 & 0.830 & 0.058 & 11.46 & $< 10^{-5}$ & Yes \\
 & GPT-3.5 & GPT-OSS & 0.887 & 0.808 & 0.079 & 10.26 & $< 10^{-5}$ & Yes \\
 & GPT-3.5 & \textcolor{blue}{Claude} Haiku 4.5 & 0.887 & 0.806 & 0.081 & 12.18 & $< 10^{-5}$ & Yes \\
 & Claude 3 Haiku & GPT-OSS & 0.830 & 0.808 & 0.022 & 2.68 & 0.0074 & Yes \\
 & Claude 3 Haiku & \textcolor{blue}{Claude} Haiku 4.5 & 0.830 & 0.806 & 0.023 & 3.33 & $8.71\times 10^{-4}$ & Yes \\
 & GPT-OSS & \textcolor{blue}{Claude} Haiku 4.5 & 0.808 & 0.806 & 0.002 & 0.25 & 0.8054 & No \\
\bottomrule
\end{tabular}%
}
\end{table}

%

\subsection{Adversarial SHAP Comparisons for the Remaining Configurations}
\label{appendix:adversarial_shap}


\begin{table}[h]
\centering
\small
\caption{Top three features by baseline SHAP importance for the Claude 3 Haiku function-level classifier under the adversarial (comment-stripping) transformation. The dominant baseline feature, Comment to Code Ratio, grows almost six-fold in SHAP magnitude after the transformation, driving the AUC from 0.681 to 0.985.}
\label{tab:adv_shap_claude_3_haiku_function}
\resizebox{\textwidth}{!}{
\begin{tabular}{clcccc}
\toprule
\textbf{Rank} & \textbf{Feature} & \textbf{Baseline SHAP} & \textbf{Adversarial SHAP} & \textbf{$\Delta$} & \textbf{$\Delta$\%} \\
\midrule
1 & Comment to Code Ratio & 0.34 & 2.01 & +1.67 & +495\% \\
2 & Blank Lines & 0.29 & 0.33 & +0.04 & +14\% \\
3 & Declarative Statements & 0.19 & 0.21 & +0.02 & +12\% \\
\bottomrule
\end{tabular}}
\end{table}

\begin{table}[h]
\centering
\small
\caption{Top three features by baseline SHAP importance for the GPT-3.5 function-level classifier under the adversarial (comment-stripping) transformation. Comment to Code Ratio was already overwhelmingly dominant in the baseline; the transformation modestly increases it further, with AUC moving from 0.961 to 0.9997.}
\label{tab:adv_shap_gpt_3_5_function}
\resizebox{\textwidth}{!}{
\begin{tabular}{clcccc}
\toprule
\textbf{Rank} & \textbf{Feature} & \textbf{Baseline SHAP} & \textbf{Adversarial SHAP} & \textbf{$\Delta$} & \textbf{$\Delta$\%} \\
\midrule
1 & Comment to Code Ratio & 3.79 & 4.40 & +0.61 & +16\% \\
2 & Maximum Nesting & 0.70 & 0.56 & -0.14 & -21\% \\
3 & Declarative Code Lines & 0.63 & 0.51 & -0.12 & -19\% \\
\bottomrule
\end{tabular}}
\end{table}

\begin{table}[h]
\centering
\small
\caption{Top three features by baseline SHAP importance for the GPT-OSS function-level classifier under the adversarial (comment-stripping) transformation. Comment Lines becomes more dominant under the transformation, with AUC moving from 0.795 to 0.850.}
\label{tab:adv_shap_gpt_oss_function}
\resizebox{\textwidth}{!}{
\begin{tabular}{clcccc}
\toprule
\textbf{Rank} & \textbf{Feature} & \textbf{Baseline SHAP} & \textbf{Adversarial SHAP} & \textbf{$\Delta$} & \textbf{$\Delta$\%} \\
\midrule
1 & Comment Lines & 0.57 & 0.95 & +0.38 & +68\% \\
2 & Essential Complexity & 0.42 & 0.33 & -0.09 & -22\% \\
3 & Comment to Code Ratio & 0.32 & 0.49 & +0.17 & +52\% \\
\bottomrule
\end{tabular}}
\end{table}


\begin{table}[h]
\centering
\small
\caption{Top three features by baseline SHAP importance for the Claude 3 Haiku class-level classifier under the adversarial (comment-stripping) transformation. The dominant features are non-comment metrics, leaving the classifier largely unaffected; AUC moves only slightly from \textcolor{blue}{0.830} to 0.858.}
\label{tab:adv_shap_claude_3_haiku_class}
\resizebox{\textwidth}{!}{
\begin{tabular}{clcccc}
\toprule
\textbf{Rank} & \textbf{Feature} & \textbf{Baseline SHAP} & \textbf{Adversarial SHAP} & \textbf{$\Delta$} & \textbf{$\Delta$\%} \\
\midrule
1 & Average Code Lines & 0.60 & 0.63 & +0.03 & +5\% \\
2 & Coupled Classes (Modified) & 0.34 & 0.32 & -0.02 & -4\% \\
3 & Average Comment Lines & 0.32 & 0.33 & +0.02 & +5\% \\
\bottomrule
\end{tabular}}
\end{table}

\begin{table}[h]
\centering
\small
\caption{Top three features by baseline SHAP importance for the Claude Haiku 4.5 class-level classifier under the adversarial (comment-stripping) transformation. The dominant feature, Average Blank Lines, is non-comment and remains discriminative; AUC moves only slightly from 0.806 to 0.788.}
\label{tab:adv_shap_claude_haiku45_class}
\resizebox{\textwidth}{!}{
\begin{tabular}{clcccc}
\toprule
\textbf{Rank} & \textbf{Feature} & \textbf{Baseline SHAP} & \textbf{Adversarial SHAP} & \textbf{$\Delta$} & \textbf{$\Delta$\%} \\
\midrule
1 & Average Blank Lines & 1.81 & 2.21 & +0.40 & +22\% \\
2 & Average Comment Lines & 0.73 & 0.81 & +0.09 & +12\% \\
3 & Comment to Code Ratio & 0.28 & 0.29 & +0.01 & +4\% \\
\bottomrule
\end{tabular}}
\end{table}

\begin{table}[h]
\centering
\small
\caption{Top three features by baseline SHAP importance for the GPT-3.5 class-level classifier under the adversarial (comment-stripping) transformation. Maximum Nesting dominates and is unaffected by comment stripping; AUC moves only slightly from 0.887 to 0.903.}
\label{tab:adv_shap_gpt_3_5_class}
\resizebox{\textwidth}{!}{
\begin{tabular}{clcccc}
\toprule
\textbf{Rank} & \textbf{Feature} & \textbf{Baseline SHAP} & \textbf{Adversarial SHAP} & \textbf{$\Delta$} & \textbf{$\Delta$\%} \\
\midrule
1 & Maximum Nesting & 1.03 & 1.03 & -0.01 & -1\% \\
2 & Average Comment Lines & 0.74 & 0.78 & +0.04 & +5\% \\
3 & Comment to Code Ratio & 0.54 & 0.25 & -0.29 & -54\% \\
\bottomrule
\end{tabular}}
\end{table}

\end{document}